\documentclass{article}

\usepackage[english]{babel}

\usepackage[letterpaper,top=2cm,bottom=2cm,left=3cm,right=3cm,marginparwidth=1.75cm]{geometry}

\usepackage{amsmath}
\usepackage{graphicx}
\usepackage{xcolor}
\usepackage{color,soul}
\usepackage[colorlinks=true, allcolors=blue]{hyperref}
\usepackage{authblk}
\usepackage[left]{lineno}
\usepackage{array}
\usepackage{float}
\usepackage{braket}
\usepackage{changepage}
\usepackage[inkscapelatex=false]{svg}
\usepackage[absolute]{textpos}
\usepackage{xcolor}
\usepackage{comment}
\usepackage[normalem]{ulem}
\usepackage{units}
\usepackage{multirow}
\usepackage{amsfonts}
\usepackage[section]{placeins}

\title{Drone- and Vehicle-Based Quantum Key Distribution}

\author[a,b]{Andrew Conrad*$^{\dagger,}$}
\author[c]{Roderick Cochran$^{\dagger,}$}
\author[c]{Daniel Sanchez-Rosales$^{\dagger,}$}
\author[a]{Samantha Isaac$^{\dagger,}$}
\author[a,b]{Timur Javid}
\author[a]{Tahereh Rezaei}
\author[a,b]{A.J. Schroeder}
\author[a]{Grzegorz Golba}
\author[c]{Akash Gutha}
\author[a,b]{Brian Wilens}
\author[a]{Kyle Herndon}
\author[a]{Alex Hill}
\author[e]{Joseph Chapman}
\author[a]{Ian Call}
\author[c]{Joseph Szabo}
\author[d]{Aodhan Corrigan}
\author[d]{Lars Kamin}
\author[d]{Norbert Lütkenhaus}
\author[c]{Daniel J. Gauthier}
\author[a,b]{Paul G. Kwiat}

\affil[a]{Department of Physics, The Grainger College of Engineering, University of Illinois at Urbana-Champaign (UIUC), 1110 W Green St. Loomis Laboratory, Urbana, IL 61801, USA}
\affil[b]{Department of Electrical and Computer Engineering (ECE), The Grainger College of Engineering, University of Illinois at Urbana-Champaign (UIUC), 306 N. Wright St., Urbana, IL 61801, USA}
\affil[c]{Department of Physics, The Ohio State University, 191 W Woodruff Ave, Columbus, OH 43210, USA}
\affil[d]{Institute for Quantum Computing and Department of Physics and Astronomy, 200 University Avenue West, Waterloo, ON, Canada}
\affil[e]{Quantum Information Science Section, Oak Ridge National Laboratory, Oak Ridge, Tennessee 37831, USA}
\affil[$\dagger$]{Equal Contribution}
\affil[*]{Corresponding Author}

\begin{document}
\maketitle

\begin{abstract}
Quantum key distribution is a point-to-point communication protocol that leverages quantum mechanics to enable secure information exchange. Commonly, the transmitter and receiver stations are at fixed locations, and the single-photon quantum states are transmitted over fiber or free space. Here, we describe a modular, platform-agnostic, quantum key distribution transmitter and receiver with reduced size, weight, and power consumption to realize a mobile quantum communication system. We deploy the system on different moving platforms, demonstrating drone-to-drone, drone-to-vehicle, and vehicle-to-vehicle quantum communication, achieving secure key rates in the finite-key regime in the range of \unit[$1.6 - 20$]{kbps}.  To prove the security of the system, we develop advanced physics models of the devices that account for non-ideal behaviors that are of greater importance in mobile platforms.  The modular system can be easily upgraded to include sources of entangled photonic quantum states, which will find application in future quantum networks.
\end{abstract}

\begin{textblock}{12.13}(1.94,15)
\noindent\fontsize{7}{7}\selectfont \textcolor{black!30}{This manuscript has been co-authored by UT-Battelle, LLC, under contract DE-AC05-00OR22725 with the US Department of Energy (DOE). The US government retains and the publisher, by accepting the article for publication, acknowledges that the US government retains a nonexclusive, paid-up, irrevocable, worldwide license to publish or reproduce the published form of this manuscript, or allow others to do so, for US government purposes. DOE will provide public access to these results of federally sponsored research in accordance with the DOE Public Access Plan (http://energy.gov/downloads/doe-public-access-plan).}
\end{textblock}

\section{Introduction}
The future global quantum internet \cite{liu2025road,li2024survey} will interface quantum computers \cite{cacciapuoti2019quantum} and quantum sensor networks \cite{khabiboulline2019quantum} that will enable new applications that are not possible with existing classical networks. The quantum internet is still under development, where proof-of-concept entanglement-based metropolitan scale   \cite{liu2024creation,stolk2024metropolitan}, and space-to-stationary ground station \cite{yin2017satellite} networks have been demonstrated.  In these systems, quantum information is encoded on photonic states and transported over fiber-optic cables or free-space links between non-reconfigurable nodes \cite{elliott2005current}.  A current challenge is to demonstrate quantum communication from mobile platforms that are ``on-the-go,'' such as drones or vehicles, to realize reconfigurable networks.

\begin{figure} [h]
    \begin{center}
    \begin{tabular}{c} 
    \includegraphics[height=10cm]{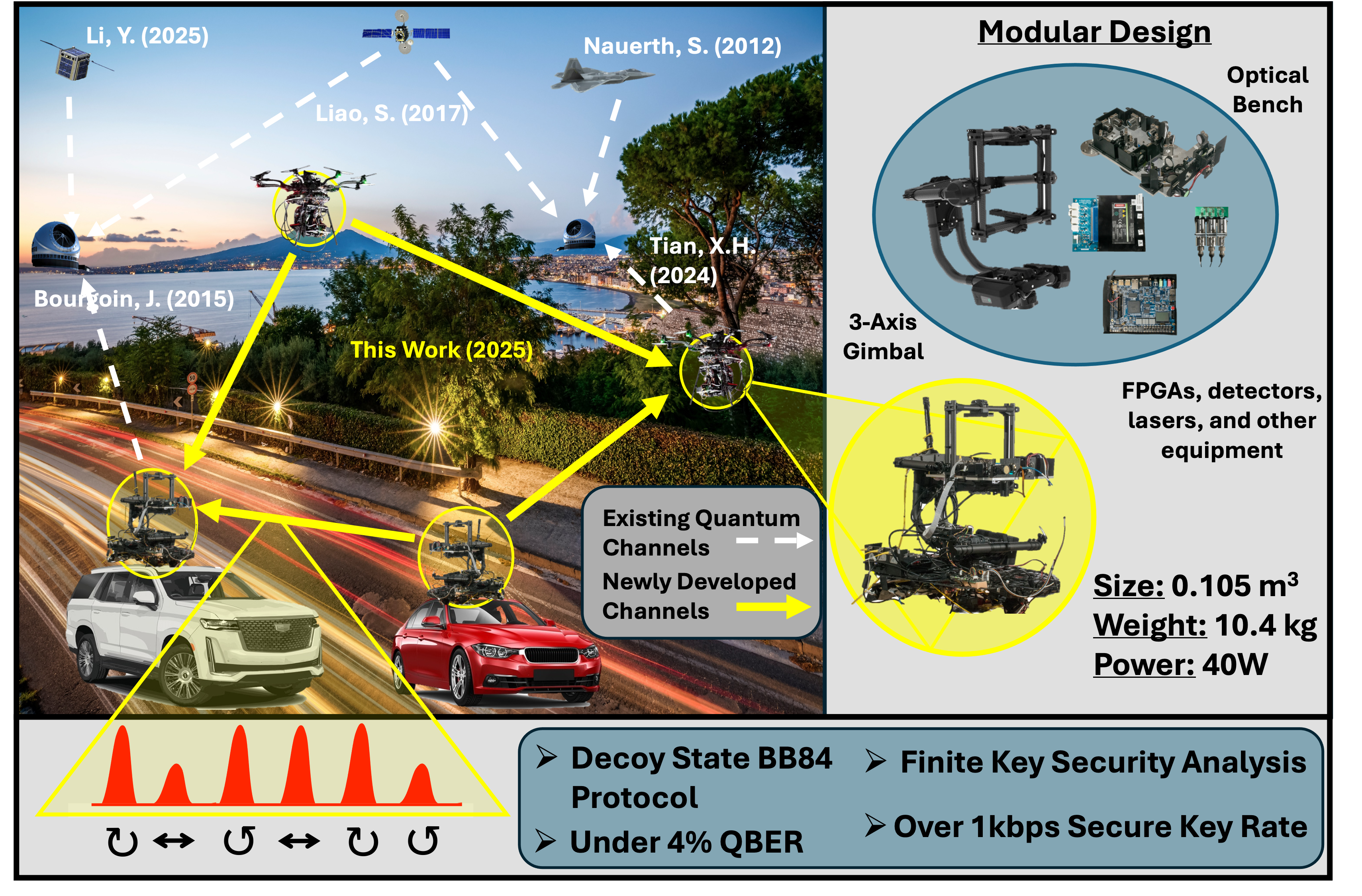} 
    \end{tabular}
    \end{center}
    \caption{\textbf{Mobile quantum communication systems.} The modular system architecture enables rapid deployment of QKD links between mobile platforms such as drones and vehicles.  The top left panel shows the configurations we demonstrate (yellow arrows) and QKD configurations demonstrated by others.  The top right panel shows the modular QKD payload. The bottom panel shows our BB84 decoy-state polarization-based protocol, where the transmitted states $\ket{R},\ket{L}$ (right and left circular polarization, respectively) encode the information, and the state $\ket{H}$ (horizontal polarization) is used for error checking. The mean photon number of the wave packets is adjusted randomly for the decoy states.} \label{fig:Overview}   
\end{figure}

Here, we demonstrate a polarization-based quantum key distribution (QKD) system between mobile platforms, enabling information-theoretic secure communication between two parties. Figure \ref{fig:Overview} illustrates our approach based on a modular transmitter and receiver, which includes the pointing-and-tracking (PAT) system for maintaining the quantum photonic link.  The devices can be swapped in under an hour between various mobile platforms, allowing rapid network reconfiguration. We characterize many device non-idealities and include them in the system security analysis, where we demonstrate secure communication, even for the brief key exchange sessions.  Quantum key distribution does not require entangled states, but the overall system design and technologies are similar \cite{liu2020drone,liu2021optical,tu2022lower}; hence, our demonstration is a stepping stone for developing an entanglement-based mobile quantum internet.  

Prior mobile QKD systems have been demonstrated for space- \cite{liao2018satellite,li2025microsatellite}, air- \cite{nauerth2012air,pugh2017airborne}, drone-ground \cite{PhysRevLett.133.200801}, drone-vehicle under relative motion \cite{zhou2026long}, and ground-vehicle \cite{bourgoin2015free} platforms.  These works highlight the difficulty in proving system security: information exchange sessions are brief, and the non-ideal behaviors of the equipment take on greater importance for mobile platforms.  The security analyses for these systems are typically simplified by predicting the secure key rate in the limit of an infinitely long exchange session and for idealized equipment, obviously both invalid assumptions in practice. 

In contrast, we report QKD links between fully mobile platforms: drone-to-drone, drone-to-vehicle, and vehicle-to-vehicle.  In the Results section below, we give the systems' operating characteristics and secure key rates for each configuration, using a custom finite key analysis that accounts for non-ideal system behaviors.  We then discuss the implications of our work and how it can be further extended.  The Methods section briefly summarizes the technology developed for this demonstration and the improved security analysis. Details of the drone platforms are in Supplementary Note \hyperref[chap:Note_1]{1}. Details of the QKD source, QKD source characterization, and optimization are presented in Supplementary Notes \hyperref[chap:Note_2]{2}, \hyperref[chap:Note_3]{3} and \hyperref[chap:Note_4]{4}, respectively. Full QKD results are presented in Supplementary Note \hyperref[chap:Note_5]{5}. The modular systems and optical payload, and PAT systems are shown in Supplementary Notes \hyperref[chap:Note_6]{6} and \hyperref[chap:Note_7]{7}, respectively. Single-photon detectors, time taggers, and post-processing are presented in Supplementary Notes \hyperref[chap:Note_8]{8}, \hyperref[chap:Note_9]{9}, and \hyperref[chap:Note_10]{10}, respectively. QKD channel modeling and safety considerations are presented in Supplementary Notes \hyperref[chap:Note_11]{11} and \hyperref[chap:Note_12]{12}, respectively. Finally, daylight operations, drone power tether, and relative motion at a simulated traffic intersection are presented in Supplementary Notes \hyperref[chap:Note_13]{13}, \hyperref[chap:Note_14]{14}, and \hyperref[chap:Note_15]{15} respectively. 

\section{Results}
We performed multiple QKD experiments with the mobile platforms in different configurations: a) tabletop, b) ground-to-ground, c) drone-to-drone, d) drone-to-vehicle, e) vehicle-to-vehicle \unit[5]{miles/hr (mph)} (\unit[8]{km/hr}), and f) vehicle-to-vehicle \unit[70]{mph} (\unit[113]{km/hr}) on a U.S. Interstate Highway.  Briefly, the quantum transmitter encodes the key and checks for eavesdropping using three quantum polarization states (see Fig.~\ref{fig:Overview}), and the receiver measures all four states of the complete basis using single-photon counting detectors, simplifying the setup while maintaining the same key rate and security \cite{tamaki2014loss}. Because we are using attenuated LEDs to create the states, i.e., not true single-photon states, we employ decoy-states \cite{lo2005decoy} to prevent the photon-number splitting attack. We choose circular polarization states $\ket{R}$ and $\ket{L}$ for the QKD key-generation basis due to their immunity from bit errors caused by potential in-flight platform rotations, and $\ket{H}$ for error checking. We transmit three different intensities: signal, decoy, and vacuum; the probability distribution of decoy intensity levels is optimized using in-flight transmission data to maximize the secret key rate in the finite-key regime as discussed in Supplementary Note \hyperref[chap:Note_4]{4}.  

Table \ref{table:Best_Results_Table} summarizes the best results for each link configuration. We produce a secure key in the finite-size regime using only a single flight/drive in every configuration. Supplementary Note \hyperref[chap:Note_5]{5} summarizes the results for all our QKD experiments.  

\begin{table}[h]
\begin{adjustwidth}{0cm}{0cm}
\begin{center}
\caption{QKD results for the highest secure key rate$\dagger$ for each link configuration. These results are averaged over the experiment runtime. 
The Ground-to-Ground data is collected while the QKD payloads are mounted to stationary non-airborne drones separated by \unit[10]{m}.} 
\label{table:Best_Results_Table}
\begin{tabular}
{||c|m{0.23\linewidth}|m{0.08\linewidth}|m{0.05\linewidth}|m{0.07\linewidth}|m{0.05\linewidth}|m{0.06\linewidth}|m{0.05\linewidth}||} 
 \hline
 Configuration & Experiment & Signal Mean Photon Number & Flight Time (s) & Average QBER (\%) & Raw Key Rate (kbps) & Secure Key Rate$\dagger$ (kbps) & Total \newline Secret Key (kb)\\ 
 \hline\hline
 Tabletop & Run 4 (Aug. 8, 2023) & 0.43 & 93.7 & 2.87 & 500.5 & 41.5 & 3,885\\ 
 \hline
 Ground-to-Ground & Run 1 (Nov. 14, 2023) & 0.44 & 209.1 & 2.83 & 164.6 & 6.1 & 1,278\\
 \hline
 Air-to-Air & Flight 2 (Nov. 2, 2022) & 0.78 & 166.4 & 2.45 & 255.5 & 8.5 & 1,414\\
 \hline
 Air-to-Vehicle & Run 1 (Nov. 14, 2023) & 0.44 & 129.8 & 3.43 & 113.0 & 1.6 & 205 \\
 \hline
 Vehicle-to-Vehicle & 5 mph, Illinois Center for Transportation Run 3 (Mar. 27, 2024) & 0.37 & 93.7 & 2.58 & 235.0 & 20.0 & 1,876 \\
 \hline
 Vehicle-to-Vehicle & 70~mph Interstate Highway Run 2 (Apr. 2, 2024) & 0.54 & 209.1 & 3.08 & 45.7 & 2.5 & 519\\
 \hline
\end{tabular}\\
$\dagger$ Custom Finite Key Analysis
\end{center}
\end{adjustwidth}
\end{table}

\subsection{Drone-to-Drone Configuration}

\begin{figure} [h]
    \begin{center}
    \begin{tabular}{c} 
    \includegraphics[height=7cm]{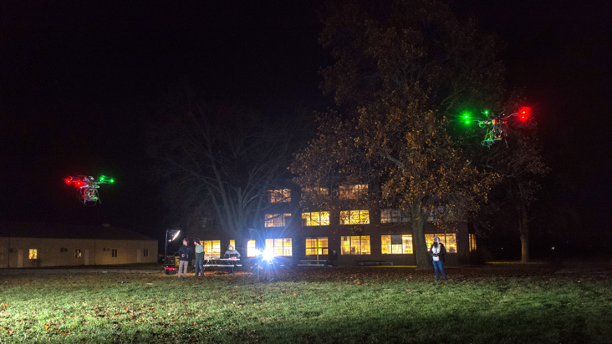} \\
    (a) \\
    \end{tabular}
    \begin{tabular}{c c} 
    \hspace*{-1cm}
    \includegraphics[height=5.5cm]{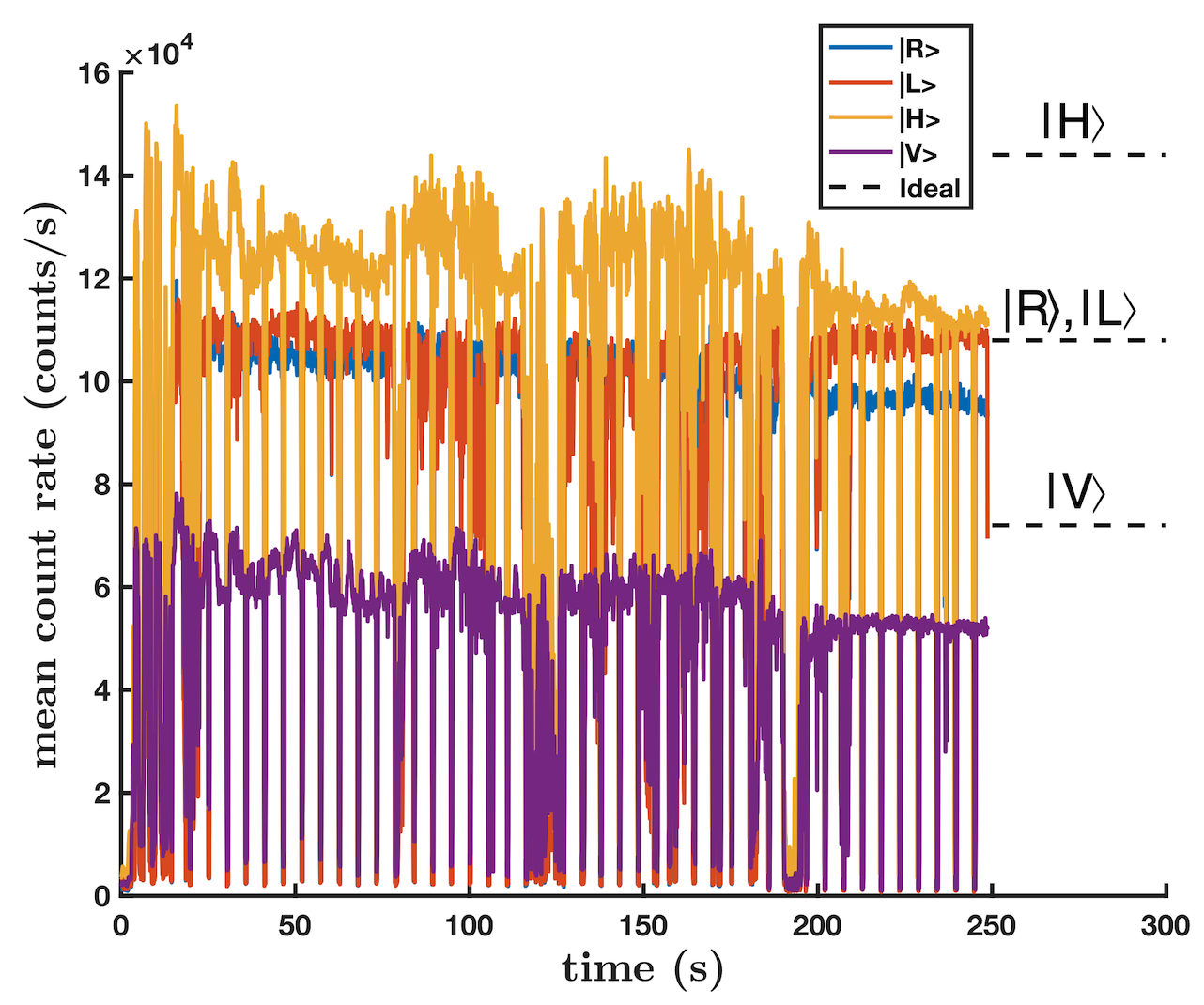} & \hspace*{-1cm} \includegraphics[height=5.5cm]{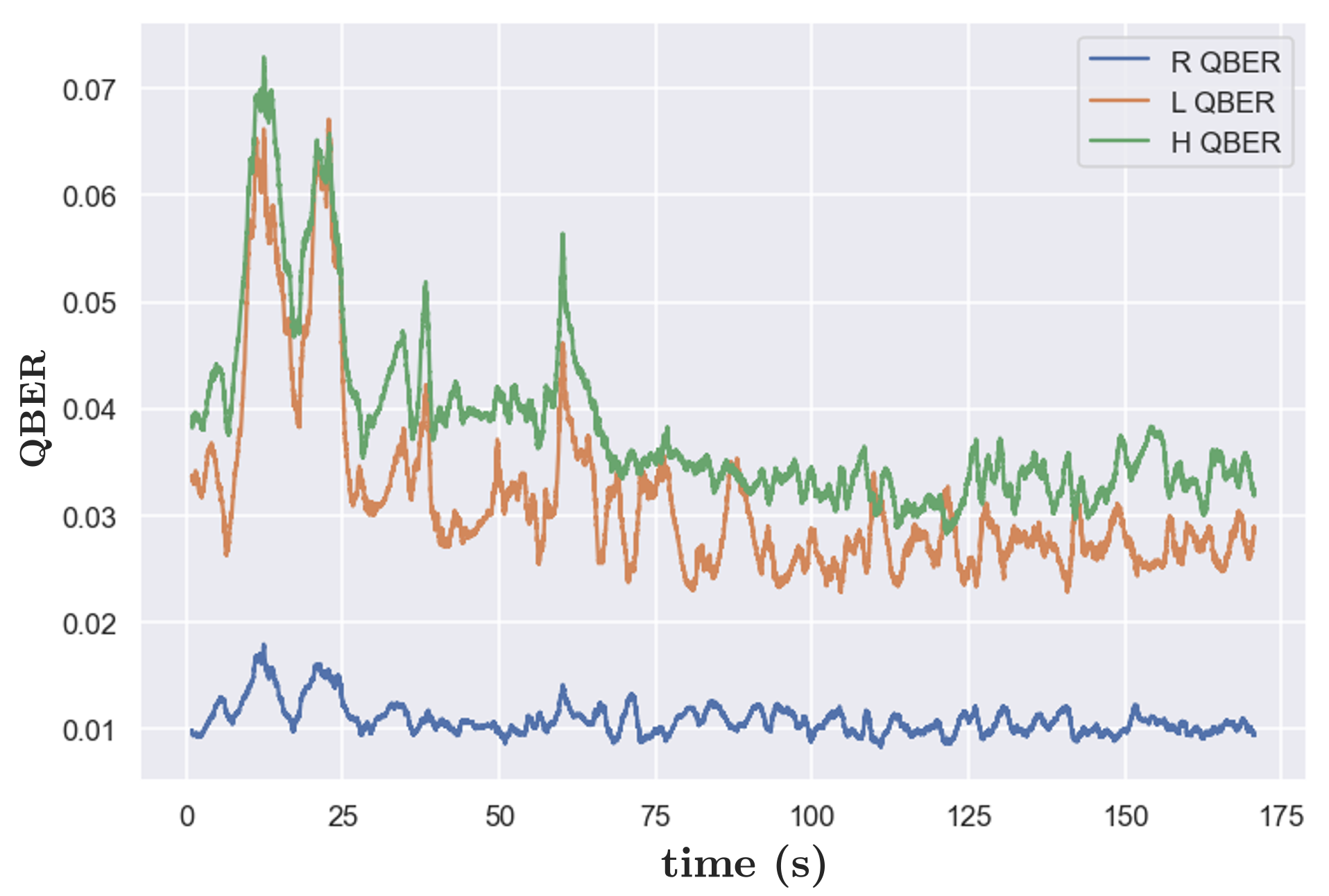}\\
    (b) & (c)\\
    \end{tabular}
    \end{center}
    \caption{\label{fig:Air2Air}\textbf{Quantum communication between drones.} Air-to-air drone-based QKD. (a) Air-to-air experimental setup, (b) temporal evolution of the QKD transmission (0.1-sec integration time) where the ideal proportions are shown in dashed lines, (c) QBER trends for the different input states over the course of one flight (different than shown in (b)). (Photo courtesy Alan Mitchell and Timur Javid.)}
\end{figure}

Figure \ref{fig:Air2Air}a) shows a photo of the drones in flight during a drone-to-drone QKD session, where they are hovering at an altitude 
of \unit[$\sim$4]{m} and separated by \unit[$\sim$10]{m}. The mean count rate averaged over \unit[0.1]{s} in each polarization channel is shown in Fig.~\ref{fig:Air2Air}(b). The QKD transmitter is designed to transmit in a pattern for \unit[4.86]{s} and then shut off for \unit[0.5]{s} to aid in time synchronization and to afford time to save the data to disk, resulting in periodic signal drops in Fig.~\ref{fig:Air2Air}b). The dashed horizontal lines show the expected count rates if the apparatus was ideal; deviations arise due to system imperfections, such as variable link transmission, mismatched detector efficiencies, different decoy-state mean photon numbers, etc. The mean count rate also shows dips, indicating decreasing overall transmission. Furthermore, the ratio of the counts from channel-to-channel varies, indicating a state-dependent detection rate. 

Variations in count rate and inefficiency with respect to the sending states are present in any QKD system. However, it is more apparent for the mobile platform and must be accounted for in the security analysis.  Supplementary Note \hyperref[chap:Note_11]{11} describes a system model that accounts for these experimental non-idealities. The analysis applies to a finite-duration key exchange session, assuming Eve performs a collective attack. Based on this analysis, the data shown in Fig.~\ref{fig:Air2Air}(b) reports an average secret key rate of \unit[8.5]{kbps} and a total secret key of \unit[$1.4$]{Mb}. Additional flight QKD results are included in Supplementary Note \hyperref[chap:Note_5]{5}.

\subsection{Drone-to-Vehicle Configuration}

\begin{figure}[h]
    \begin{center}
    \begin{tabular}{c} 
    \includegraphics[height=5cm]{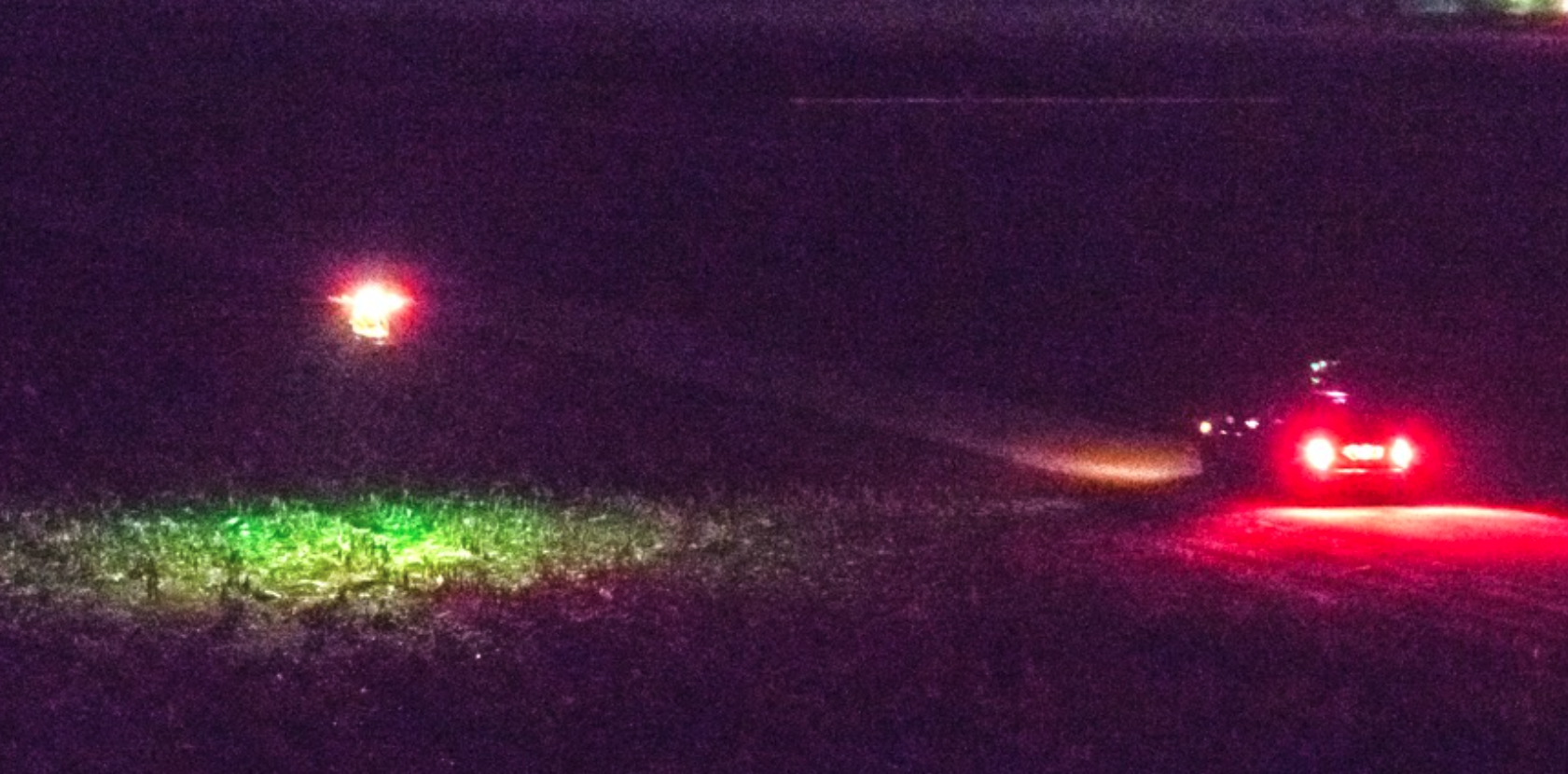} \\
    (a)
    \end{tabular}
    \begin{tabular}[c]{c c} 
    \includegraphics[height=6.5cm] {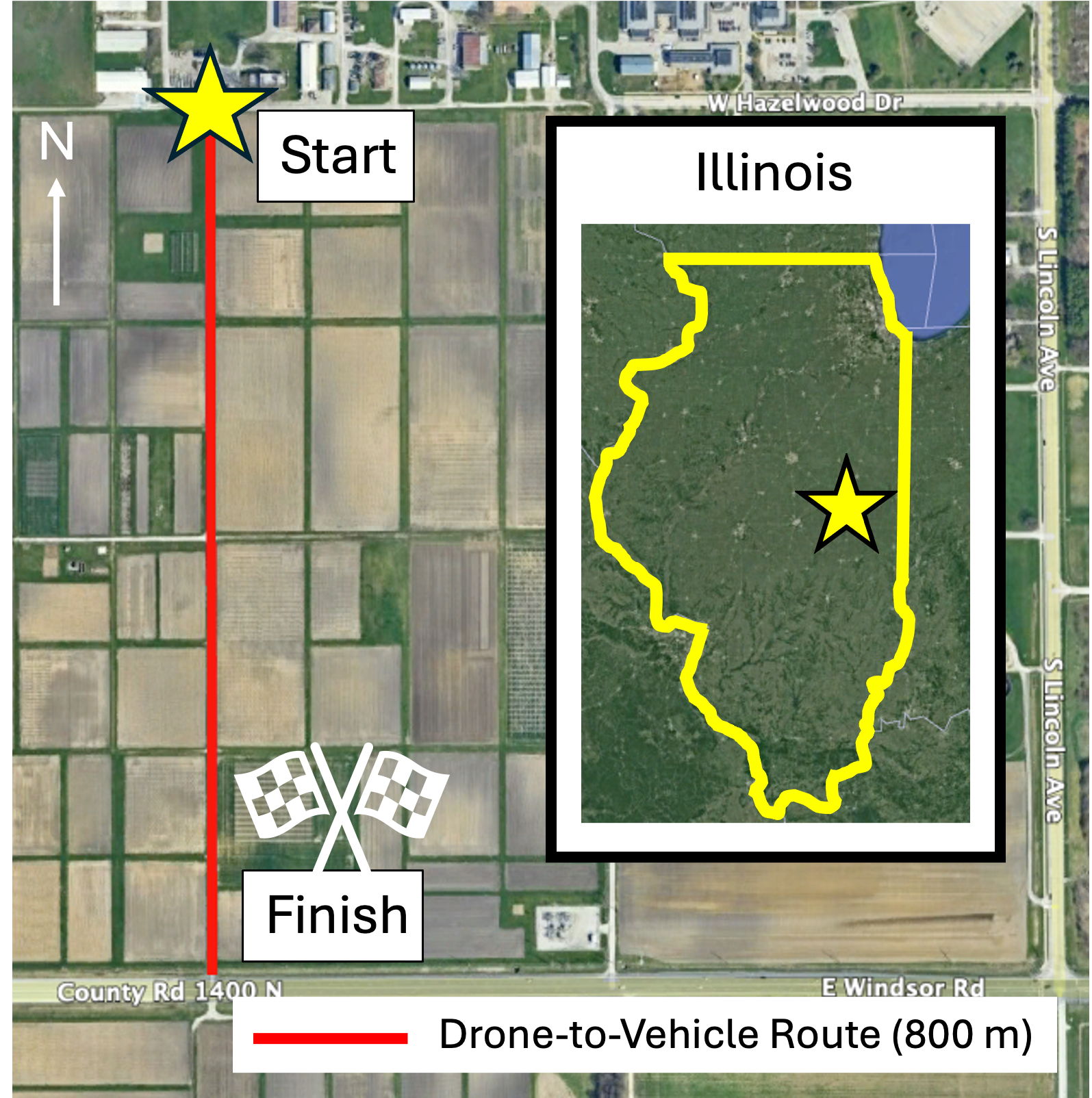} & \includegraphics[height=6.5cm]{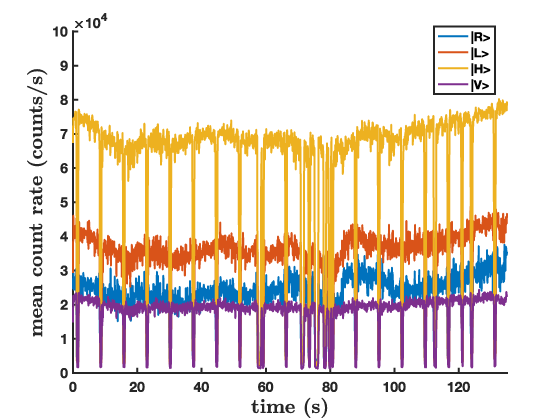}\\
    (b) & (c)
    \end{tabular}
    \end{center}
    \caption{ \label{fig:Air2Vehicle} \textbf{Quantum communication between drones and cars.} Air-to-Vehicle QKD. (a). Air-to-vehicle setup (photo courtesy of Timur Javid), (b) air-to-vehicle route map, and (c) air-to-vehicle QKD transmission (0.1-sec integration time) where the ideal proportions are shown in dashed lines.}
\end{figure}

We collected air-to-vehicle QKD data using a flying drone and a moving ground vehicle traveling parallel to each other at speeds of approximately \unit[10]{mph} (\unit[16]{km/hr}); see Fig.~\ref{fig:Air2Vehicle}(a). The QKD transmitter is mounted on the drone, and the QKD receiver is placed on the moving ground vehicle. The air-to-vehicle tests are conducted on a private road located in the agronomy research fields on the campus of UIUC. The drone and car traveled \unit[806]{m} along the ground as illustrated in Fig.~\ref{fig:Air2Vehicle}(b). Quantum transmission data is presented in Fig.~\ref{fig:Air2Vehicle}(c). The count rate data is close to ideal with high throughput, although the relative proportion of the right-circular polarized state $\ket{R}$ count rate is lower than expected, which our custom finite-key model takes into account. The overall stability of the quantum transmission is higher for the drone-to-vehicle configuration than for drone-to-drone due to the lower-intensity vibrations on the car at lower speeds. For this configuration and using the advanced security analysis, we achieve a secret key rate of \unit[1.6]{kbps} and a total key of \unit[204.5]{kb}. 

\subsection{Vehicle-to-Vehicle Configuration}

\begin{figure} [h!]
    \begin{center}
    \begin{tabular}{c} 
    \includegraphics[height=6cm]{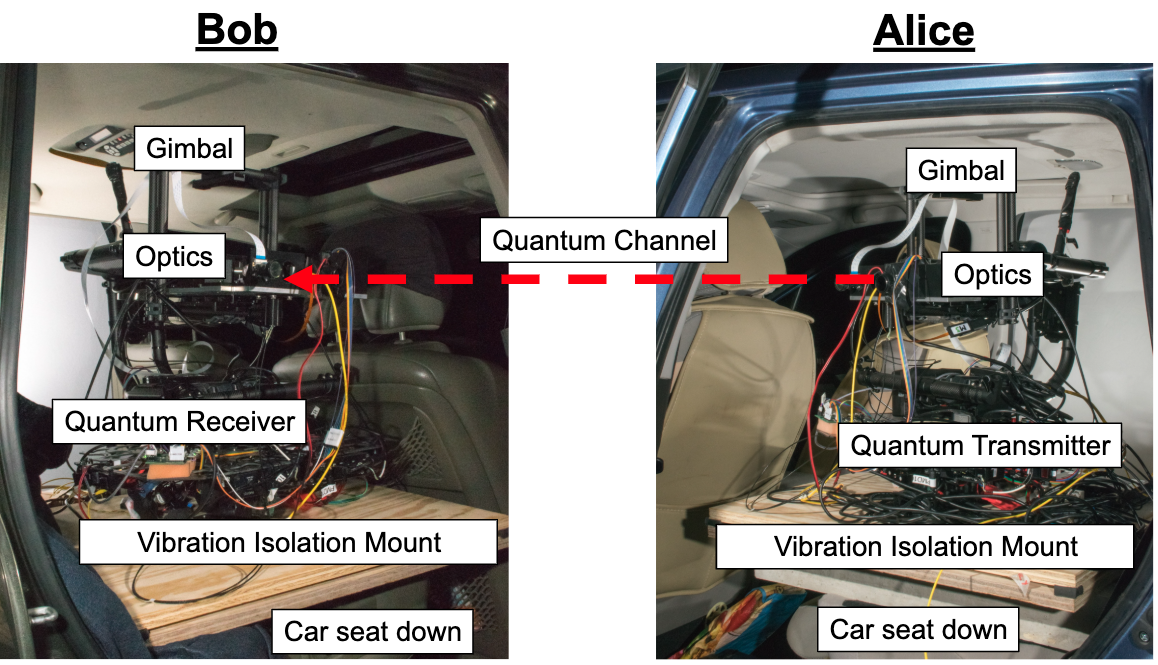}\\
    (a) \\
    \includegraphics[height=8cm]{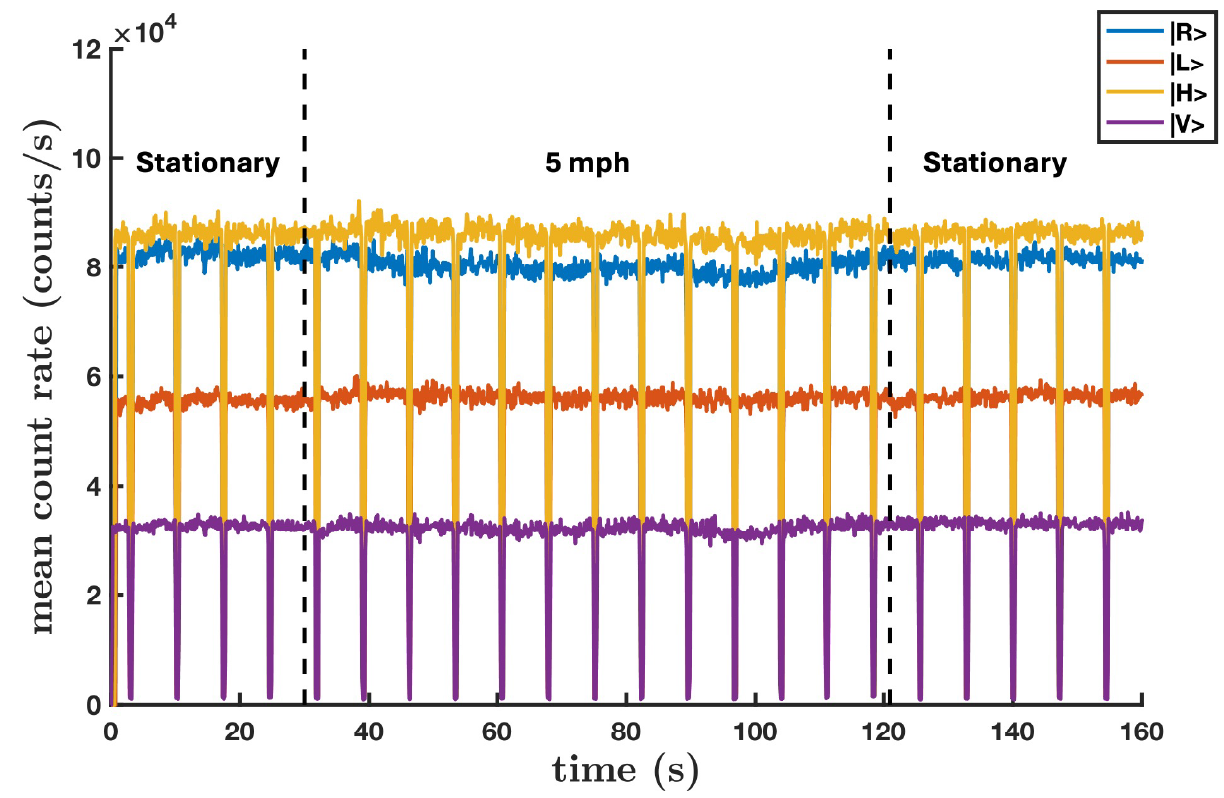} \\
    (b) 
    \end{tabular}
    \end{center}
    \caption{ \label{fig:Vehicle2Vehicle} \textbf{Quantum communication between cars.} Vehicle-to-Vehicle QKD. (a) QKD Transmitter and QKD Receiver setups, (b) 5-mph Vehicle-to-Vehicle QKD (0.1-sec integration time) where the ideal proportions are shown in dashed lines, at Illinois Center for Transportation (ICT), Mar. 27, 2024.}   
\end{figure}

We performed vehicle-to-vehicle QKD in two different test configurations. In the first, the vehicles traveled parallel to each other at a speed of \unit[5]{mph} (\unit[8]{km/hr}) on a closed test-track at the Illinois Center for Transportation (ICT). In the second, the vehicles traveled parallel to each other at a speed of \unit[70]{mph} (\unit[113]{km/hr}) on U.S. Interstate 57 in Illinois, USA.  In all test cases, the cars are separated by approximately one standard U.S. lane width (\unit[12]{ft}/\unit[3.6]{m}) \cite{Lane_Width_2014}. The modular quantum transmitter and receiver sets are placed on the rear seats of the vehicles and oriented to face each other as shown in Fig.~\ref{fig:Vehicle2Vehicle}(a). The back seat car windows are rolled down to avoid distorting or attenuating the quantum signal. 

During a key exchange session for the slow-vehicle experiments, both vehicles are initially stationary, then accelerate to a speed of \unit[$\sim$5]{mph} (\unit[8]{km/hr}), which is maintained for \unit[$\sim$91]{s} before decelerating and stopping. The temporal evolution of the QKD rate is presented in Fig. \ref{fig:Vehicle2Vehicle} for this test. We see that the quantum transmission is more stable than in the drone experiments, which is expected due to the lower-intensity vibrations of the vehicle compared to drones. Additionally, the quantum transmission while the vehicles are moving is nearly ideal, indicated by the constant count rate when transitioning between the stationary and moving cases. For this session, we obtain a secret key rate of \unit[20.0]{kbps} and a total key of \unit[1.88]{Mb}.  

For the high-speed experiments, while driving on a public highway, the pointing and acquisition alignment laser beacons are modified to operate at eye-safety levels; see Supplementary Note \hyperref[chap:Note_12]{12} for safety calculations. We only used a single 850-nm beacon emanating from the transmitter and sent to the receiver, allowing for three out of four control loops to operate; the pointing stabilization systems are described in Supplementary Note \hyperref[chap:Note_7]{7}. We achieve a secret key rate of \unit[2.5]{kbps} and \unit[519]{kb} for this experimental run. 

\section{Discussion}

There are several limitations of our study. Our current system only operates at night, but this will be mitigated in the future using improved spectral, spatial, and temporal filtering. For example, we can use a 0.7-nm width spectral filter, single-spatial-mode optical fiber for collecting the received signal, and a shorter $\sim$1-ns time window as implemented in \cite{zhang2025daylight} to achieve daylight operations.  Combined, we predict a received signal-to-noise ratio improvement \unit[$>38$]{dB} during daylight, which will allow continuous operation; see Supplementary Note \hyperref[chap:Note_13]{13}, where we measured a daylight background noise rate of only \hbox{\unit[3000]{$s^{-1}$}}, comparable to the night time background rate in the current system.

Another limitation is that our current PAT system has an effective range of \unit[100]{m} or less, primarily due to reduced visible-wavelength beacon power; we are upgrading the system to use higher power beacon lasers (at an eye-safe wavelength of \unit[$\geq 1400$]{nm}) and adding alignment detectors that have higher spatial sensitivity to improve the operational range. Environmental conditions such as fog and rain contribute additional attenuation and transverse spatial-mode aberrations to the quantum optical signal and indeed to \emph{any} free-space optical communication system. These effects can potentially be reduced by working in the long-wave infrared part of the electromagnetic spectrum \cite{sua2017direct}, but low SWaP single-photon detectors are not currently available at these wavelengths, and thermal electromagnetic background is much higher there. For mobile systems, we see no alternative to using free-space links, and hence they will likely be limited to shorter distance links under these conditions, unless they  can operate at higher altitudes, e.g., above the fog.

Finally, for drone-based mobile platforms, typical commercial batteries have relatively limited energy capacity, restricting the useful flight time to several minutes, which constrains the maximum altitude and travel distance. The flight time limit can be extended by using a Class 3 or higher drone, available in military systems, and by further reducing the system SWaP. Table \ref{table:Power_Budget_Table} shows the estimated power draw for the various subsystems in our mobile platform transmitters and receivers. Note that another option that completely avoids the battery energy limitations is to employ a tether to supply power to the in-flight drone, although all of our secure key results in the finite-key regime are achieved without using such a tether. We have previously designed and implemented such a solution \cite{conrad2021drone}, see Supplementary Note \hyperref[chap:Note_14]{14}. One possible benefit of a tether, in addition to essentially unlimited  flight times, is it could additionally include a fiber optic link so that, e.g., a more sophisticated (and larger SWaP) quantum communication source could be located off the transmitter drone, or a cryogenically cooled detector, e.g., superconducting nanowire detector could be located off the receiver drone. Also, for many applications it is likely the drone would not be moving, e.g., deployed straight up from a ship to allow longer line-of-sight and greatly reduced turbulence when connecting to a similarly elevated receiver.

\begin{table}[h]
\begin{adjustwidth}{0cm}{0cm}
\begin{center}
\caption{Drone Power Budget.} 
\label{table:Power_Budget_Table}
\begin{tabular}
{||c|c|c||} 
 \hline
 Subsystem & QKD Transmitter (Alice) & QKD Receiver (Bob) \\
 \hline\hline 
 Drone Flight Motors$^{\dagger}$ & \unit[$1.6$]{kW} & \unit[$1.7$]{kW} \\ 
 \hline
 Pointing and Tracking & \unit[$38$]{W} & \unit[$34$]{W} \\
 \hline
 Quantum Hardware & \unit[$6$]{W} & \unit[$27$]{W} \\
 \hline
\end{tabular}\\
$^{\dagger}$Hovering thrust in-flight.
\end{center}
\end{adjustwidth}
\end{table}

Looking to a future upgraded system, we anticipate operation for link distances up to \unit[15]{km} using higher power laser beacons, upgraded beam position sensors, larger send and receive optical apertures (e.g., with diameters \unit[$>2.5$]{cm}, corresponding to a Rayleigh length of \unit[$\sim$1]{km}) which will result in a link efficiency $>\unit[-20]{dB}$. Furthermore, we estimate that such a system could achieve a raw key rate of 500 kbps using a $5\%$ photon pair per pulse probability and GHz-repetition rate sources. Additionally, a Key Management System (KMS) and encryptors such as AES-256 are required to field a complete end-to-end encryption solution; such capabilities could be achieved using a Hardware Security Module (HSM) connected to the onboard single-board computer. 

We envision future heterogeneous quantum networks that have mobile as well as fixed nodes. Here, we take the first step of demonstrating quantum communication where both sender and receiver platforms are mobile. There are several applications of our current system in which the mobile systems approach each other, exchange a key, and separate.  This will allow them to use the key for subsequent long-distance secure communication. Additionally, such on-the-fly engagements may be useful for other protocols, e.g., sending quantum signals bidirectionally for secure clock synchronization \cite{lee2019symmetrical}. One important feature of our demonstration is developing an advanced security analysis that accounts for the variation in the received count rates for each communication system. Also, our analysis is valid for the relatively short communication sessions --- known as the finite-size regime \cite{tomamichel2012tight} --- that are more likely in a mobile network. We suggest that QKD links using mobile platforms should produce a secret key in the finite-size regime using only a single-pass, which we achieved in all link configurations. This demonstrates that free-space QKD is practical for moving platforms. Moreover, all the same supporting system hardware –  pointing, time synchronization, detection and quantum state analysis  – will be necessary and sufficient for transmitting entanglement; thus, our demonstration is relevant for broader applications in quantum networking. Finally, our work underscores the utility of developing modular systems that can be easily swapped between platforms.

\section{Methods}

We use a modular system design approach in which the quantum transmitter and receiver systems do not share \emph{communication, control, or power} with the host mobile platforms. This allows for rapid re-configurability. The QKD modules include all subsystems needed for secure QKD between mobile platforms as described in Supplementary Note \hyperref[chap:Note_6]{6}. 

\subsection{Drone Platform}

\begin{figure}[h] 
    \centering 
\includegraphics[height=12cm]{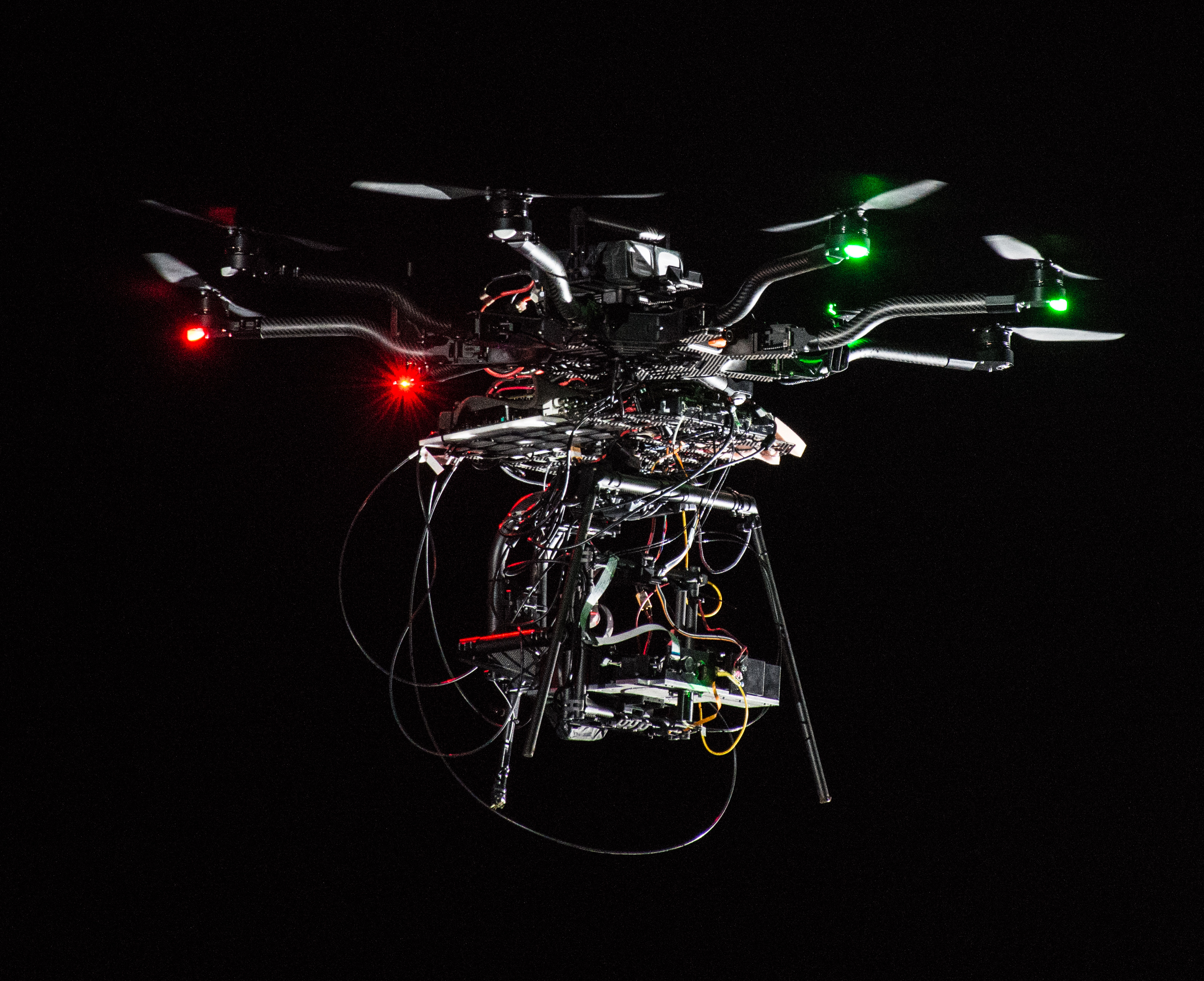} 
    \caption{The drone-based QKD platform in flight. Photo courtesy of Alan Mitchell.} 
    \label{fig:my_fig1} 
\end{figure}

The drone platform is a commercial octocopter (Freefly Systems, PN: Alta 8 Pro); see Fig. \ref{fig:my_fig1} of the drone in flight with the QKD receiver payload.  Each has a payload capacity of approximately \unit[9]{kg}, and is powered by two \unit[10,000]{mA-hr} lithium polymer (LiPo) batteries, which provide approximately \unit[5]{minutes} of flight time while carrying the QKD equipment. The QKD payload connects to the drones using a single mechanical connector (Freefly Systems, PN: 910-00625) featuring a quick-release button without requiring tools to add or remove the quantum payload. More detailed system information is included in the Supplementary Information.

We operate the drones using Futaba radio controllers, and we monitor the in-flight status using QGround Control Software \cite{QGroundControl}. The Alta 8 Pro drone uses the PX4 flight controller \cite{PX4_Firmware} and supports manual, altitude-, and position-hold flight modes. Following best practices, we perform take-offs and landings in manual control mode using one pilot per drone, and switch the drones to position-hold mode to collect QKD data while the drones hover in flight. Drone operation is regulated in the United States by the U.S. FAA Small UAS Rule (Part 107) \cite{Part_107} and requires a Remote Pilot Certificate, waivers to fly at night, and authorization to fly in Class-C restricted airspace, which we received for all QKD flights. 

\subsection{Information Reconciliation}

Information reconciliation, also known as error correction, is a protocol performed by two parties, Alice and Bob, over a classical public channel to ensure that the final keys are identical. Errors in Bob's received key can be introduced by channel noise or by an eavesdropper Eve. Both of these sources of error are, in principle, indistinguishable, so must be attributed to Eve. Moreover, because error correction is performed over a public channel, it necessarily leaks information about the secret keys, and thus we must minimize the use of the public channel to reduce the amount of information gained by Eve. This leakage, along with any information the eavesdropper might have acquired during the quantum exchange, is accounted for in the privacy-amplification step. In classical error correction, linear block codes with low-density parity-check matrices (known as LDPC codes) can be applied for forward error correction \cite{gallager1962low,mackay1999good}. More recently, it has been shown that LDPC codes can be adapted for use in QKD \cite{elkouss2010information}. For the experimental data, we use a symmetric blind information reconciliation protocol \cite{kiktenko2017symmetric}, incorporating standard LDPC codes provided by IEEE Standard for Information Technology \cite{5307322}.

\subsection{Finite Key Analysis}
To extract secure key in the case of limited amounts of raw key, we use the finite-size analysis from \cite{kamin2026improved}, which works in the composable \(\varepsilon\)-security framework of \cite{renner2008security} and applies the numerical framework of \cite{WLC18} to evaluate the resulting finite-size secret key rates, after incorporating source and detector imperfections. First, sources typically do not emit perfect states but, \textit{e.g.}, a unitarily rotated version of them, which we determine by performing state tomography to characterize the polarization state of the signals; see Supplementary Note \hyperref[chap:Note_3]{3}. The descriptions of the resulting imperfect states are then directly inserted into the security proof.

We also account for detector imperfections, following Ref.~\cite{Kamin2024Phys.Rev.Res.}, which allows us to include arbitrary passive linear optical detection setups as unitary transformations on modes. Following App.~C of \cite{Kamin2024Phys.Rev.Res.}, we use coherent states to characterize the detection setup, and together with a least-squares estimation, to find the unitary transformations. Then, using Corollary 4 of Ref.~\cite{Kamin2024Phys.Rev.Res.}, we convert the model of the detection setup to an equivalent lossless setup, thus effectively attributing all losses, including transmission losses, to the eavesdropper. Furthermore, we recover the positive operator-value measurement elements for Bob’s detection events from the characterization of the detection setup. This allows us to include experimental imperfections, such as a misalignment of the beam paths, imperfect optics, and mismatched detector efficiencies, etc., in the security analysis. The imperfections contribute, for example, to the non-ideal ratio of the counts from channel-to-channel. See Supplementary Note \hyperref[chap:Note_11]{11} for further details.

Finally, because the setup uses different light sources for each signal state, they will naturally emit signals at different intensities (beyond the intentional differences to implement decoy state QKD). Section VI of Ref.~\cite{Kamin2024Phys.Rev.Res.} shows how intensities differing with each signal state can be included in the security proof. We extended this approach to the finite-size regime with the methods from \cite{kamin2026improved}. Because the intensities sent out by the setup can change during a key exchange session, we use the characterization data of the detection setup and search for self-consistent intensities given the observations by performing another least-squares estimation. 

Furthermore, we note that \emph{any} experimental setup naturally possesses these kinds of imperfections to some extent; hence, device modeling is necessary in general. Moreover, accounting for these imperfections results in a much stronger security statement because the devices satisfy the assumptions of the security proof.

\section*{Data availability}
All relevant data are available from the corresponding author upon reasonable request.

\section*{Code availability}
All relevant code is available from the corresponding author upon reasonable request. The code used to prepare the secret key rates in this paper is available at \href{https://openqkdsecurity.wordpress.com/repositories-for-publications/}{Secret Key Rate Code}.

\bibliographystyle{naturemag}
\bibliography{references}

@article{tamaki2014loss,
  title={Loss-tolerant quantum cryptography with imperfect sources},
  author={Tamaki, Kiyoshi and Curty, Marcos and Kato, Go and Lo, Hoi-Kwong and Azuma, Koji},
  journal={Physical Review A},
  volume={90},
  number={5},
  pages={052314},
  year={2014},
  publisher={APS}
}

@article{liao2018satellite,
  title={Satellite-relayed intercontinental quantum network},
  author={Liao, Sheng-Kai and Cai, Wen-Qi and Handsteiner, Johannes and Liu, Bo and Yin, Juan and Zhang, Liang and Rauch, Dominik and Fink, Matthias and Ren, Ji-Gang and Liu, Wei-Yue and others},
  journal={Physical Review Letters},
  volume={120},
  number={3},
  pages={030501},
  year={2018},
  publisher={APS}
}

@inproceedings{nauerth2012air,
  title={Air to ground quantum key distribution},
  author={Nauerth, Sebastian and Moll, Florian and Rau, Markus and Horwath, Joachim and Frick, Stefan and Fuchs, Christian and Weinfurter, Harald},
  booktitle={Quantum Communications and Quantum Imaging X},
  volume={\textbf{8518}},
  pages={71--76},
  year={2012},
  organization={SPIE}
}

@article{pugh2017airborne,
  title={Airborne demonstration of a quantum key distribution receiver payload},
  author={Pugh, Christopher J and Kaiser, Sarah and Bourgoin, Jean-Philippe and Jin, Jeongwan and Sultana, Nigar and Agne, Sascha and Anisimova, Elena and Makarov, Vadim and Choi, Eric and Higgins, Brendon L and others},
  journal={Quantum Science and Technology},
  volume={2},
  number={2},
  pages={024009},
  year={2017},
  publisher={IOP Publishing}
}

@article{liu2020drone,
  title={Drone-based entanglement distribution towards mobile quantum networks},
  author={Liu, Hua-Ying and Tian, Xiao-Hui and Gu, Changsheng and Fan, Pengfei and Ni, Xin and Yang, Ran and Zhang, Ji-Ning and Hu, Mingzhe and Guo, Jian and Cao, Xun and others},
  journal={National Science Review},
  volume={7},
  number={5},
  pages={921--928},
  year={2020},
  publisher={Oxford University Press}
}

@article{liu2021optical,
  title={Optical-relayed entanglement distribution using drones as mobile nodes},
  author={Liu, Hua-Ying and Tian, Xiao-Hui and Gu, Changsheng and Fan, Pengfei and Ni, Xin and Yang, Ran and Zhang, Ji-Ning and Hu, Mingzhe and Guo, Jian and Cao, Xun and others},
  journal={Physical Review Letters},
  volume={126},
  number={2},
  pages={020503},
  year={2021},
  publisher={APS}
}

@article{tu2022lower,
  title={A Lower Size, Weight Acquisition and Tracking System for Airborne Quantum Communication},
  author={Tu, Chengxiang and Shen, Jiayi and Dai, Jiansheng and Zhang, Liang and Wang, Jianyu},
  journal={IEEE Photonics Journal},
  volume={14},
  number={6},
  pages={1--8},
  year={2022},
  publisher={IEEE}
}

@article{gallager1962low,
  title={Low-density parity-check codes},
  author={Gallager, Robert},
  journal={IRE Transactions on Information Theory},
  volume={8},
  number={1},
  pages={21--28},
  year={1962},
  publisher={IEEE}
}

@article{mackay1999good,
  title={Good error-correcting codes based on very sparse matrices},
  author={MacKay, David JC},
  journal={IEEE Transactions on Information Theory},
  volume={45},
  number={2},
  pages={399--431},
  year={1999},
  publisher={IEEE}
}

@article{elkouss2010information,
  title={Information reconciliation for quantum key distribution},
  author={Elkouss, David and Martinez-Mateo, Jesus and Martin, Vicente},
  journal={arXiv preprint arXiv:1007.1616},
  year={2010}
}

@article{kiktenko2017symmetric,
  title={Symmetric blind information reconciliation for quantum key distribution},
  author={Kiktenko, Evgeniy O and Trushechkin, Anton S and Lim, Charles Ci Wen and Kurochkin, Yury V and Fedorov, Aleksey K},
  journal={Physical Review Applied},
  volume={8},
  number={4},
  pages={044017},
  year={2017},
  publisher={APS}
}

@ARTICLE{5307322,
  author={},
  journal={IEEE Std 802.11n-2009 (Amendment to IEEE Std 802.11-2007 as amended by IEEE Std 802.11k-2008, IEEE Std 802.11r-2008, IEEE Std 802.11y-2008, and IEEE Std 802.11w-2009)}, 
  title={{IEEE} Standard for Information technology-- Local and metropolitan area networks-- Specific requirements-- Part 11: Wireless {LAN} Medium Access Control ({MAC}) and Physical Layer ({PHY}) Specifications Amendment 5: Enhancements for Higher Throughput}, 
  year={2009},
  volume={},
  number={},
  pages={1-565},
  doi={10.1109/IEEESTD.2009.5307322}
}

@article{lim2014concise,
  title={Concise security bounds for practical decoy-state quantum key distribution},
  author={Lim, Charles Ci Wen and Curty, Marcos and Walenta, Nino and Xu, Feihu and Zbinden, Hugo},
  journal={Physical Review A},
  volume={89},
  number={2},
  pages={022307},
  year={2014},
  publisher={APS}
}

@article{kamin2026improved,
  title={Improved finite-size effects in quantum key distribution with applications to decoy-state protocols},
  author={Kamin, Lars and Tupkary, Devashish and L{\"u}tkenhaus, Norbert},
  journal={Physical Review Research},
  volume={8},
  number={1},
  pages={013332},
  year={2026},
  publisher={APS}
}

@article{Kamin2024Phys.Rev.Res.,
  title = {Improved Decoy-State and Flag-State Squashing Methods},
  author = {Kamin, Lars and L{\"u}tkenhaus, Norbert},
  year = {2024},
  month = dec,
  journal = {Physical Review Research},
  volume = {6},
  number = {4},
  pages = {043223},
  publisher = {American Physical Society},
  doi = {10.1103/PhysRevResearch.6.043223}
}

@article{WLC18,
  title={{Reliable numerical key rates for quantum key distribution}},
  author={Winick, A. and L{\"u}tkenhaus, N. and Coles, P. J.},
  journal={Quantum},
  volume={2},
  pages={77},
  year={2018},
  doi={10.22331/q-2018-07-26-77}
}

@article{cochran2021qubit,
  title={Qubit-based clock synchronization for QKD systems using a Bayesian approach},
  author={Cochran, Roderick D and Gauthier, Daniel J},
  journal={Entropy},
  volume={23},
  number={8},
  pages={988},
  year={2021},
  publisher={MDPI}
}

@misc{Lane_Width_2014,
  author       = "",
  title        = "{Lane Width, U.S. Department of Transportation Federal Highway Administration}",
  howpublished = "",
  month        = "Oct",
  year         = "2014",
  note         = "Available Online: \url{https://safety.fhwa.dot.gov/geometric/pubs/mitigationstrategies/chapter3/3_lanewidth.cfm} (accessed Feb 20, 2025)",
}

@misc{QGroundControl,
  author       = "",
  title        = "{QGroundControl Software}",
  howpublished = "Available Online: \url{https://qgroundcontrol.com} (accessed Feb 19, 2025)",
  month        = "",
  year         = "",
  note         = "",
}

@misc{PX4_Firmware,
  author       = "",
  title        = "{PX4 Flight Control Firmware}",
  howpublished = "Available Online: \url{https://github.com/PX4/PX4-Autopilot} (accessed Feb 20, 2025)",
  month        = "",
  year         = "",
  note         = "",
}

@article{cacciapuoti2019quantum,
  title={Quantum internet: Networking challenges in distributed quantum computing},
  author={Cacciapuoti, Angela Sara and Caleffi, Marcello and Tafuri, Francesco and Cataliotti, Francesco Saverio and Gherardini, Stefano and Bianchi, Giuseppe},
  journal={IEEE Network},
  volume={34},
  number={1},
  pages={137--143},
  year={2019},
  publisher={IEEE}
}

@article{khabiboulline2019quantum,
  title={Quantum-assisted telescope arrays},
  author={Khabiboulline, Emil T and Borregaard, Johannes and De Greve, Kristiaan and Lukin, Mikhail D},
  journal={Physical Review A},
  volume={100},
  number={2},
  pages={022316},
  year={2019},
  publisher={APS}
}

@inproceedings{elliott2005current,
  title={Current status of the {DARPA} quantum network},
  author={Elliott, Chip and Colvin, Alexander and Pearson, David and Pikalo, Oleksiy and Schlafer, John and Yeh, Henry},
  booktitle={Quantum Information and Computation III},
  volume={5815},
  pages={138--149},
  year={2005},
  organization={SPIE}
}

@article{lo2005decoy,
  title={Decoy state quantum key distribution},
  author={Lo, Hoi-Kwong and Ma, Xiongfeng and Chen, Kai},
  journal={Physical Review Letters},
  volume={94},
  number={23},
  pages={230504},
  year={2005},
  publisher={APS}
}

@misc{Part_107,
    title={{14 C.F.R. §107. PART 107—SMALL UNMANNED AIRCRAFT SYSTEMS}},
    url ={https://www.ecfr.gov/current/title-14/chapter-I/subchapter-F/part-107}
}

@article{PhysRevLett.133.200801,
  title = {Experimental Demonstration of Drone-Based Quantum Key Distribution},
  author = {Tian, Xiao-Hui and Yang, Ran and Liu, Hua-Ying and Fan, Pengfei and Zhang, Ji-Ning and Gu, Changsheng and Chen, Mengwen and Hu, Mingzhe and Lu, Feng-Yu and Zhu, Chunxi and Yin, Zhen-Qiang and Yin, Zhi-Jun and Yuan, Mo and Wang, Shuang and Chen, Wei and Gong, Yan-Xiao and Zhu, Shi-Ning and Xie, Zhenda},
  journal = {Phys. Rev. Lett.},
  volume = {133},
  issue = {20},
  pages = {200801},
  numpages = {5},
  year = {2024},
  month = {Nov},
  publisher = {American Physical Society},
  doi = {10.1103/PhysRevLett.133.200801},
  url = {https://link.aps.org/doi/10.1103/PhysRevLett.133.200801}
}

@phdthesis{rosales2024design,
  title={Design and Development of a Quantum Key Distribution System for Highly Mobile Platforms and Its Implementation on Drones and Cars},
  author={Rosales, Daniel E Sanchez},
  year={2024},
  school={The Ohio State University}
}

@article{liu2024creation,
  title={Creation of memory--memory entanglement in a metropolitan quantum network},
  author={Liu, Jian-Long and Luo, Xi-Yu and Yu, Yong and Wang, Chao-Yang and Wang, Bin and Hu, Yi and Li, Jun and Zheng, Ming-Yang and Yao, Bo and Yan, Zi and others},
  journal={Nature},
  volume={629},
  number={8012},
  pages={579--585},
  year={2024},
  publisher={Nature Publishing Group UK London}
}

@article{stolk2024metropolitan,
  title={Metropolitan-scale heralded entanglement of solid-state qubits},
  author={Stolk, Arian J and van der Enden, Kian L and Slater, Marie-Christine and te Raa-Derckx, Ingmar and Botma, Pieter and van Rantwijk, Joris and Biemond, JJ Benjamin and Hagen, Ronald AJ and Herfst, Rodolf W and Koek, Wouter D and others},
  journal={Science advances},
  volume={10},
  number={44},
  pages={eadp6442},
  year={2024},
  publisher={American Association for the Advancement of Science}
}

@article{yin2017satellite,
  title={Satellite-based entanglement distribution over 1200 kilometers},
  author={Yin, Juan and Cao, Yuan and Li, Yu-Huai and Liao, Sheng-Kai and Zhang, Liang and Ren, Ji-Gang and Cai, Wen-Qi and Liu, Wei-Yue and Li, Bo and Dai, Hui and others},
  journal={Science},
  volume={356},
  number={6343},
  pages={1140--1144},
  year={2017},
  publisher={American Association for the Advancement of Science}
}

@article{tomamichel2012tight,
  title={Tight finite-key analysis for quantum cryptography},
  author={Tomamichel, Marco and Lim, Charles Ci Wen and Gisin, Nicolas and Renner, Renato},
  journal={Nature communications},
  volume={3},
  number={1},
  pages={634},
  year={2012},
  publisher={Nature Publishing Group UK London}
}

@article{li2025microsatellite,
  title={Microsatellite-based real-time quantum key distribution},
  author={Li, Yang and Cai, Wen-Qi and Ren, Ji-Gang and Wang, Chao-Ze and Yang, Meng and Zhang, Liang and Wu, Hui-Ying and Chang, Liang and Wu, Jin-Cai and Jin, Biao and others},
  journal={Nature},
  pages={1--8},
  year={2025},
  publisher={Nature Publishing Group UK London}
}

@article{bourgoin2015free,
  title={Free-space quantum key distribution to a moving receiver},
  author={Bourgoin, Jean-Philippe and Higgins, Brendon L and Gigov, Nikolay and Holloway, Catherine and Pugh, Christopher J and Kaiser, Sarah and Cranmer, Miles and Jennewein, Thomas},
  journal={Optics express},
  volume={23},
  number={26},
  pages={33437--33447},
  year={2015},
  publisher={Optical Society of America}
}

@book{american2022ansi,
  title={ANSI Z136.1 Safe Use of Lasers - 2022},
  author={American National Standards Institute and Laser Institute of America},
  isbn={9781940168296},
  url={https://books.google.com/books?id=8ZmUzwEACAAJ},
  year={2022},
  publisher={Laser Institute of America}
}

@misc{AF_Debrief,
    title={{The Art of the Debrief}},
    author={{Col. Todd Dyer}},
    year={2019},
    month={7},
    day={25},
    url ={https://www.aetc.af.mil/News/Article-Display/Article/1917581/the-art-of-the-debrief/}
}

@article{rosin2015ultra,
  title={Ultra-fast physical generation of random numbers using hybrid boolean networks},
  author={Rosin, David P. and Rontani, Damien and Gauthier, Daniel J.},
  journal={Dynamics of Complex Autonomous Boolean Networks},
  pages={57--79},
  year={2015},
  publisher={Springer}
}

@article{rusca2018finite,
  title={Finite-key analysis for the 1-decoy state QKD protocol},
  author={Rusca, Davide and Boaron, Alberto and Gr{\"u}nenfelder, Fadri and Martin, Anthony and Zbinden, Hugo},
  journal={Applied Physics Letters},
  volume={112},
  number={17},
  year={2018},
  publisher={AIP Publishing}
}

@article{altepeter2005photonic,
  title={Photonic state tomography},
  author={Altepeter, Joseph B and Jeffrey, Evan R and Kwiat, Paul G},
  journal={Advances in atomic, molecular, and optical physics},
  volume={52},
  pages={105--159},
  year={2005},
  publisher={Elsevier}
}

@article{Daniel_QKD_Source,
  title={A Quantum Key Distribution System for Mobile
    Platforms with Highly Indistinguishable States},
  author={Rosales, Daniel Sanchez and Cochran, Roderick D and Isaac, Samantha D and Kwiat, Paul G and Gauthier, Daniel J},
  journal = {arXiv:2411.19880},
  year = {2024}
}

@phdthesis{cochran2024full,
  title={Full-System Design, Implementation, and Analysis for Quantum Key Distribution on Mobile Platforms},
  author={Cochran, Roderick D},
  year={2024},
  school={The Ohio State University}
}

@phdthesis{conrad2024full,
  title={Drone and Vehicle-based Quantum Communication and Towards Practical Quantum Position Verification},
  author={Conrad, Andrew P},
  year={2024},
  school={University of Illinois Urbana-Champaign (UIUC)}
}

@inproceedings{conrad2021drone,
  title={Drone-based quantum key distribution (QKD)},
  author={Conrad, Andrew and Isaac, Samantha and Cochran, Roderick and Sanchez-Rosales, Daniel and Wilens, Brian and Gutha, Akash and Rezaei, Tahereh and Gauthier, Daniel J and Kwiat, Paul},
  booktitle={Free-space laser communications XXXIII},
  volume={11678},
  pages={177--184},
  year={2021},
  organization={SPIE}
}

@inproceedings{conrad2023drone,
  title={Drone-based quantum communication links},
  author={Conrad, Andrew and Isaac, Samantha and Cochran, Roderick and Sanchez-Rosales, Daniel and Rezaei, Tahereh and Javid, Timur and Schroeder, AJ and Golba, Grzegorz and Gauthier, Daniel and Kwiat, Paul},
  booktitle={Quantum Computing, Communication, and Simulation III},
  volume={12446},
  pages={99--106},
  year={2023},
  organization={SPIE}
}

@misc{FlightVideo,
    title={{Drone Flight Video}},
    author={Andrew Conrad},
    year={2022},
    howpublished = {\url{https://www.amazon.com/photos/shared/5z-9OlawRZy0jY2HWPlcAA.Ctndiv8JtfGqC1J3rLb_-F}}
}

@article{zhou2026long,
  title={Long Distance Daylight Drone-based Quantum Key Distribution under Relative Motion},
  author={Zhou, Chun and Zhou, Yanyang and Wang, Ping and Li, Tan and Zhou, Yu and Wang, Hao and Li, Yanmei Zhao Huan and Li, Dawei and Wang, Fangxiang and Zheng, Diyuan and others},
  journal={arXiv preprint arXiv:2603.18934},
  year={2026}
}

@article{lee2019symmetrical,
  title={Symmetrical clock synchronization with time-correlated photon pairs},
  author={Lee, Jianwei and Shen, Lijiong and Cer{\`e}, Alessandro and Troupe, James and Lamas-Linares, Antia and Kurtsiefer, Christian},
  journal={Applied Physics Letters},
  volume={114},
  number={10},
  year={2019},
  publisher={AIP Publishing}
}

@article{liu2025road,
  title={The road to quantum internet: Progress in quantum network testbeds and major demonstrations},
  author={Liu, Jianqing and Le, Thinh and Ji, Tingxiang and Yu, Ruozhou and Farfurnik, Demitry and Byrd, Greg and Stancil, Daniel},
  journal={Progress in Quantum Electronics},
  volume={99},
  pages={100551},
  year={2025},
  publisher={Elsevier}
}

@article{li2024survey,
  title={A survey of quantum internet protocols from a layered perspective},
  author={Li, Yuan and Zhang, Hao and Zhang, Chen and Huang, Tao and Yu, F Richard},
  journal={IEEE Communications Surveys \& Tutorials},
  volume={26},
  number={3},
  pages={1606--1634},
  year={2024},
  publisher={IEEE}
}

@article{zhang2025daylight,
  title={Daylight quantum key distribution over free-space optics for future security networks},
  author={Zhang, Peide and Oliveira, Romerson D and Davidson, Zoe CM and Hugues Salas, Emilio and Kosmatos, Evangelos A and Stavdas, Alexandros and Lord, Andrew and Rarity, John and Nejabati, Reza and Simeonidou, Dimitra},
  journal={Journal of Optical Communications and Networking},
  volume={17},
  number={6},
  pages={B61--B70},
  year={2025},
  publisher={Optica Publishing Group}
}

@article{renner2008security,
  title={Security of quantum key distribution},
  author={Renner, Renato},
  journal={International Journal of Quantum Information},
  volume={6},
  number={01},
  pages={1--127},
  year={2008},
  publisher={World Scientific}
}

@article{sua2017direct,
  title={Direct generation and detection of quantum correlated photons with 3.2 um wavelength spacing},
  author={Sua, Yong Meng and Fan, Heng and Shahverdi, Amin and Chen, Jia-Yang and Huang, Yu-Ping},
  journal={Scientific reports},
  volume={7},
  number={1},
  pages={17494},
  year={2017},
  publisher={Nature Publishing Group UK London}
}
\section*{Acknowledgments}

This work has been funded in part by U.S. Army Grant \# W911NF-24-2-0097, by the ONR MURI program on Wavelength-Agile Quantum Key Distribution in a Marine Environment, Grant \# N00014-13-1-0627, by the U.S. Department of Defense (DoD) through the National Defense Science Engineering Graduate (NDSEG) Fellowship Program, the U.S. Naval Air Warfare Center Aircraft Division (NAWCAD) Internship Program, and by the
Air Force Research Laboratory (AFRL) under agreement FA8650-19-2-9300. This work was also supported in part by the KRISS Korea NRF Grant Number 112168, through the Southwestern Ohio Council for
Higher Education (SOCHE) Fellowship Program. 

LK, AC and NL have been funded by Canada through the NSERC discovery under the Discovery Grants Program, Grant No. 341495 and Alliance Grant QUINT. Furthermore, their research has been conducted at the Institute for Quantum Computing at the University of Waterloo, which is supported by Innovation, Science, and Economic Development Canada.

Some of this work was performed at Oak Ridge National Laboratory, operated by UT-Battelle for the U.S. Department of Energy under contract no. DE-AC05-00OR22725. This work was supported in minor part by U.S. Department of Energy, Office of Cybersecurity Energy Security and Emergency Response (CESER) through the Risk management Tools and Technologies (RMT) Program and U.S. Department of Energy, Office of Science, Advanced Scientific Computing Research, under the Quantum Internet to Accelerate Scientific Discovery program (Field Work Proposal ERKJ381).

We thank Alan Mitchell and Timur Javid for professional photographs, Matteo Stefanini and University Laboratory High School Students (Nate Jones, Jacquelyn Butts, Seyed-Ahmad Dastgheib, Bruce Tang) for assisting with flight testing, Benjamin Cochran for providing post-processing code improvements. We thank Shuen Wu and Max Gold for assisting with vehicle-based QKD testing, John Benson and Patrick Mugg for assisting with drone and vehicle vibration testing, and Jaedeok Park for assisting with daytime background testing. Additionally, we thank Luke Prunkard and Jake Tammen for providing drone testing and storage access on campus, as well as Prof. Imad L. Al-Qadi and Greg Renshaw for providing access to the Illinois Center for Transportation's closed test track for Vehicle-to-Vehicle testing.  

\section*{Author contributions}
PGK and DJG conceived the initial concept. AC, SI, IC, TR, TJ, AJS, AH, BW, DSR, and RC contributed to drone flight and vehicle operations and testing. DSR, RC, AG, JS, SI, and DJG contributed to the QKD Decoy-State Source design and development. SI, RC, AC, DSR, AH, GG, JS and DJG contributed to the design and development of the custom optical benches. AC, TJ, AH, JC, DSR, BW, KH, and SI contributed to the Pointing, Acquisition, and Tracking system design, development and testing. DSR, RC, AG, and DJG contributed to the FPGA-based timetagger. RC, DSR, and TJ contributed to the data post-processing. JS, LK, RC, AC, and NL contributed to the security analysis. AC, SI, RC, and DSR performed drone-to-drone, drone-to-car, and car-to-car QKD data collection. Authors AC, SI, RC, DSR contributed to writing the manuscript, while AC, RC, DSR, IC, LK, JS, JC, NL, DJG, and PGK contributed to editing the manuscript.  

\section*{Competing interests}
The authors declare no competing interests. 

\subsection*{Correspondence}
Correspondence to Andrew Conrad (aconrad5@illinois.edu)


\newpage

\hspace{-0.7cm} 
\begin{minipage}{\textwidth}
    \begin{center}
        {\Large \textbf{Supplementary Information: Drone- and Vehicle-Based Quantum Key Distribution}} \\ 
        \vspace{0.5cm}
        \text{Andrew Conrad*$^{\dagger,a,b}$, 
        Roderick Cochran$^{\dagger,c}$,
        Daniel Sanchez-Rosales$^{\dagger,c}$,
        Samantha}\\
        \text{Isaac$^{\dagger,a}$,
        Timur Javid$^{a,b}$, 
        Tahereh Rezaei$^{a}$, 
        A.J. Schroeder$^{a,b}$, 
        Grzegorz Golba$^{a}$, 
        Akash} \\
        \text{Gutha$^{c}$, 
        Brian Wilens$^{a,b}$, 
        Kyle Herndon$^{a}$, 
        Alex Hill$^{a}$, 
        Joseph Chapman$^{e}$, 
        Ian Call$^{a}$,} \\
        \text{Joseph Szabo$^{c}$, 
        Aodhan Corrigan$^{d}$, 
        Lars Kamin$^{d}$, 
        Norbert Lütkenhaus$^{d}$, 
        Daniel J. }\\ 
        \text{Gauthier$^{c}$, 
        Paul G. Kwiat$^{a,b}$} \\
        \vspace{0.5cm}
        $^{a}$Department of Physics, The Grainger College of Engineering, University of Illinois at Urbana-Champaign (UIUC), 1110 W Green St. Loomis Laboratory, Urbana, IL \\ 61801, USA \\
        $^{b}$Department of Electrical and Computer Engineering (ECE), The Grainger College of \\ Engineering, University of Illinois at Urbana-Champaign (UIUC), 306 N. Wright St., \\ Urbana, IL 61801, USA \\
        $^{c}$Department of Physics, The Ohio State University, 191 W Woodruff Ave, Columbus, \\OH 43210, USA \\
        $^{d}$Institute for Quantum Computing and Department of Physics and Astronomy, 200 \\ University Avenue West, Waterloo, ON, Canada \\
        $^{e}$Quantum Information Science Section, Oak Ridge National Laboratory, Oak Ridge, \\ Tennessee 37831, USA \\
        $^{\dagger}$Equal Contribution \\
        {$^{*}$Corresponding Author}
    \end{center}
\end{minipage}

\section*{Supplementary Note 1. Drone Platform and Flight Operations}
\label{chap:Note_1}

\begin{figure} [hbt!]
    \centering 
    \includegraphics[height=7cm]{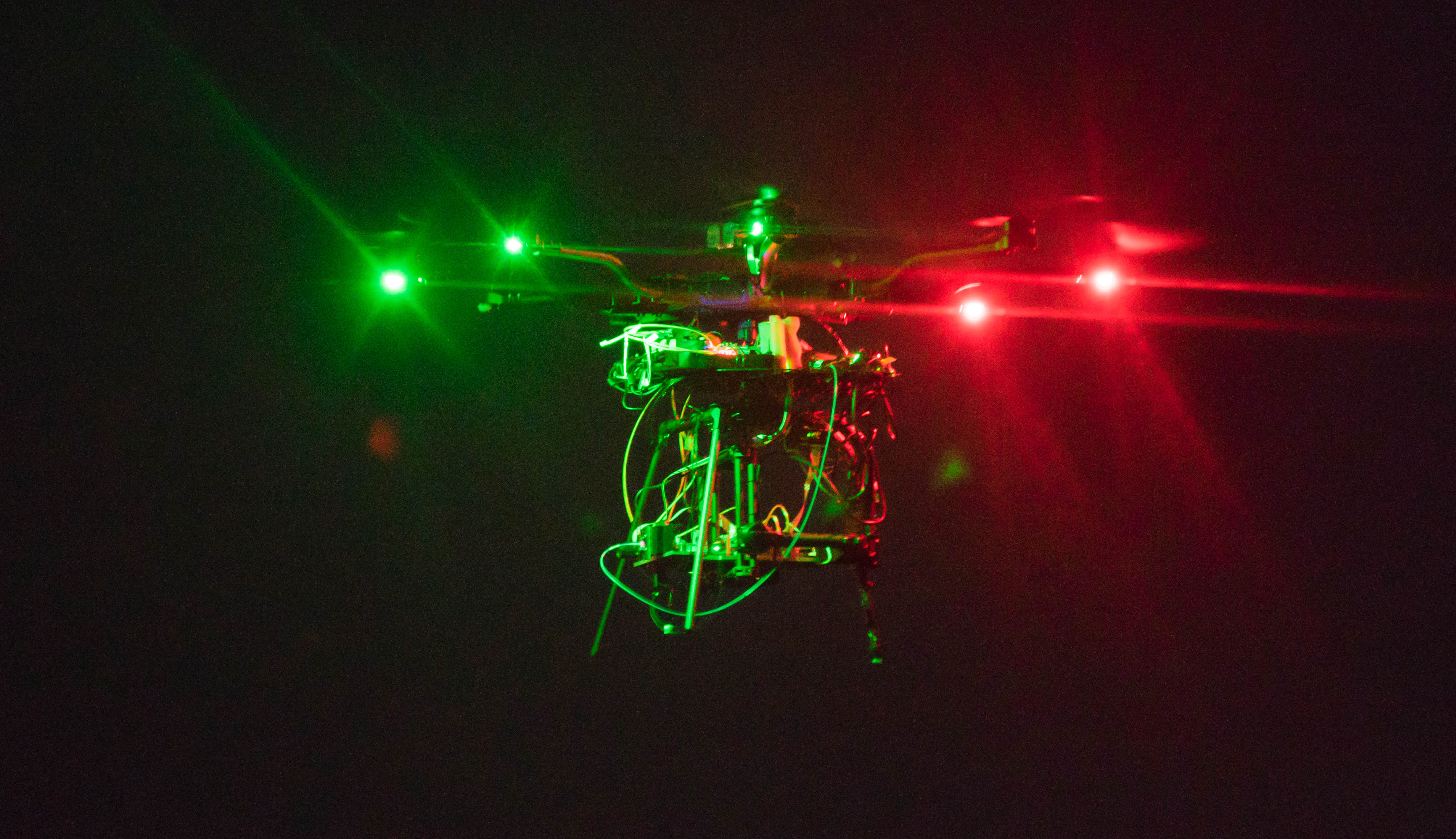} 
    \caption{\label{fig:Drone_Flight} Drone QKD Flight Operations  (Photo Courtesy Timur Javid).}   
\end{figure}

\begin{figure} [hbt!]
    \centering 
    \includegraphics[height=8.05cm]{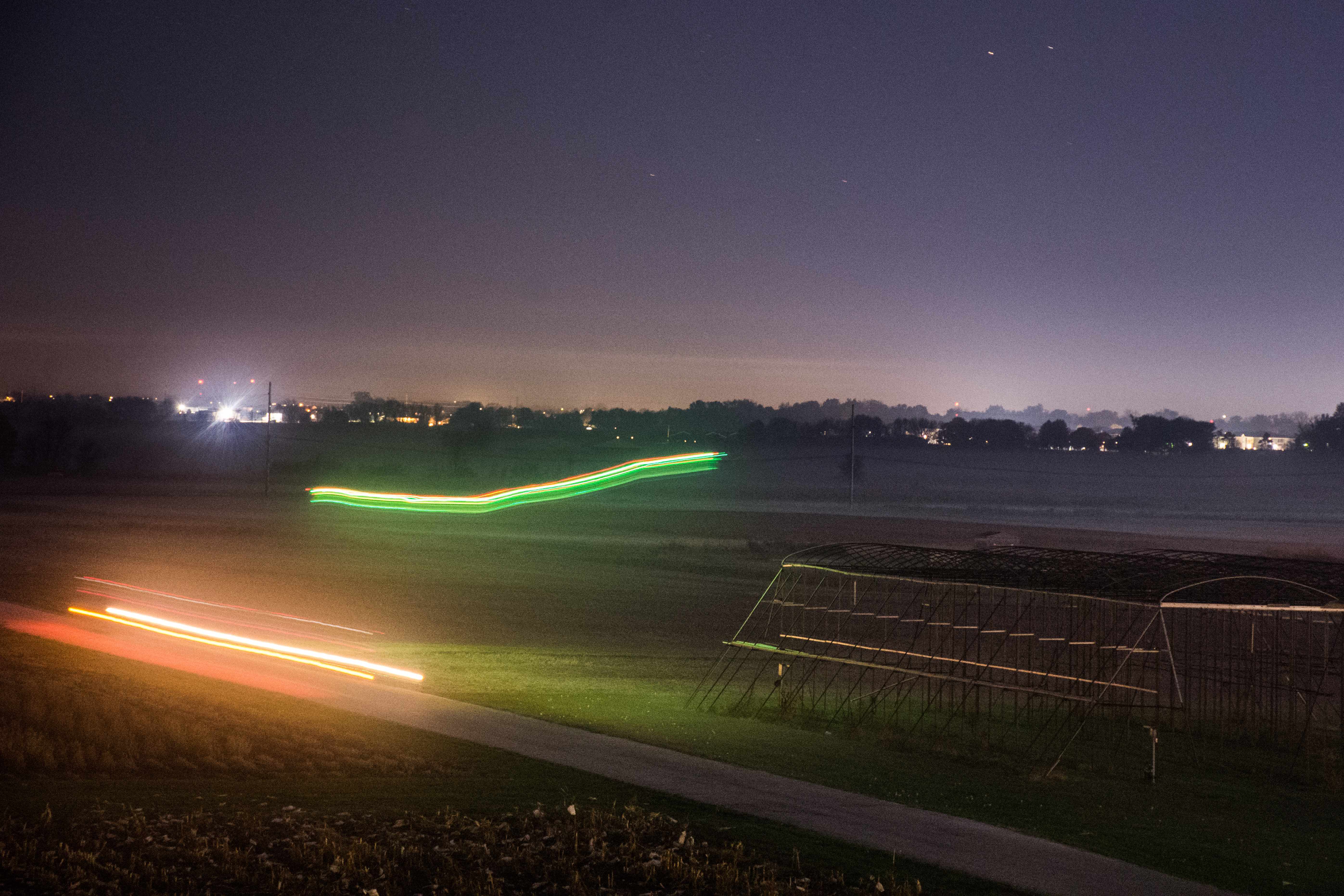} 
    \caption{\label{fig:Drone_to_Vehicle} Drone-to-Vehicle QKD Flight Operations Time-lapse (Photo Courtesy Timur Javid).}   
\end{figure}

We use octocopters (Freefly Systems, Alta 8 Pro) with a payload capacity of \unit[20]{lbs} (\unit[9.07]{kg}), and a gross takeoff weight of \unit[40]{lbs} (\unit[18.1]{kg}). The drones are powered by two \unit[10,000]{mA-hr} Lithium Polymer (LiPo) batteries, which provide a typical flight time of \unit[2.5]{minutes}. Our Quantum Key Distribution (QKD) system does not share power, communication, or control signals with the drone platform. The Pointing, Acquisition, and Tracking (PAT) system automatically detects takeoff and landing using a pressure sensor and begins initial pointing and acquisition after takeoff. After initial acquisition, the gimbal and fast-steering-mirror (FSM) control loops automatically perform closed-loop tracking of the optical line-of-sight (LOS), until landing is detected. Figure~\ref{fig:Drone_Flight} is a photo of one of the drones in flight with the QKD package, while Fig.~\ref{fig:Drone_to_Vehicle} shows the drone and vehicle paths during a Drone-to-Vehicle session. A video of one of our QKD Drone-to-Drone flights is available here: \href{https://doi.org/10.6084/m9.figshare.33313020}{Flight Video} \cite{FlightVideo}.  

Our flight operations are governed by U.S. Federal Aviation Administration (FAA) Part 107 federal regulations \cite{Part_107} because the weight of our drone is greater than \unit[250]{g}. These regulations stipulate that drone pilots obtain a remote pilot certificate from the FAA. Additionally, flights in restricted airspace near airports require FAA authorization via the low-altitude authorization and notification capability (LAANC) notification system. The Drone-to-Drone QKD flight sessions take place in an agronomy field (\unit[40° 5'29.59"]{N}, \unit[88°13'40.06"]{W}) at the University of Illinois Urbana-Champaign (UIUC), which is in Class-C restricted airspace of the Champaign Willard Airport (CMI). We obtain LAANC authorization before each drone flight, using the \textit{Air Control} smartphone application, and comply with altitude restrictions. We use one licensed drone pilot for each drone per FAA regulations during each drone-to-drone QKD test.

We use a custom preflight checklist to support QKD drone flights, which combines standard best-practices from the drone and aviation communities, and custom procedures to improve safety and minimize risk of damage to the critical quantum hardware. For example, we calibrate all flight sensors, measure the gross takeoff weight, and adjust the hovering thrust percentage flight parameter based on previous data for both drones before each flight. After each day of flights, we perform a flight debrief adopted from the U.S. Air Force \cite{AF_Debrief}, which systematically stores knowledge from past experiences and is used to update the pre-flight checklist as needed.

\section*{Supplementary Note 2. Resonant-Cavity QKD Source}
\label{chap:Note_2}

\begin{figure} [ht]
    \centering 
    \includegraphics[height=3.5cm]{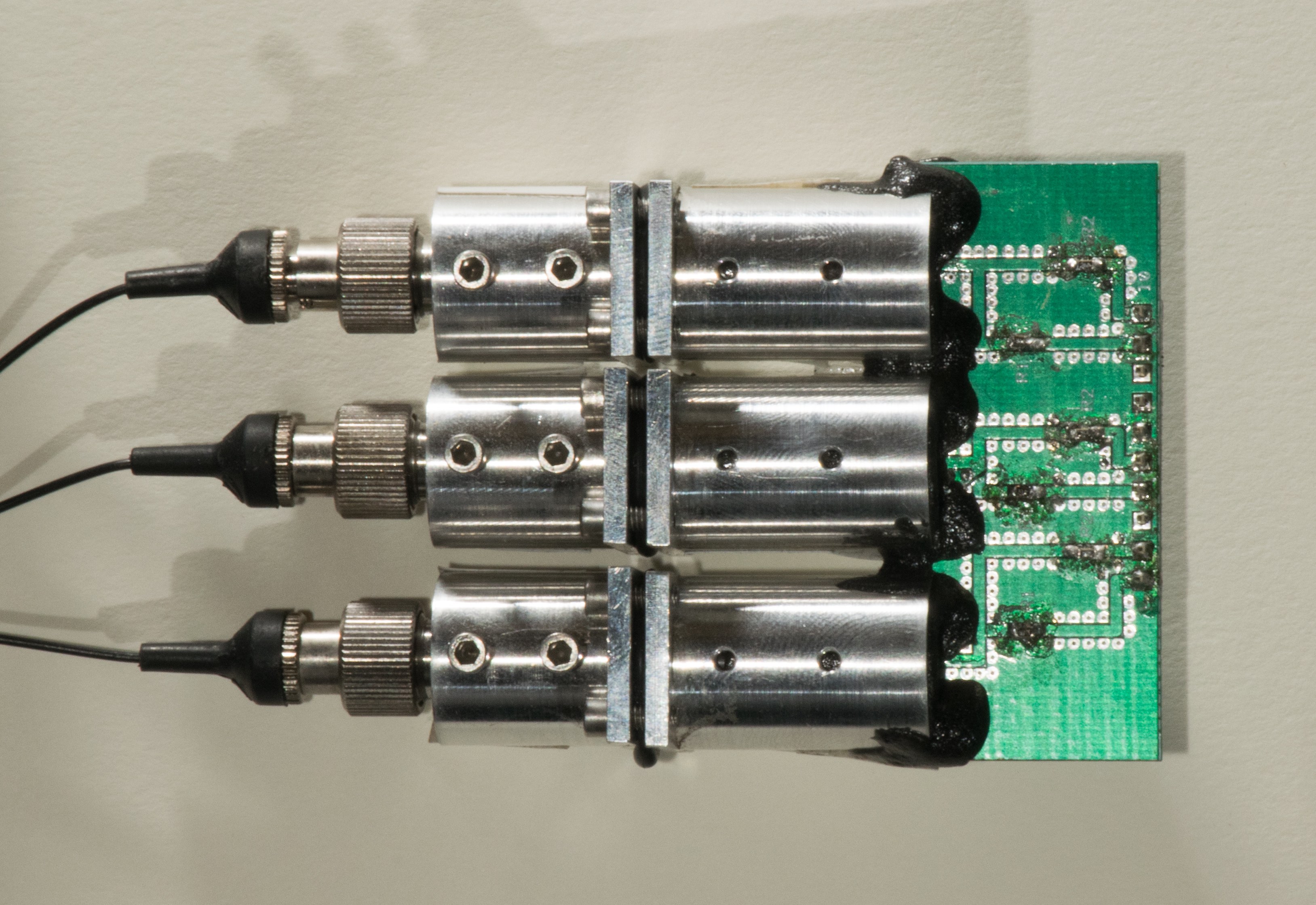} 
    \caption{\label{fig:Resonant_Cavity_LEDs} The QKD source consisting of three RC-LEDs coupled to single-mode optical fibers. (Photo Courtesy Timur Javid.)}   
\end{figure}

\begin{figure} [ht]
    \centering 
    \includegraphics[height=8cm]{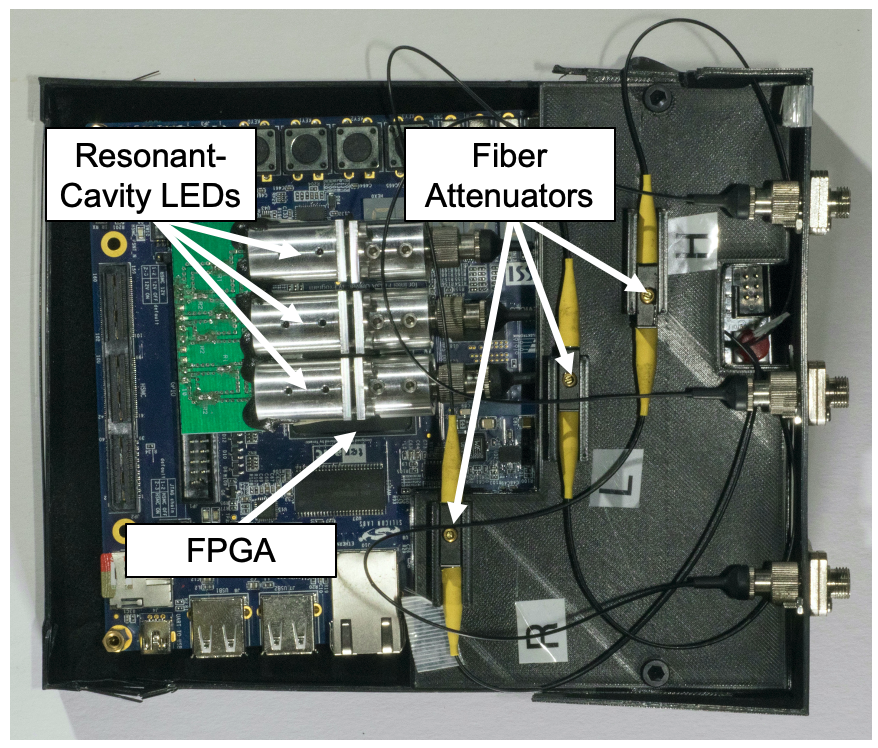} 
    \caption{\label{fig:QKD_Source} The QKD source package includes the RC-LEDs, FPGA, and fiber-based optical attenuators. (Photo Courtesy Timur Javid.)}   
\end{figure}

The QKD source is described in detail in Refs.~\cite{Daniel_QKD_Source,cochran2024full,rosales2024design}.  It produces polarization-encoded states: $\ket{R},\ket{L},\ket{H}$. The QKD sources are three independent resonant-cavity light-emitting diodes (RC-LED, Roithner Lasertechnik, PN: RC-LED-650-02, 7-nm bandwidth), which are mounted in a custom machined case as seen in Fig.~\ref{fig:Resonant_Cavity_LEDs}. The RC-LEDs are electrically driven by an FPGA (Terasic, PN: DE10 standard) at a clock rate of \unit[12.5]{MHz}, and light produced by the RC-LEDs is attenuated to the single-photon level using fiber-based attenuators seen in Fig.~\ref{fig:QKD_Source}. 

Signal (decoy) intensities for the protocol are realized using low- (high-) impedance paths on a custom printed circuit board, which interfaces the FPGA and RC-LEDs shown in Fig.~\ref{fig:Resonant_Cavity_LEDs}. Vacuum states are produced by not driving the RC-LEDs. The selection of polarization state and intensity level is determined randomly using a custom FPGA-based true random number generator \cite{rosin2015ultra}. 

To prevent potential side-channel attacks arising from the distinguishability of the independent sources, light generated by each RC-LED is filtered through a common single-mode fiber and a narrow-band (1-nm-wide) spectral filter (Andover, PN: 656fs02-25). Finally, the electrical pulses generated by the FPGA and driving the RC-LEDs are adjusted to produce temporally indistinguishable QKD states \cite{Daniel_QKD_Source,rosales2024design}.

Using the measurements of the spectral and temporal waveforms of the different states, we compute the classical mutual information fraction an eavesdropper could gain by performing measurements on these degrees of freedom.  This allows us to quantify the vulnerability of the QKD system to potential side-channel attacks.  We estimate $2.4\times10^{-5}$ and $4.3\times10^{-5}$ mutual information fraction leaked in the spectral and temporal waveforms, respectively.  Details on this calculation and measurement bias are presented in \cite{Daniel_QKD_Source}.

\section*{Supplementary Note 3. Quantum State Tomographies}
\label{chap:Note_3}

We use standard quantum state tomography techniques \cite{altepeter2005photonic} to characterize the states transmitted from the QKD source. Ideally, the states should be $\ket{R},\ket{L}$, and $\ket{H}$; however, source imperfections in the optical setup produced quantum polarization states that deviate from their ideal values. We perform quantum state tomography and measured the state purity, defined by
\begin{equation}
    \mathrm{Purity} = \mathrm{Tr}[\rho^2], \label{eq:Purity}
\end{equation}
where $\rho$ is the density operator, and fidelity between the measured state $\rho_m$ and the target state $\rho_1$, is defined by
\begin{equation}
\text{F}(\rho_1,\rho_2) = \left(\mathrm{Tr}\left[\sqrt{\sqrt{\rho_1}\rho_2\sqrt{\rho_1}}\right]\right)^2. \label{eq:Fidelity}
\end{equation}

We measure the states $\ket{R},\ket{L}$, and $\ket{H}$ using weak laser light that passes through each input port of the TX optical bench (propagating one input port at a time). This light passes through the entire QKD optical path, including the 1-nm bandwidth (FWHM) spectral filter. We then project the output state onto $\ket{H},\ket{V}, \ket{D},\ket{A},\ket{R}$, and $\ket{L}$ using a half-wave plate (HWP), quarter-wave plate (QWP), and polarizing beam splitter (PBS). We reconstruct the quantum state using a least-squares optimization function, which allows for the case of an impure density matrix, \textit{e.g.}, $\mathrm{Tr}[\rho^2]\neq 1$, and for the case when we restrict the estimated state to be pure $\mathrm{Tr}[\rho^2] = 1$. The estimated transmitted state (constrained to a pure state) is required as input to the custom finite-key model described in Supplementary Note \hyperref[chap:Note_11]{11}. QKD Channel and Non-Ideal System Modeling.   

The quantum state tomographies also allow us to estimate the quantum bit error rate (QBER) contribution from source imperfections $\text{QBER}_{\text{Source}}$. This is achieved by projecting the estimated states onto ideal projectors using the relation
\begin{equation}
\mathrm{QBER}_{\mathrm{Source}} = \frac{1}{3}\left( \mathrm{Tr}\left[\ket{L}\bra{L}\rho_R\right] + \mathrm{Tr}[\ket{R}\bra{R}\rho_L] + \mathrm{Tr}[\ket{V}\bra{V}\rho_H]\right).
\end{equation}
The $\text{QBER}_{\text{Source}}$ serves as a lower bound to the overall QBER for the QKD system. 

\begin{figure} [ht]
    \centering 
    \includegraphics[height=12cm]{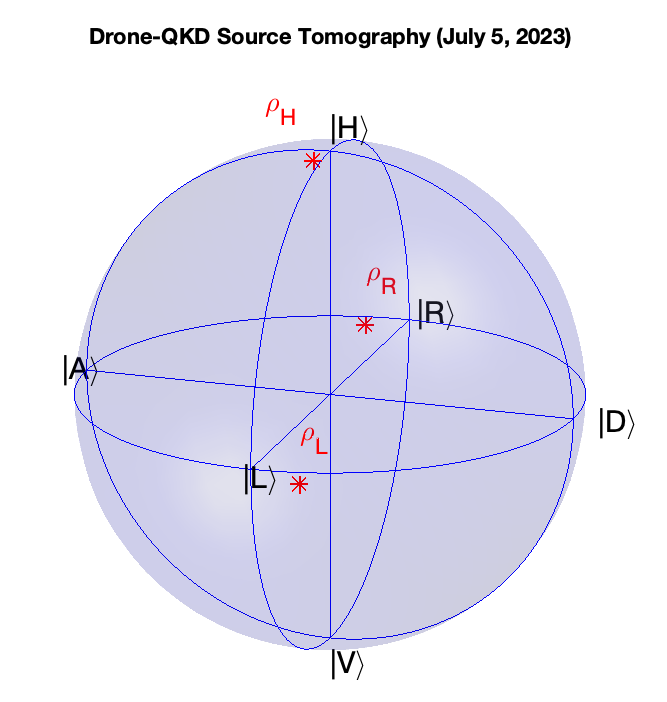} 
    \caption{\label{fig:Tomography} Quantum state tomography of the QKD Source.}   
\end{figure}

For the November 2022 data, the estimated quantum states for $\ket{R}$, $\ket{L}$,  and $\ket{H}$, are given by
\begin{align}
    \label{eq:rho_R_2023}
    \rho_R &= \begin{pmatrix}
                0.4846 + 0.0000i& -0.0863 - 0.4913i\\
                -0.0863 + 0.4913i&0.5154 - 0.0000i
            \end{pmatrix},\\
            \notag\\
    \label{eq:rho_L_2023}
    \rho_L &= \begin{pmatrix}
                0.4730 + 0.0000i   &0.0961 + 0.4853i\\
                0.0961 - 0.4853i&0.5270 + 0.0000i
            \end{pmatrix},\\
            \notag\\
    \label{eq:rho_H_2023}
    \rho_H &= \begin{pmatrix}
                0.9963 + 0.0000i &-0.0142 + 0.0587i\\
                -0.0142 - 0.0587i&0.0037 - 0.0000i
            \end{pmatrix},
\end{align} 
respectively.  The purity and fidelity with the target state are given in Table \ref{table:Tomography_2023}, where it is seen that the purities are greater than $99\%$. Additionally, the QKD source has high fidelities with the target states: We obtain a fidelities between  $98.5\%$ and $99.6\%$, and $\text{QBER}_{\text{Source}} = 1.17\%$. The states are shown on a Bloch sphere in Fig. \ref{fig:Tomography}.

\begin{table}[h!]
\begin{center}
\caption{Quantum state tomography (July 2023 data)}
\label{table:Tomography_2023}
\begin{tabular}
{||c|c|c||} 
 \hline
 Density Matrix & Purity & Fidelity with Target State\\ 
 \hline\hline
 $\rho_R$ & 99.8\%  & 99.1\% \\ 
 \hline
 $\rho_L$ & 99.1\%  & 98.5\% \\ 
 \hline
 $\rho_H$ & 1 & 99.6\% \\ 
 \hline
\end{tabular}
\end{center}
\end{table}

\section*{Supplementary Note 4. Secret Key Rate Optimization}
\label{chap:Note_4}

We use the finite-key model for decoy-state QKD with a signal and two decoy intensity levels of Ref.~\cite{lim2014concise} using security parameters $\epsilon_{sec} = \epsilon_{cor} = 1\times 10^{-12}$, where the probability that Alice and Bob's key is not secret from an eavesdropper $\leq \epsilon_{sec}$, and the probability that Alice and Bob's key are not the same $\leq \epsilon_{cor}$.

The sending fractions of decoy-state QKD protocols are optimized based on channel loss of the link to maximize secret key rate (SKR) in the finite-key regime \cite{rusca2018finite}. We follow a similar approach to Ref.~\cite{rusca2018finite} to find the relative sending fractions of signal, decoy \#1, and decoy \#2 (vacuum) states in simulation using the Drone-to-Drone in-flight QKD transmission data.

\begin{figure} [ht]
    \centering 
    \includegraphics[height=8cm]{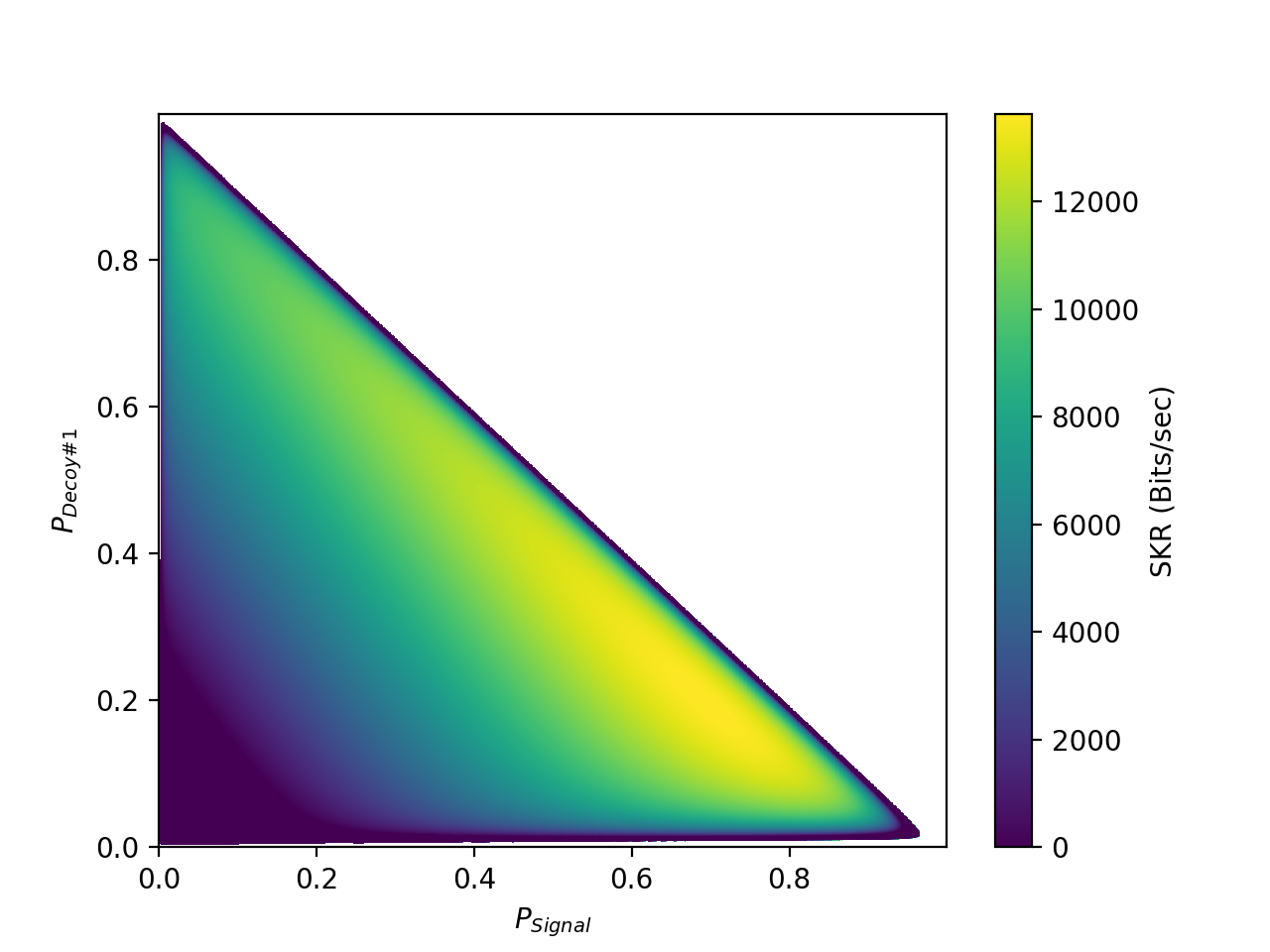} 
    \caption{\label{fig:SKR_Optimization} Dependence of the SKR on the decoy-state protocol on the sending fractions, where $P_{Signal}+P_{Decoy\#1}+P_{Decoy\#2} = 1$.}
\end{figure}

Consider the initial sending fractions of $P_{Signal} = 75\%$, $P_{Decoy \#1} = 18.8\%$, and $P_{Decoy \#2} = 6.3\%$ and drone-to-drone QKD data (Nov 2, 2022), which gives a \unit[$\mathrm{SKR} = 12.8$] {kilobits per second (kbps)} in the finite-key regime using Ref.~\cite{lim2014concise}. The estimated SKR is shown in Fig. \ref{fig:SKR_Optimization} as a function of $P_{Signal}$ and $P_{Decoy \#1}$, where $P_{Decoy \#2} = 1 - P_{Signal} - P_{Decoy \#1}$.  There are some regions where no secret key is achievable in the finite-key regime, which emphasizes the importance of this optimization task.  The optimization yields a maximum SKR when $P_{Signal} = 70\%$, $P_{Decoy \#1} = 19.2\%$, and $P_{Decoy \#2} = 10.8\%$, leading to \unit[$\text{SKR} = 13.6$]{kbps}, which is a 6.2\% improvement from the starting distribution shown in Fig. \ref{fig:SKR_Optimization}. We implement the optimal sending fraction distributions using an 8-bit true random number generator in the transmitter FPGA \cite{rosin2015ultra}, corresponding to a resolution of $1/2^8 \approx 0.4\%$. 

The QKD receiver (RX) optical setup uses a passive 50:50 beam splitter to select between the measurement bases.  The measured non-ideal splitting ratio of the beam splitter at the source wavelength is 56.6/43.4 (\textit{i.e.}, 56.6\% of the light is transmitted towards the R/L measurement and 43.4\% is reflected towards the H/V measurement). Therefore, to maximize the SKR rate, the sending probabilities for each basis are adjusted to $P_{R/L} = 56.6\%$ and $P_{H/V} = 43.4\%$, where $P_{R/L} + P_{H/V} = 1$.  Table \ref{table:Sending_Fraction_Table} gives the ideal and attainable sending fractions.  

\begin{table}[h!]
\begin{center}
\caption{Optimized decoy-state sending fractions to maximize the SKR. Using these attainable sending fractions, we predicted \unit[$\text{SKR} = 13.6$]{kbps}, corresponding to a $6.2\%$ improvement in the finite-key regime.}
\label{table:Sending_Fraction_Table}
\begin{tabular}
{||c|c|c|c||} 
 \hline
 Intensity & State & Ideal Fraction & Implemented Fraction\\ 
 \hline\hline
 Signal & $\ket{R}$ & 50.7/256 & 51/256\\ 
 \hline
 Signal & $\ket{L}$ & 50.7/256 & 51/256\\  
 \hline
 Signal & $\ket{H}$ & 77.7/256 & 78/256\\ 
 \hline
 Decoy \#1 & $\ket{R}$ & 13.9/256 & 14/256\\ 
 \hline
 Decoy \#1 & $\ket{L}$ & 13.9/256 & 14/256\\  
 \hline
 Decoy \#1 & $\ket{H}$ & 21.3/256 & 21/256\\ 
 \hline
 Decoy \#2 & $\ket{\text{Vacuum}}$ & 27.6/256 & 27/256\\ 
 \hline
\end{tabular}
\end{center}
\end{table}

\section*{Supplementary Note 5. Full QKD Results Table}
\label{chap:Note_5}

Table \ref{table:Results_Table_Full} lists results for all quantum key distribution (QKD) experiments that achieved a positive secret key rate (SKR) in the finite-size regime. We obtained a positive SKR with only a single exchange session in each link configuration. To explore future systems with longer sessions, we combined runs with similar statistics to compute the SKR for some configurations as indicated in the table. These combined data sets produce a greater SKR than the weighted average of individual runs, highlighting the importance of longer exchange sessions.

For each exchange session, we adjusted the mean photon number (MPN) $\mu$ to be equal for each signal polarization state; however, the MPN was not consistent from session to session, while the system's QBER was nearly the same at the $\sim2\%-3\%$ level. The link operation time for each session is the total duration when the clock synchronization confidence is greater than $95\%$, which depends on the time the drones (vehicles) are in the air (driving) and the link stability.

\begin{table}[h!]
\caption{Results of all our QKD sessions with a non-zero SKR.}
\label{table:Results_Table_Full}
\begin{center}
\begin{tabular}
{||c|m{0.23\linewidth}|m{0.08\linewidth}|m{0.05\linewidth}|m{0.07\linewidth}|m{0.05\linewidth}|m{0.06\linewidth}|m{0.05\linewidth}||} 
 \hline
 Configuration & Experiment & Signal Mean Photon Number & Flight Time (s) & Average QBER (\%) & Raw Key Rate (kbps) & Secure Key Rate$\dagger$ (kbps) & Total \newline Secret Key (kb)\\ 
 \hline\hline
 \multirow{3}{4em}{Tabletop} & Run 3 (Aug. 8, 2023) & 0.43 & 93.7 & 2.94 & 500.7 & 33.5 & 3,139\\ 
 \cline{2-8}
 & Run 4 (Aug. 8, 2023)$^{\dagger\dagger}$ & 0.43 & 93.70 & 2.87 & 500.5 & 41.5 & 3,885\\ 
 \cline{2-8}
 & Runs 3 and 4 Combined (Aug. 8, 2023) & 0.43 & 187.4 & 2.91 & 500.6 & 55.0 & 10,308\\ 
 \hline
 Ground-to-Ground & Run 1 (Nov. 14, 2023) & 0.44 & 209.1 & 2.83 & 164.60 & 6.1 & 1,278\\ 
 \hline
 \multirow{3}{5em}{Air-to-Air} & Flight 1 (Nov. 2, 2022) & 0.78 & 171.8 & 2.71 & 291.8 & 2.3 & 389\\
 \cline{2-8}
 & Flight 2 (Nov. 2, 2022) & 0.78 & 166.4 & 2.45 & 255.5 & 8.5 & 1,414\\
 \cline{2-8}
 & Flights 1 and 2 Combined (Nov. 2, 2022) & 0.78 & 338.2 & 2.59 & 274.0 & 5.4 & 1,827\\ 
 \hline
 Air-to-Vehicle & Run 1 (Nov. 14, 2023) & 0.44 & 129.8 & 3.43 & 113.0 & 1.6 & 205 \\
 \hline
 \multirow{6}{5em} & 5 mph, Run 1 Illinois Center for Transportation (ICT) (Mar. 27, 2024) & 0.37 & 86.5 & 2.52 & 231.6 & 14.9 & 1,289 \\
 \cline{2-8}
 & 5 mph, Run 3 ICT (Mar. 27, 2024) & 0.37 & 93.7 & 2.58 & 235.0 & 20.0 & 1,876 \\
 \cline{2-8}
 {Vehicle-to-Vehicle} & 5 mph, Run 4 ICT (Mar. 27, 2024) & 0.37 & 86.5 & 2.60 & 215.5 & 14.0 & 1,216 \\ 
 \cline{2-8}
 & 70 mph Interstate Highway Run 1 (Apr. 2, 2024) & 0.54 & 209.1 & 3.35 & 26.1 & 0.7 & 154\\
 \cline{2-8}
 & 70 mph Interstate Highway Run 2 (Apr. 2, 2024) & 0.54 & 209.1 & 3.08 & 45.7 & 2.5 & 519\\
 \cline{2-8}
 & 70 mph Interstate Highway Combined Runs 1 and 2 (Apr. 2, 2024) & 0.54 & 418.2 & 3.18 & 35.9 & 1.6 & 685\\
 \hline
\end{tabular}

$\dagger$ Custom Finite Key Analysis, $\dagger\dagger$ The Run 4 data is presented in Fig. 12 of Ref. \cite{Daniel_QKD_Source}.
\end{center}
\end{table}

Additional details, including a complete list of parameters to calculate secure key rates, are included with the source code: \href{https://openqkdsecurity.wordpress.com/repositories-for-publications/}{Secret Key Rate Code}.

\section*{Supplementary Note 6. Modular QKD Systems and Optical Payload}
\label{chap:Note_6}

\begin{figure}
    \begin{center}
    \begin{tabular}{c c} 
    \includegraphics[height=8cm]{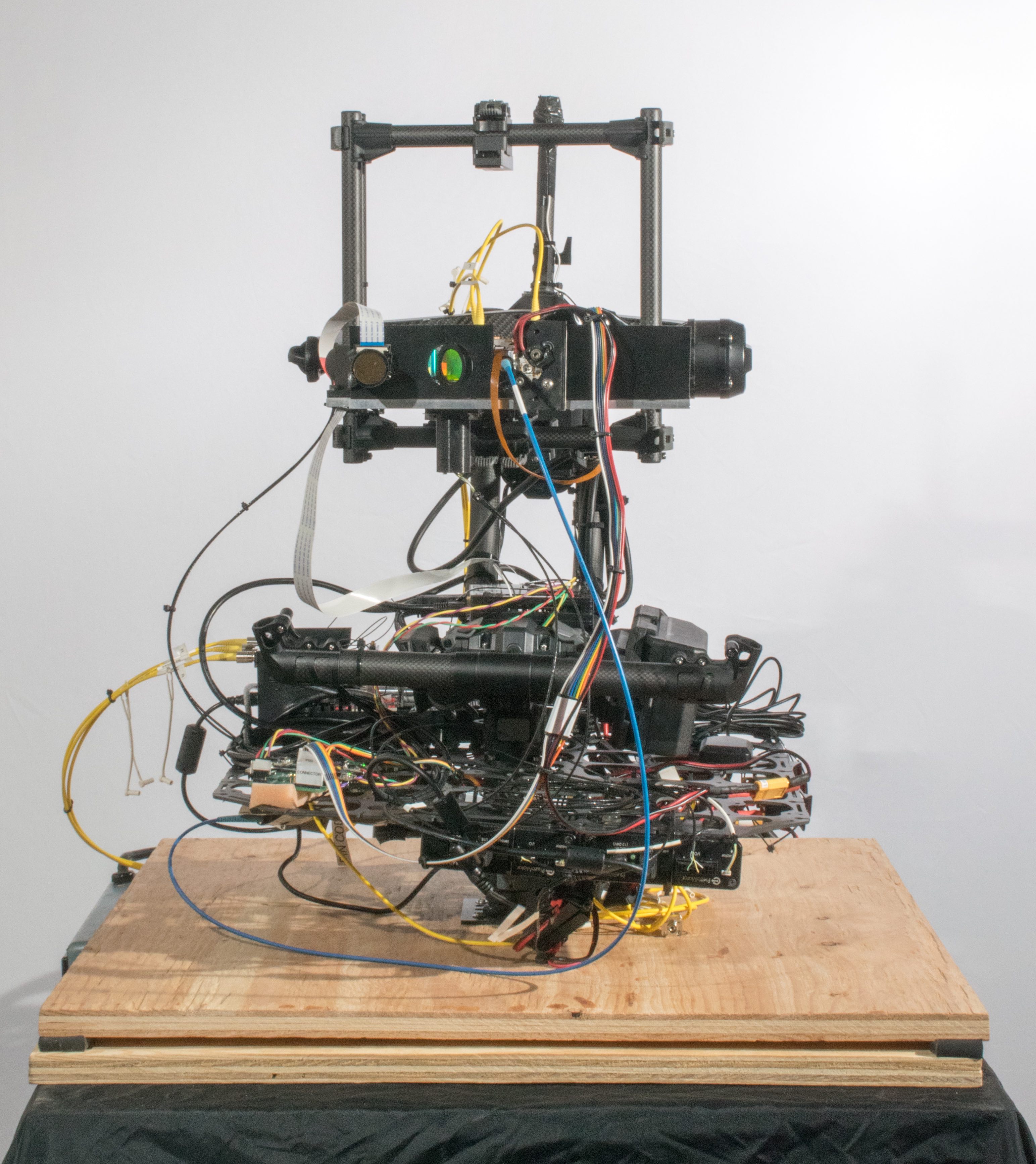} & \includegraphics[height=8cm]{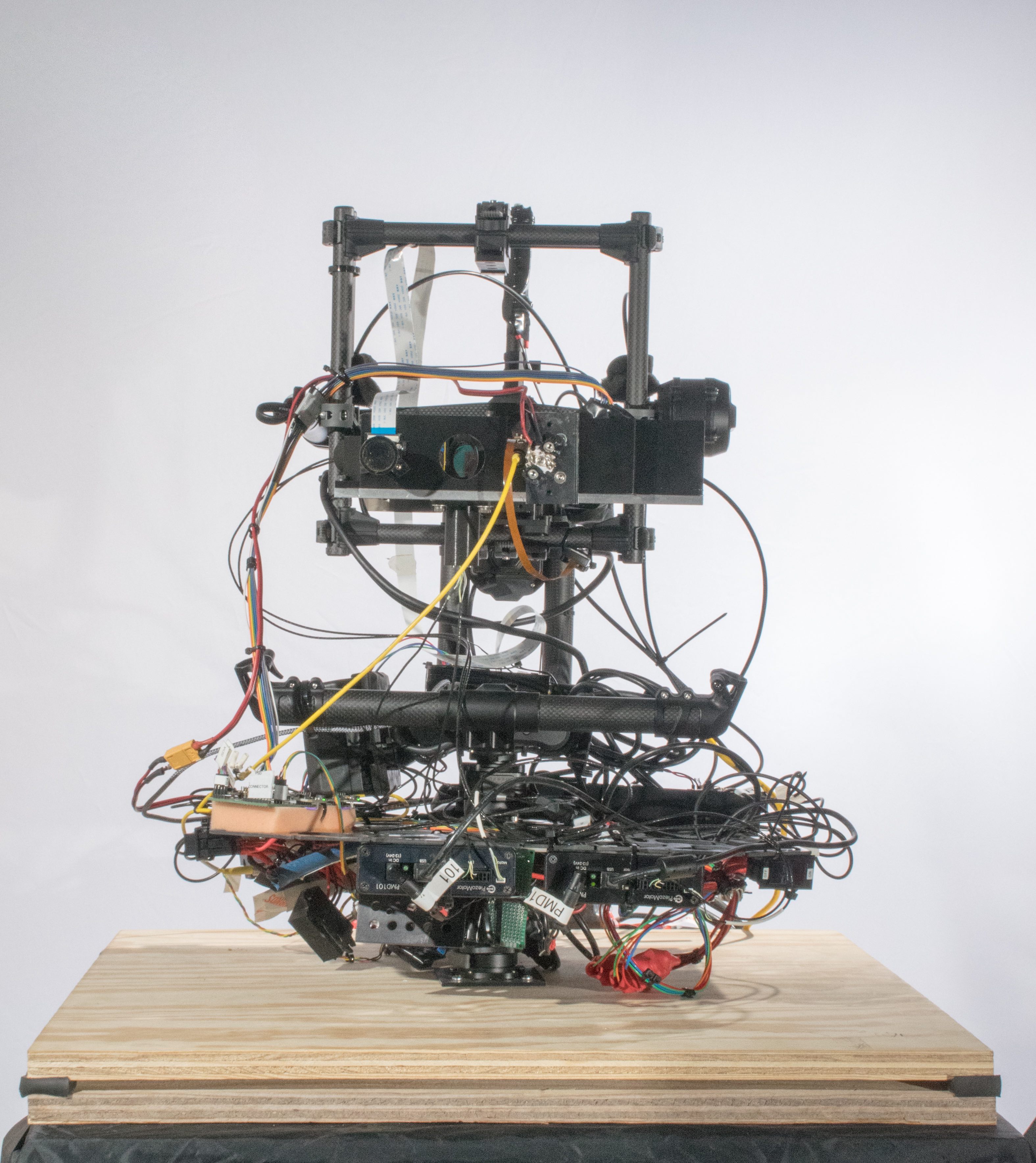}\\
    (a) & (b)
    \end{tabular}
    \end{center}
    \caption{\label{fig:Modular} Modular QKD System. (a) QKD transmitter platform, (b) QKD receiver platform. (Photos courtesy of Timur Javid.)}   
\end{figure}

Figure~\ref{fig:Modular} shows photos of the modular QKD transmitter and receiver.  It has quick disconnects allowing to be easily swapped between mobile platforms and is self-contained.

\begin{figure} [hbt!]
    \centering 
    \includegraphics[width=\textwidth]{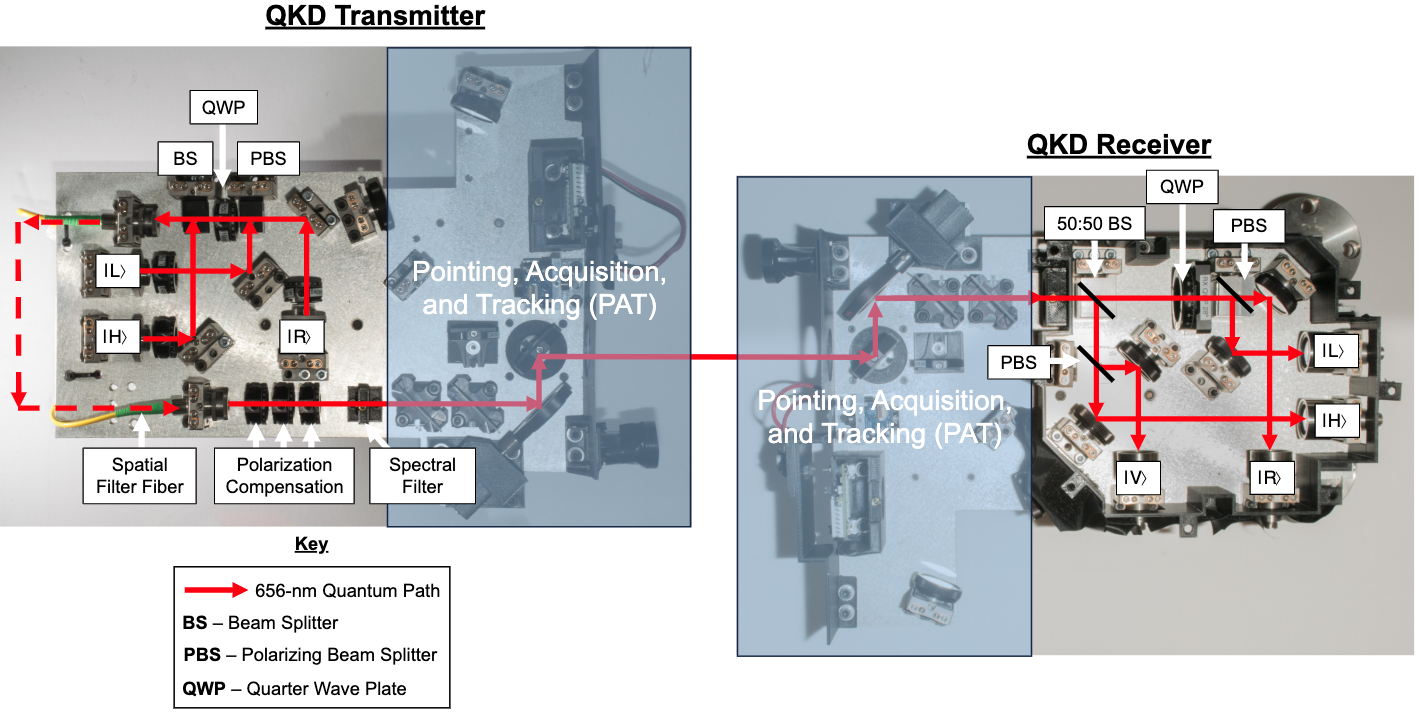}
    \caption{\label{fig:QKD_State_Prep_Measurement_Optics} QKD transmitter and receiver optics. (Photos courtesy Timur Javid.)}   
\end{figure}

We use custom optical benches for state preparation, PAT components, and state projection consisting of an aluminum base platform and 3D-printed mounts as shown in Fig.~\ref{fig:QKD_State_Prep_Measurement_Optics}. The transmitter (TX) optical bench includes the state preparation optics and the PAT subsystem, while the RX optical bench includes state measurement optics, detectors, and the PAT subsystem.    

We first discuss the TX optical bench. Light from the source package shown in Fig.~\ref{fig:QKD_Source} is routed to the TX optical bench via single-mode fibers (SMFs). The $\ket{R}$ and $\ket{L}$ states are produced by passing the RC-LED light through linear polarizers (Thorlabs, PN: LPVISC050) oriented to $\ket{H}$ and $\ket{V}$, respectively, and are combined using a PBS; the output of the PBS then passes through a QWP transforming the states $\ket{H} \rightarrow \ket{R}$ and $\ket{V} \rightarrow \ket{L}$. The final state $\ket{H}$ is produced by passing its RC-LED light through a linear polarizer oriented to $\ket{H}$ and combined with the R/L optical path using a nominal 50:50 BS, which incurs a 50\% loss in the optical path. After the 50:50 BS, the states $\ket{R}$, $\ket{L}$, and $\ket{H}$ are coupled into a short length of Single-Mode Fiber (SMF) to ensure that the spatial modes of each transmitted state are identical.  The light emerging from the SMF is coupled to the free-space channel using an adjustable collimator (Thorlabs, PN: CFC11A-A), passes through a $1$-nm-wide (FWHM) bandpass filter (Andover, PN: 656FS02-25) centered at a wavelength of \unit[656]{nm}, and is routed through the dichroic BS in the PAT system. A detailed analysis of the indistinguishability of the states is given in Ref.~\cite{Daniel_QKD_Source}.

The TX single-mode spatial filter fiber and the dichroic mirrors in the PAT optics perform an unknown unitary transformation on the polarization states, which we compensate using a QWP, half-wave plate (HWP) and another QWP. During outdoor tests, the temperature of the setup drifted over time, where the birefringence of the spatial filter fiber was the most sensitive to a change in temperature. Therefore, we perform polarization compensation before each set of data collection flights. 

We now discuss the QKD receiver. After traveling through free-space, the received QKD states pass through the PAT system and spectral filtering (Andover, PN: 656FS02-25; Thorlabs, PN: FESH0700; Thorlabs, PN: FBH660-10) on the RX optical bench before their polarization is measured. We use a passive 50:50 BS to randomly select the measurement basis, where the R/L (H/V) basis is on the transmitted (reflected) paths.  To measure the R/L states, we use a QWP to transform $\ket{R} \rightarrow \ket{H}$, and $\ket{L} \rightarrow \ket{V}$, which are sorted using a final PBS.  For the other basis, a PBS sorts the states $\ket{H}$ and $\ket{V}$ directly. 

Each sorted state is coupled into a 200-$\mu$m-core diameter, 0.39-numerical aperture black-jacketed multi-mode fiber (MMF, Thorlabs, PN: FT200UMT) using 1"-diameter optical collimators (Thorlabs, PN: F810FC-635). The collimators are modified to mate them with a tip/tilt mount (Newport, PN: HVM-05i).  We find that these collimators have higher coupling efficiency and stability compared to 1/2"-diameter collimators. Finally, the MMFs are coupled to the single-photon counting module described in Supplementary Note \hyperref[chap:Note_8]{8}. Single Photon Detectors below. 

\section*{Supplementary Note 7. Pointing, Acquisition, and Tracking (PAT)}
\label{chap:Note_7}

The PAT system is described in detail in Refs. \cite{conrad2021drone, conrad2023drone, conrad2024full}, and transmits single-photon QKD states from the transmitter platform to the receiver platform while the platforms are in motion.  It consists of two subsystems: a) gimbal control loop, and b) FSM control loop. The gimbal control loop provides initial pointing, acquisition, and slower-speed coarse tracking of the target, while the FSM control loop provides faster fine tracking. A summary of the PAT system specifications is presented in Table \ref{table:PAT_Summary}, and the typical pointing accuracy is presented in Table \ref{table:PAT_Performance_Summary}. 

\begin{table}[h!]
\begin{center}
\caption{Pointing, Acquisition, and Tracking (PAT) hardware summary for Gimbal and Fast Steering Mirror (FSM) control loops for the transmitter (TX) and receiver (RX).}
\label{table:PAT_Summary}
\begin{tabular}
{||m{0.32\linewidth}|c|c||} 
 \hline
 \textbf{Parameter} & \textbf{Gimbal control Loop} & \textbf{FSM control loop}\\ 
 \hline\hline
 Beacon Type & Light Emitting Diode & Fiber-Coupled Laser \\ 
 \hline
 Beacon Wavelength & \unit[850]{nm} & TX: \unit[520]{nm} \\ 
 & & RX: \unit[705]{nm}\\ 
 \hline
 Max Beacon Power & \unit[6]{W} & TX: \unit[15]{mW}\\ 
 & & RX: \unit[15]{mW}\\ 
 \hline
 Angle-of-Arrival Sensor Type & CMOS & Position Sensitive Detector\\
 \hline
 Sensor Field-of-View & $\pm5^{\circ}$ & $\pm0.95^{\circ}$\\ 
 \hline
 Sample Rate & \unit[90]{Hz} & \unit[800]{Hz}\\ 
 \hline
 Angle-of-Arrival Sensor Resolution & 640 x 480 pixels & 16-bit Analog-to-Digital Converter\\
 \hline
 Angle-of-Arrival Sensor Angular Resolution & \unit[152]{$\mu$rad} & \unit[3.33]{$\mu$rad}\\
 \hline
 Control Hardware & 3-axis Gimbal & 1-axis FSM for tip\\
 & & 1-axis FSM for tilt\\
 \hline
 Open-Loop Step Size & \unit[$\sim 95.9$]{$\mu$rad} & \unit[$<1$]{$\mu$rad}\\
 \hline
\end{tabular}
\end{center}
\end{table}

\begin{table}[h!]
\begin{center}
\caption{Typical PAT system closed-loop pointing accuracy (one standard deviation).}
\label{table:PAT_Performance_Summary}
\begin{tabular}
{||c|c|c||} 
 \hline
 \textbf{Link Configuration} & \textbf{Gimbal control Loop} & \textbf{FSM control loop}\\ 
 \hline\hline
 Drone-to-Drone & \unit[700]{$\mu$rad} & \unit[107]{$\mu$rad}\\
 \hline
 Drone-to-Vehicle & \unit[1,200]{$\mu$rad} & \unit[190]{$\mu$rad}\\
 \hline
 Vehicle-to-Vehicle (5 mph) & \unit[900]{$\mu$rad} & \unit[195]{$\mu$rad}\\
 \hline
 Vehicle-to-Vehicle (70 mph)& \unit[3,700]{$\mu$rad} & \unit[549]{$\mu$rad}\\
 \hline
\end{tabular}
\end{center}
\end{table}

\setcounter{section}{7} 
\setcounter{subsection}{0} 
\subsection{Gimbal Control Loop}

The gimbal control loops on the TX and RX system operate simultaneously and independently. The goal of the gimbal control loop is to establish two-way locking (\textit{i.e.}, the transmitter is locked onto the receiver \emph{and} the receiver is locked to the transmitter). After two-way locking is achieved, the QKD signals are sent from the TX to the RX.  Using only the gimbal control loop, we achieve only moderate coupling efficiency and stability, thus requiring a faster and finer control loop, as described in the next section.

The gimbal control loop consists of a gimbal (Freefly Systems, PN: M$\bar{\text{o}}$vi Pro), near-infrared (NIR, 850-nm wavelength) beacons, and a camera system (Kuman, PN: Raspberry Pi Camera Module). The custom software-based control system uses a Raspberry Pi (Version 4 with \unit[4]{GB} RAM, RT-Patch) single-board computer. After takeoff, the control system automatically begins target acquisition, which moves the gimbal in a spiral pattern until the NIR beacon of the opposite drone is detected using image processing software running on the Raspberry Pi. The gimbal receives commands from the Raspberry Pi at a rate of \unit[50]{Hz} over a serial connection. Once the NIR beacon is detected, the gimbal control loop begins tracking the target using a proportional, integral, derivative (PID) controller. Both TX and RX gimbal control loops operate independently without requiring communication or interaction with each other. 

\begin{figure} [h!tbp]
    \centering
    \begin{tabular}{c c} 
    \includegraphics[height=4.5cm]{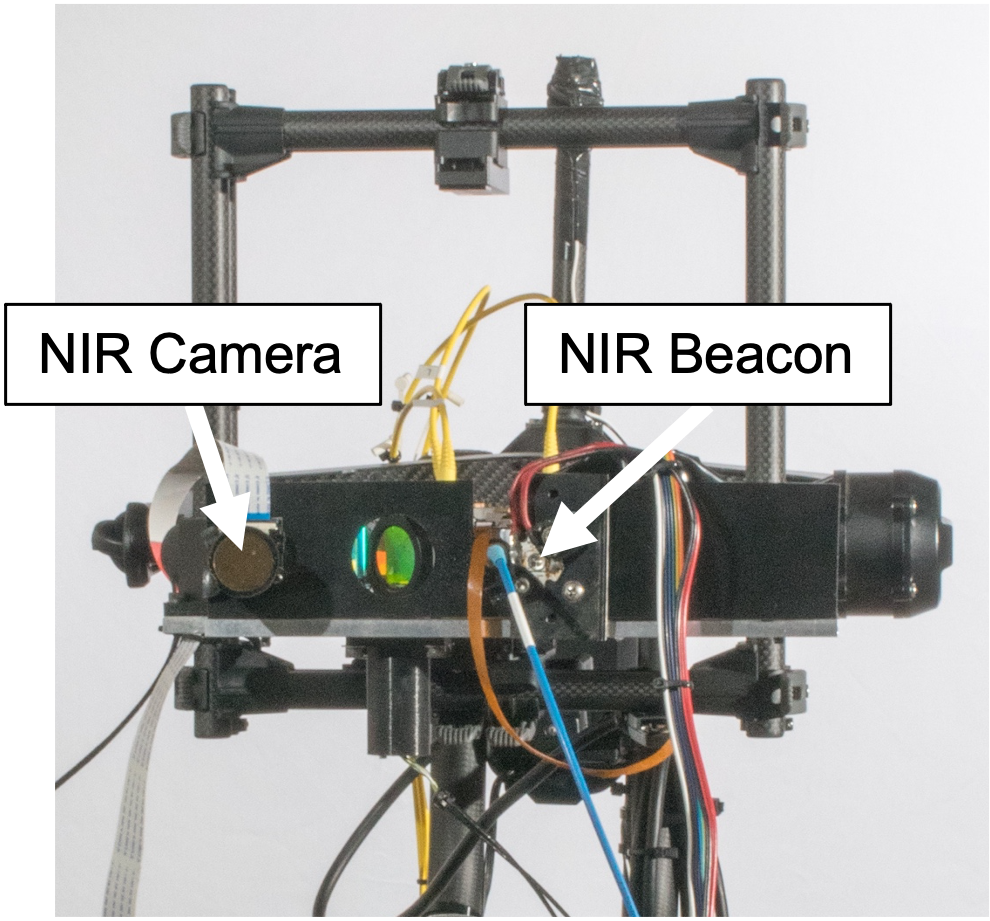} &  \includegraphics[height=4.5cm]{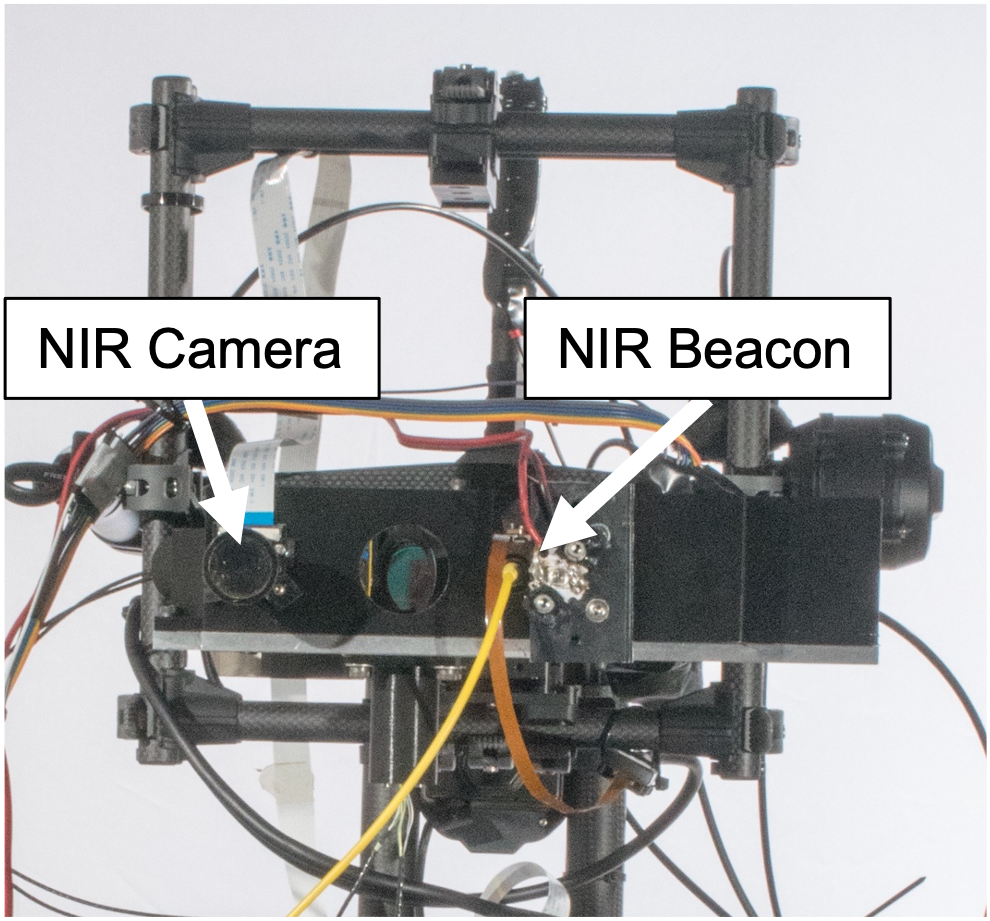}\\
    (a) & (b)\\
    \includegraphics[height=4.5cm]{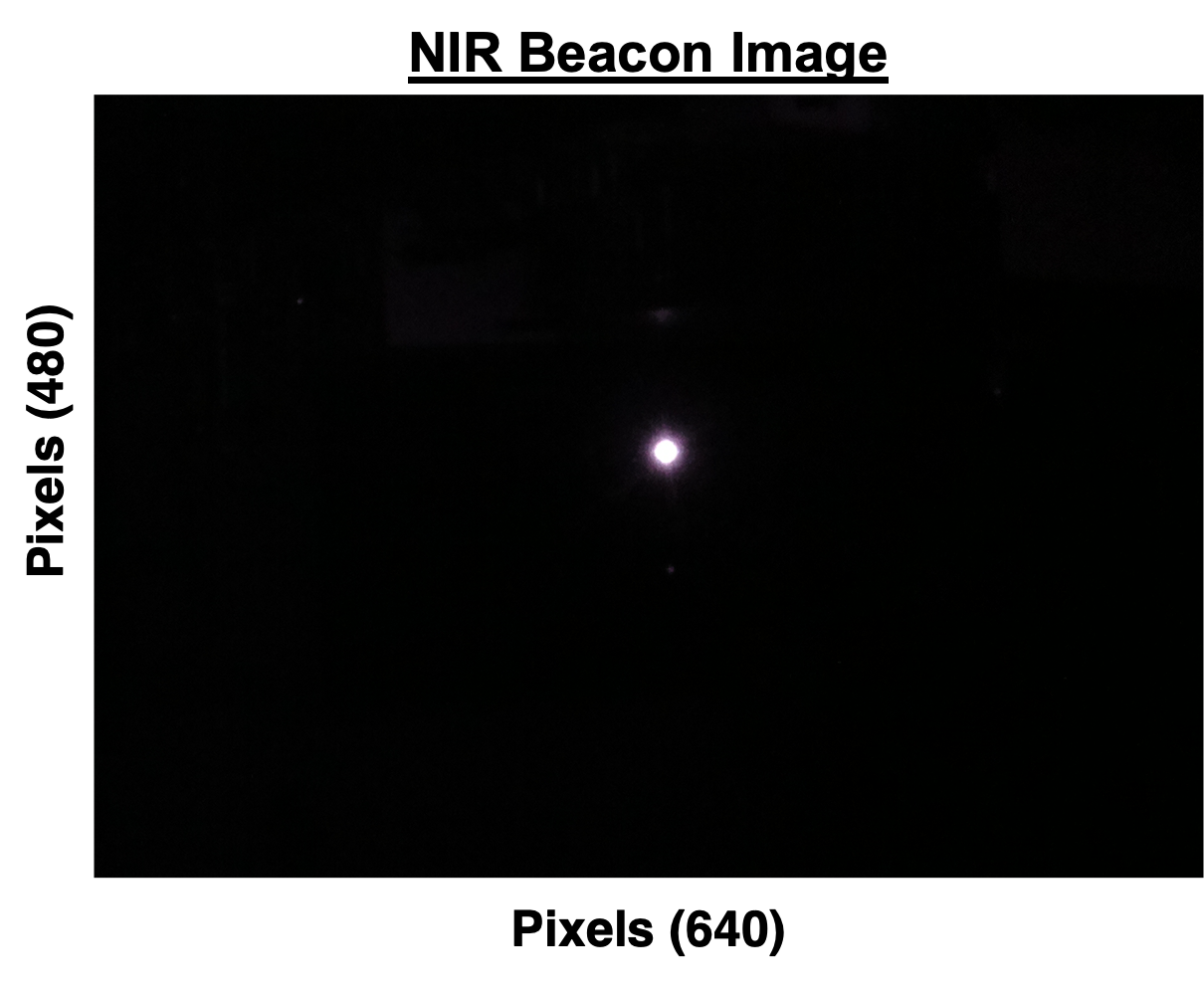} &  \includegraphics[height=5cm]{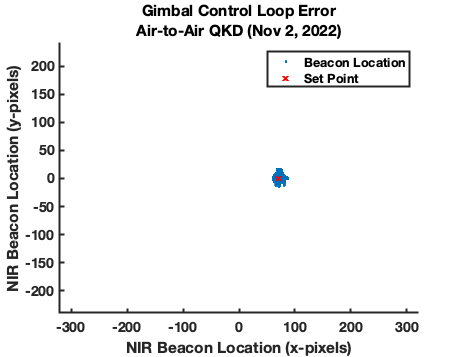}\\
    (c) & (d)
    \end{tabular}
    \caption{\label{fig:Gimbal_Control_Loop} Gimbal control loop: (a) transmitter gimbal; (b) receiver gimbal; (c) Near Infrared (NIR) beacon image in camera's frame; and (d) Drone-to-Drone tracking performance in closed-loop, (Nov 2, 2022), where blue dot represents the NIR beacon's location near the target set point for each video frame. (Photos courtesy Timur Javid and Andrew Conrad.)}   
\end{figure}

The gimbal control loop architecture is presented in Fig.~\ref{fig:Gimbal_Control_Loop}. We place a bandpass filter centered at \unit[$850$]{nm} ($40$-nm bandwidth) (Thorlabs, PN: FB850-40) in front of the NIR camera to reduce stray background light. The image of the opposite NIR beacon as seen by the NIR camera is presented in Figure \ref{fig:Gimbal_Control_Loop}(c). Ideally, the beacon from the other drone appears as a bright disk whose centroid is determined using image-processing software.  The difference between the measured and desired centroids is used to generate an error signal for controlling the gimbal. The performance of the gimbal control loop is given in Fig.~\ref{fig:Gimbal_Control_Loop}(d) for one of our Drone-to-Drone QKD flights, which shows the beacon's location (blue dot) in each video frame around the set point (red dot). The tracking error is measured in pixels (and converted to degrees using the camera's field of view). The resulting accuracies for the gimbal-control loop are presented in Table \ref{table:PAT_Performance_Summary}.

\subsection{FSM Control Loop}

\begin{figure} [hbt!]
    \centering 
    \includegraphics[height=6.5cm]{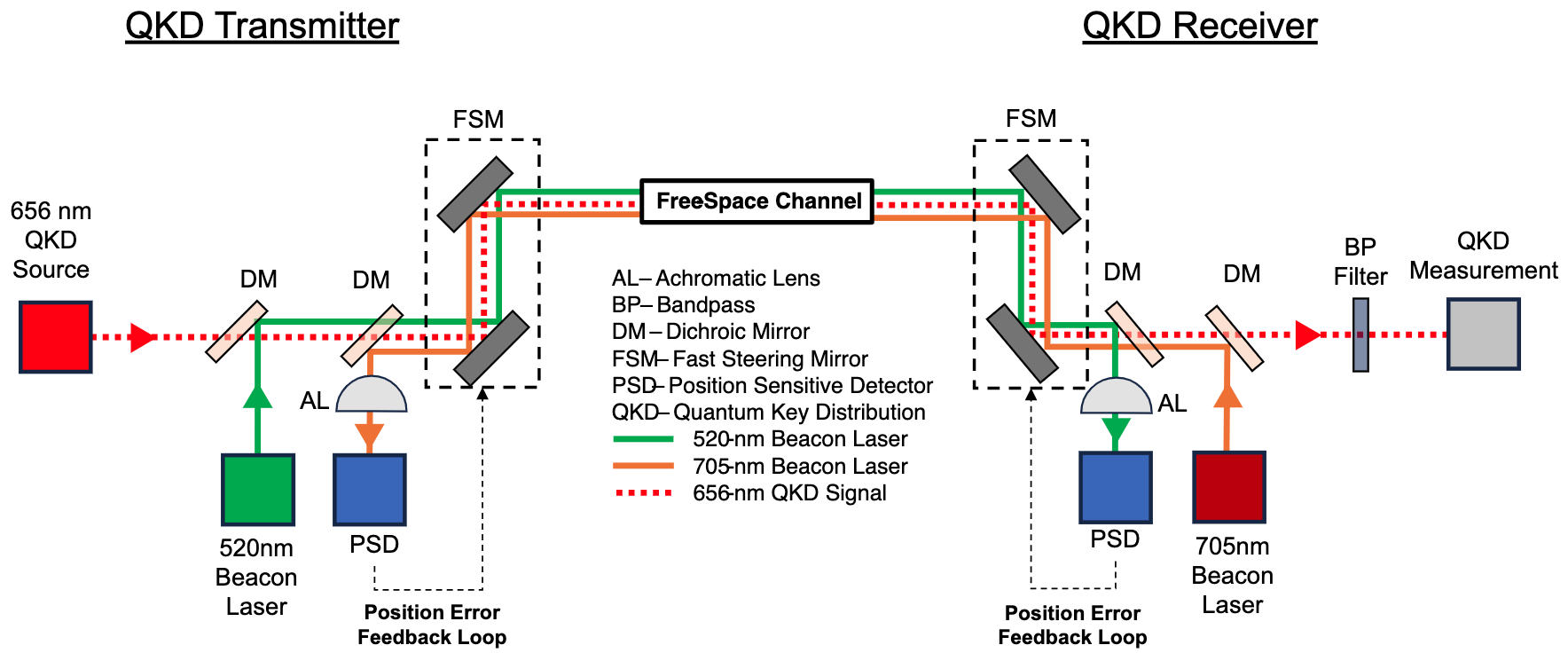} 
    \caption{\label{fig:FSM_Loop_Overview} FSM control loop optical schematic.} 
\end{figure}

\begin{figure} [hbt!]
    \centering 
    \includegraphics[height=7.5cm]{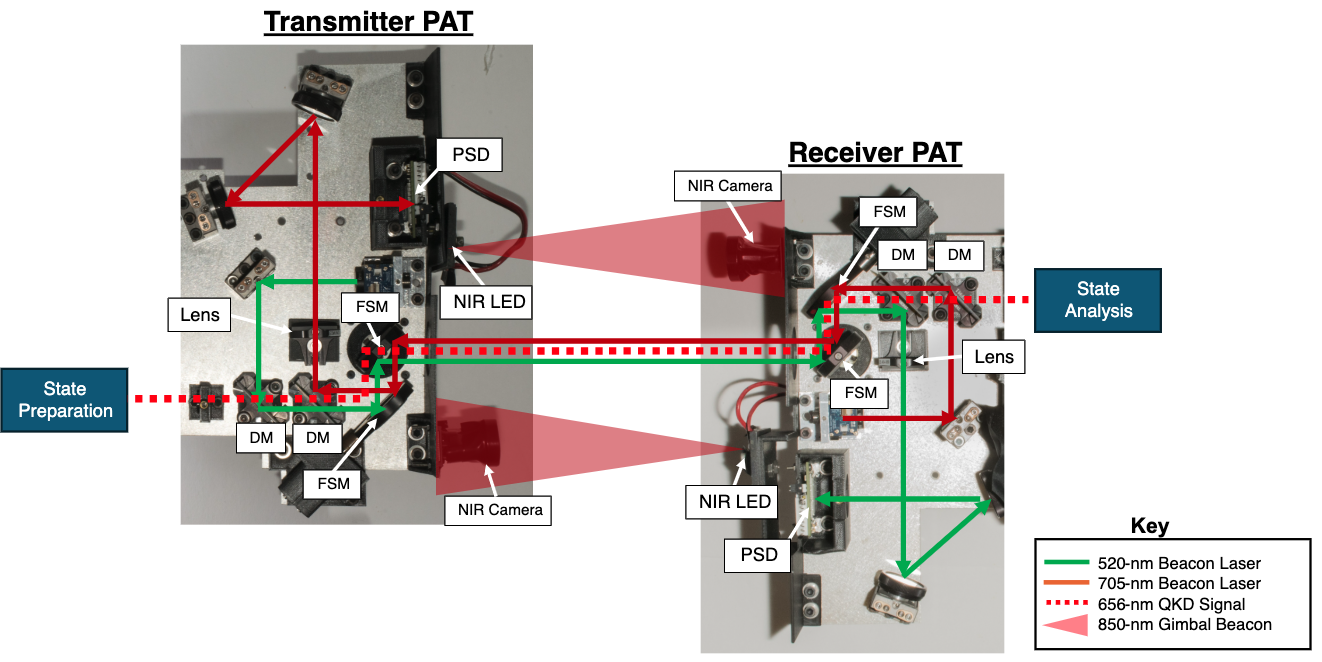} 
    \caption{\label{fig:PAT_In_Practice} FSM control loop consists of laser beacons, Fast Steering Mirrors (FSM), Dichroic Mirrors (DM), and Position Sensitive Detectors (PSD). (Photos courtesy Timur Javid.)}   
\end{figure}

A schematic of the FSM control loop overview is presented in Fig.~\ref{fig:FSM_Loop_Overview} with photos and additional details in Fig.~\ref{fig:PAT_In_Practice}.  Each drone has a position-sensitive detector (PSD) to measure the incoming angle-of-arrival (AoA) of the other drone's beacon. Based on the AoA, the FSM are adjusted so that the outgoing line-of-sight matches the incoming AoA. 

The laser beacon beams are added or dropped from the quantum optical path using dichroic mirrors (DM) and are aligned with each other. To measure the AoA, we use an achromatic lens (Thorlabs, PN: AC254-300-C, focal length $f=$ \unit[300]{mm}) to focus the incoming beacon beam onto a position-sensitive detector (PSD, First Sensor, PN: 5000011) located in the focal plane of the lens. The PSD generates an analog voltage that is processed to determine the beam's transverse location on the PSD. The position resolution of the PSD depends on the optical power level and bias voltage, with a base-scale resolution of \unit[1]{$\mu$m}. The angular resolution is given by $\theta_r \approx d/f$, where $d$ is the position error on the PSD and has a best-case resolution of \unit[3.33]{$\mu$rad}. The ideal target setpoint on the PSD is aligned to its center. Therefore, the beam's position on the PSD is equal to the error signal, which is used to control the FSMs (Piezo Motor, PN: LR17) to direct the beam's location to the origin.

The FSM control loop uses two counter-propagating beacon lasers and two single-axis FSM's (PiezoMotor, PN: LR17). Alice transmits an alignment laser (520-nm wavelength, Thorlabs, PN: LP520-SF15) to Bob, which is received and focused onto Bob's PSD located in the focal plane of the AoA sensor. Bob uses a locally generated error signal to control his FSMs, where the goal of the control loop is to center the incoming beacon on his PSD.  Simultaneously, Bob transmits a counter-propagating beacon laser (705-nm wavelength, Thorlabs, PN: LP705-SF15) to Alice, which is focused on her PSD. Alice locally generates an error signal based on the incoming beam's position on her PSD, which is used to control her FSMs.  The control loop update rate is \unit[800]{Hz}.  Using both the gimbal and FSM control loops, we obtain a typical channel loss of \unit[-2.23]{dB} (60\% transmission) through the RX optics into the MMFs while both drones are in flight.

Since operating the original PAT system described here, we have shifted to using NIR laser beacons (at \unit[1,425]{nm} and \unit[1,550]{nm}). This was done to make the laser beacons eye-safe and increase the beacon's output power to extend the operational range to longer distances.  

\subsection{Vibration Performance}

Platform vibrations from the drone or vehicle degrade the stability of the free-space link, and must be compensated by the PAT system. Our system has several vibration mitigation systems in place, including vibration dampers (Freefly Systems, teal isolator cartridges) that isolate the drone rotors from drone vehicle. Additionally, we use a stabilized gimbal system to isolate platform vibrations from the optical payload in both the drone and vehicle configurations. We also use Sorbothane vibrational dampers for the base plates when using the vehicle configuration.

\begin{figure} [hbt!]
    \begin{center}
    \begin{tabular}{c} 
    \includegraphics[height=5cm]{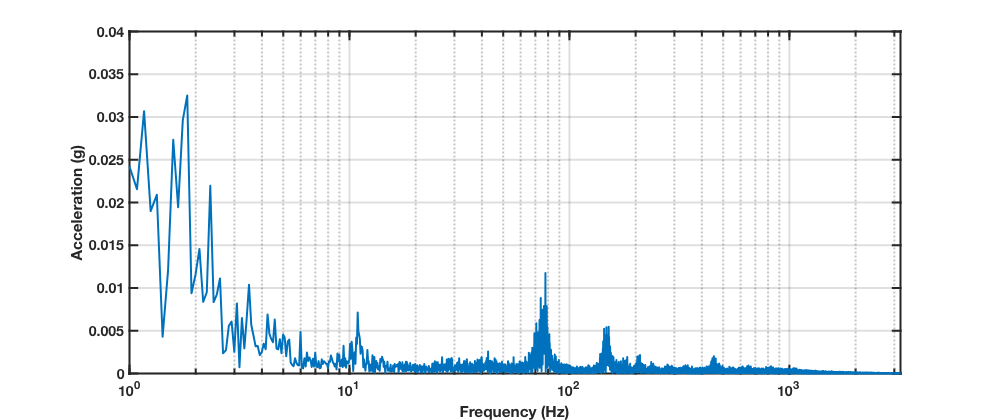}\\
    (a) \\
    \includegraphics[height=5cm]{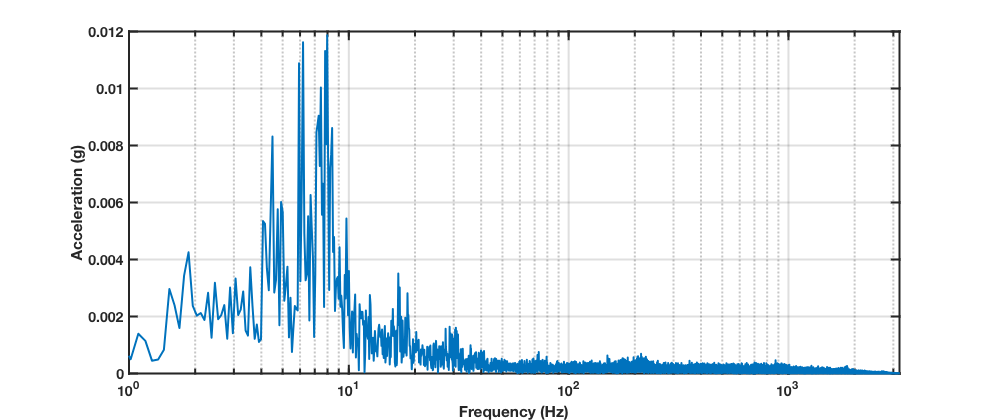} \\
    (b) \\
    \includegraphics[height=5cm]{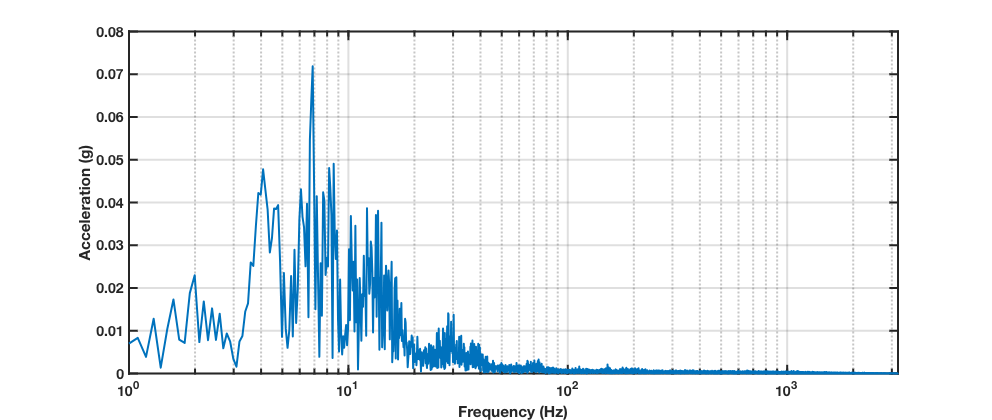} \\
    (c) \\
    \includegraphics[height=5cm]{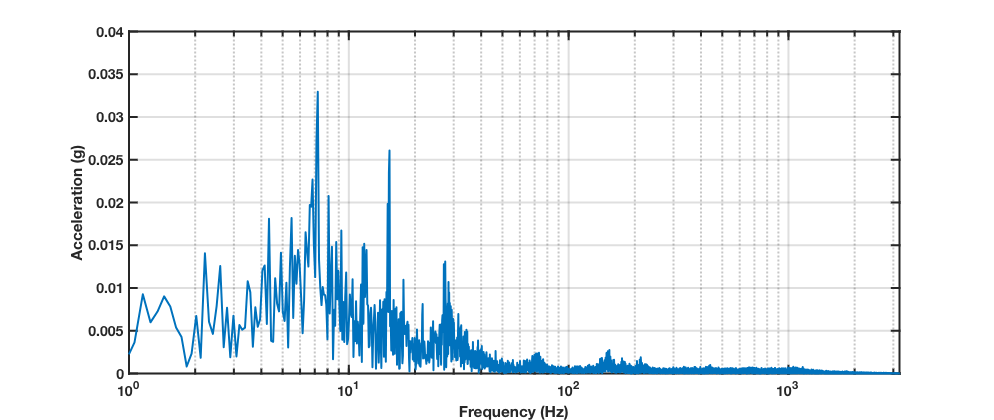} \\
    (d) 
    \end{tabular}
    \end{center}
    \caption{ \label{fig:Vibrations} Platform vibration frequency domain (a) flying drone, (b) 5-mph vehicle on pavement, (c) 10-mph vehicle on gravel, and (d) 70-mph vehicle on an Interstate Highway. Note that the vertical scales are different in each plot.}   
\end{figure}

Fig. \ref{fig:Vibrations} shows the vibrational profile of the host drone and vehicle platforms in several configurations, including (a) hovering drone, and a vehicle traveling (b) \unit[5]{mph}, (c) \unit[10]{mph}, and (d) \unit[70]{mph}. The vibration data in the vehicle in the latter cases is collected with the closest window rolled down (as in the QKD runs). Additionally, the 5-mph and 70-mph vibration datasets are collected on pavement (as in our QKD runs), while the 10-mph dataset is collected on gravel to match our drone-to-vehicle QKD run. The control bandwidth of our FSM-loop is \unit[$\sim$100]{Hz}, which is sufficient to compensate for most platform vibrations.

\section*{Supplementary Note 8. Single-Photon Detectors}
\label{chap:Note_8}

\begin{figure} [hbt!]
    \centering 
    \includegraphics[height=5.5cm]{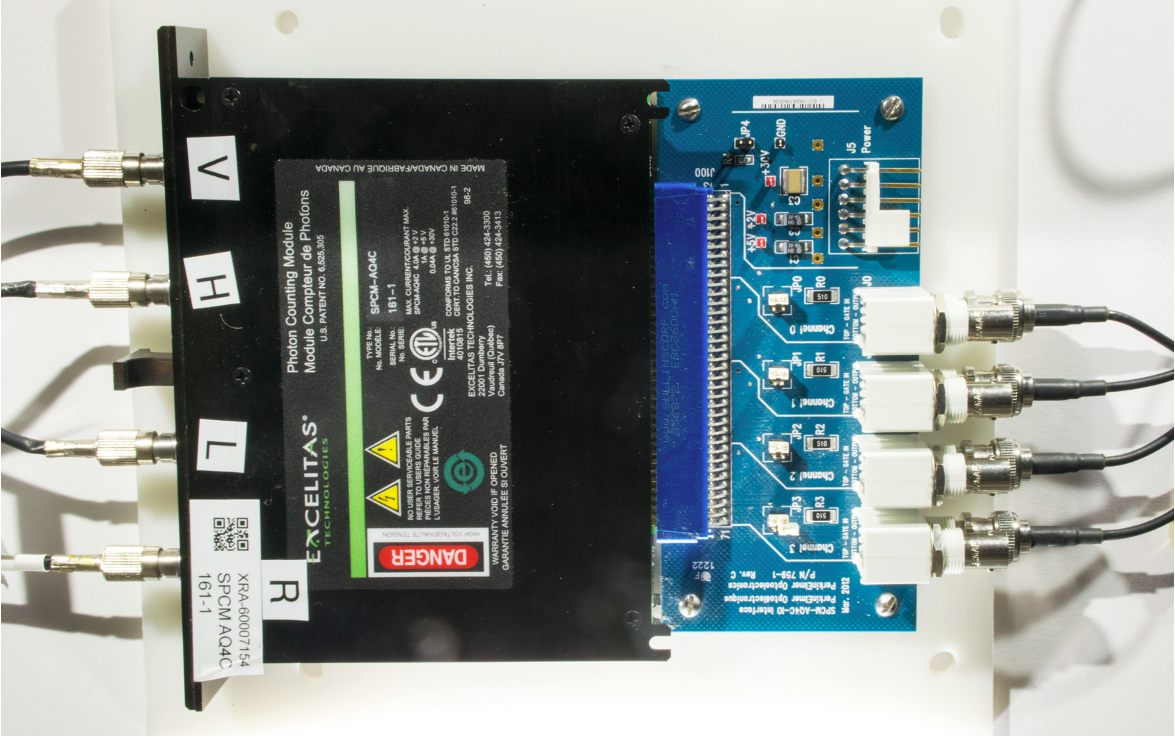} 
    \caption{\label{fig:Single_Photon_Detectors} Excelitas 4-Channel Single-Photon Counting Module. (Photo courtesy Timur Javid.)}   
\end{figure}

To measure the quantum states, we use a four-detector single-photon counting module (Excelitas, PN: SPCM-AQ4C) shown in Fig.~\ref{fig:Single_Photon_Detectors} with one detector for each quantum state: $\ket{R}, \ket{L}, \ket{H}, \ket{V}$. The peak quantum efficiency is $60\%$ and occurs at a wavelength of \unit[650]{nm}, which is close to the QKD source wavelength.  The detectors each have a dark count rate of $\sim$500 counts/s, a 50-ns-long dead-time, require electrical power (\unit[+2]{V}, \unit[+5]{V}, and \unit[+30]{V}) that is generated using buck and boost voltage converters tapped into the \unit[+25]{V} power bus. The SPCM-AQ4C has an output impedance of \unit[50]{$\Omega$} and generates a 50-ns-long high transistor-transistor logic (TTL) pulse for each event. The outputs are routed via coaxial cable to a custom time tagger described in the next section.  

\section*{Supplementary Note 9. Time Tagger}
\label{chap:Note_9}

\begin{figure}[hbt!] 
    \centering 
    \includegraphics[height=5cm]{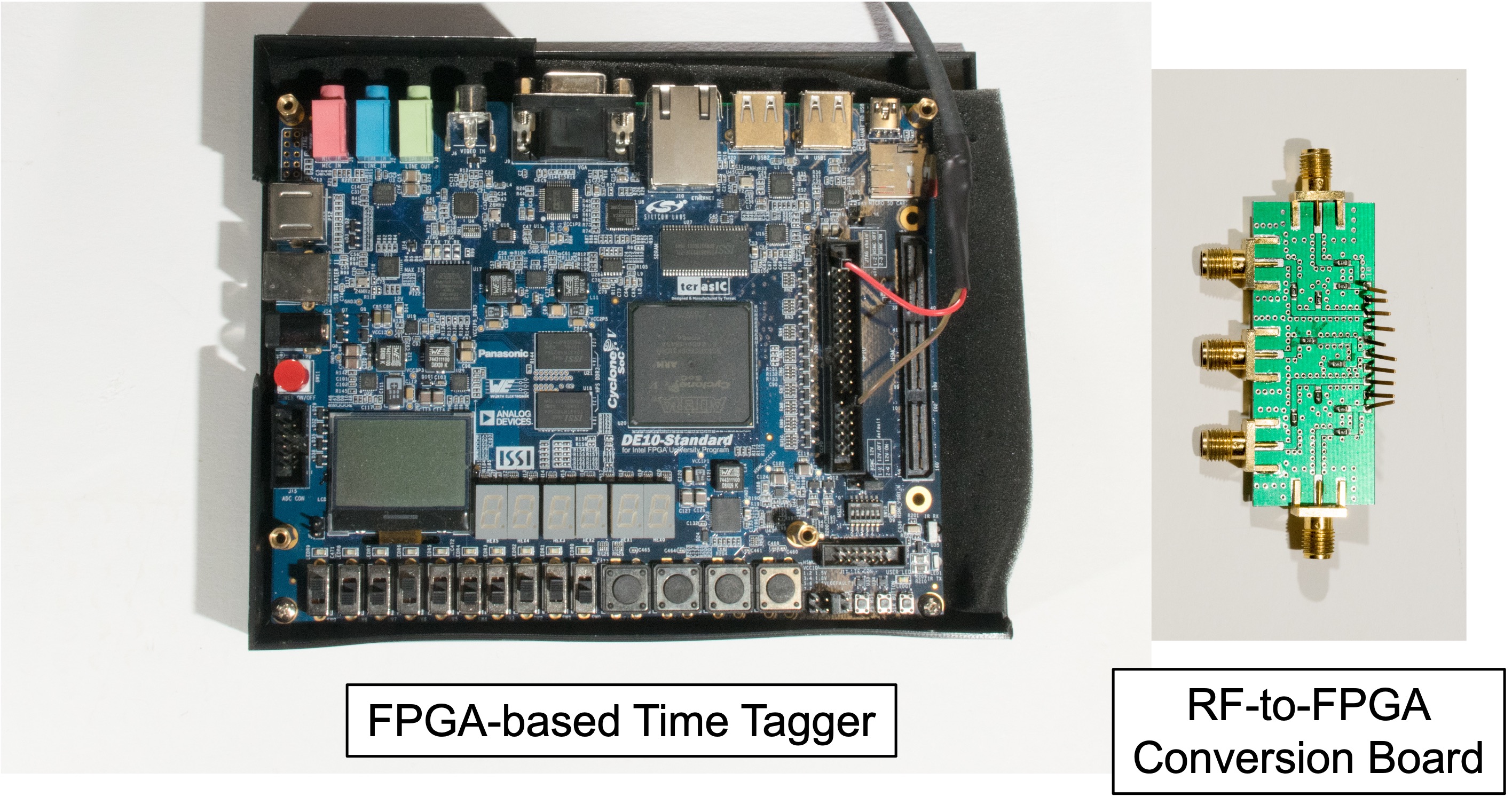} 
    \caption{Time-tagging system. (Photos courtesy Timur Javid.)} 
    \label{fig:Time_Tagger} 
\end{figure}

We created a custom low-SWaP time tagger to record the photon detection events from each quantum channel shown in the left panel of Fig.~\ref{fig:Time_Tagger}.  It is based on an FPGA evaluation board (Terasic, PN: DE10-Standard). The coaxial cables from the single-photon detectors are attached to a custom circuit board that reduces the TTL pulse voltage to \unit[3.3]{V} using a 50-$\Omega$ impedance $\Pi$ resistive network and connects the signals to the general-purpose input/output (GPIO) connects on the FPGA, as shown in Fig.~\ref{fig:Time_Tagger}, right panel.  The time tagger operates using a $100$-MHz system clock and saves the value of a 30-bit counter (with an additional 2-bits allocated to save channel ID) to a micro-SD card in blocks to speed up the transfer time.  The received-event time tags for each detector are written to a separate file in the order they are received. Additional details about the FPGA architecture used for both transmitter and receiver are given in Ref.~\cite{rosales2024design}.   

\section*{Supplementary Note 10. Post-Processing Time Synchronization}
\label{chap:Note_10}

Since the timetags are 30 bits long and we are time tagging at a rate of \unit[100]{MHz}, our counter rolls over every \unit[10.74]{s}. To reconstruct the absolute time of every received event, we use the difference between successive time tags mod $2^{30}$ to infer the elapsed time.  However, if there are no detected events in successive timing windows, this would result in a timing ambiguity.  We resolve this potential issue by writing an artificial event at the beginning of each time loop when taking data. After the absolute timing is reconstructed for all events, the artificial events are removed as part of the post-processing.

To synchronize the events from each detector, we establish a common time point across each channel and then stitch them together into one $4 \times N$ array of the presence or absence of a detection event at each detector at each of $N$ time steps.  The QKD source generates signals for a period of \unit[4.86]{s} and then sends nothing for \unit[0.5]{s}, to aid in time synchronization, for drone-to-drone flights. The period was updated to \unit[6.7]{s} on and \unit[0.5]{s} off for drone-to-vehicle and vehicle-to-vehicle tests, to provide a higher transmission duty cycle. The transitions between these parts of the cycle provide an identifiable data pattern occurring in all four detection data streams, allowing us to establish a common time point to stitch the arrays together.  By analyzing the probability of the gaps between successive time tags during the source active and source inactive times, we can pinpoint this transition with precision to within one counting cycle, which is enough to remove the timing ambiguity. 

Because we do not share a classical signal between the sender and receiver, the relative timing is subject to an initial offset and drift over time, which we must correct to perform the key analysis.  The drift between the TX and RX clocks is due to small differences in the crystal oscillator frequencies on the FPGAs, resulting in a linear timing drift.  We compensate the frequency mismatch during data post-processing by examining the received data to measure the true period, and then rectifying the data to force the period to be 8 time bins (there are 8 time bins for each QKD clock period).  

To find the relative timing offset between the sender and receiver, we use a qubit-based Bayesian synchronization protocol \cite{cochran2021qubit}.  The information Alice already publishes for basis and decoy-state reconciliation has significant correlations with Bob's measurement outcomes.  In our protocol, Bob computes the probabilities of the various event pairings (\textit{e.g.}, Alice sends decoy in the H/V basis, Bob receives a detection in R) and counts the numbers of all such pairing for the different possible timing offsets using fast Fourier transforms.  He finds the offset with the maximum correlation using a Bayesian statistical algorithm, the value of which constitutes the confidence in that finding. We typically use a string of 100,000 time bins (\unit[1]{ms}-interval), which gives us near unity synchronization confidence for our data. 

Some of the data sets only include $\sim$90\% of the total received data because of an error in an earlier version of the FPGA code related to a bottleneck in writing data to the SD card.  These losses interrupted the synchronization process by introducing unexpected offsets between the transmitted and received data. This data loss occurred in large enough blocks that the induced offset exceeded a computationally reasonable search range for our synchronization algorithm, but infrequently enough that we seldom saw more than one lost block in a \unit[4.86]{s} period.  By identifying how much data was lost from a given \unit[4.86]{s} signal period, we infer the length of the lost block(s) and identify the moment when the synchronization begins to fail.  With this knowledge, we add the appropriate offset at the location of the lost block, allowing us to process the remainder of the dataset correctly.

For all datasets, we filter out the lowest quality data by only keeping the synchronization blocks that surpass a 95\% synchronization confidence and a better than 0.2 noise fraction averaged across the 4 detectors.  Once the data is synchronized, we perform sifting, calculate the quantum bit error rates, run the security analysis, and perform privacy amplification to distill a secure key.

\section*{Supplementary Note 11. QKD Channel and Non-Ideal System Modeling}
\label{chap:Note_11}

Our custom finite-key model accounts for non-ideal properties of our QKD transmitter and receiver setups, and the free-space link. Many finite-key approaches assume system symmetries that do not necessarily exist, \textit{e.g.}, perfect beam splitters, equal detector efficiencies between channels, or equal transmitted MPNs \cite{rusca2018finite}. However, these assumptions are not valid for our system or, for that matter, most practical quantum systems. We seek to directly model non-ideal characteristics of our system and incorporate these limitations into the finite-key model. 

\begin{figure} [hbt!]
    \centering 
    \includegraphics[height=5cm]{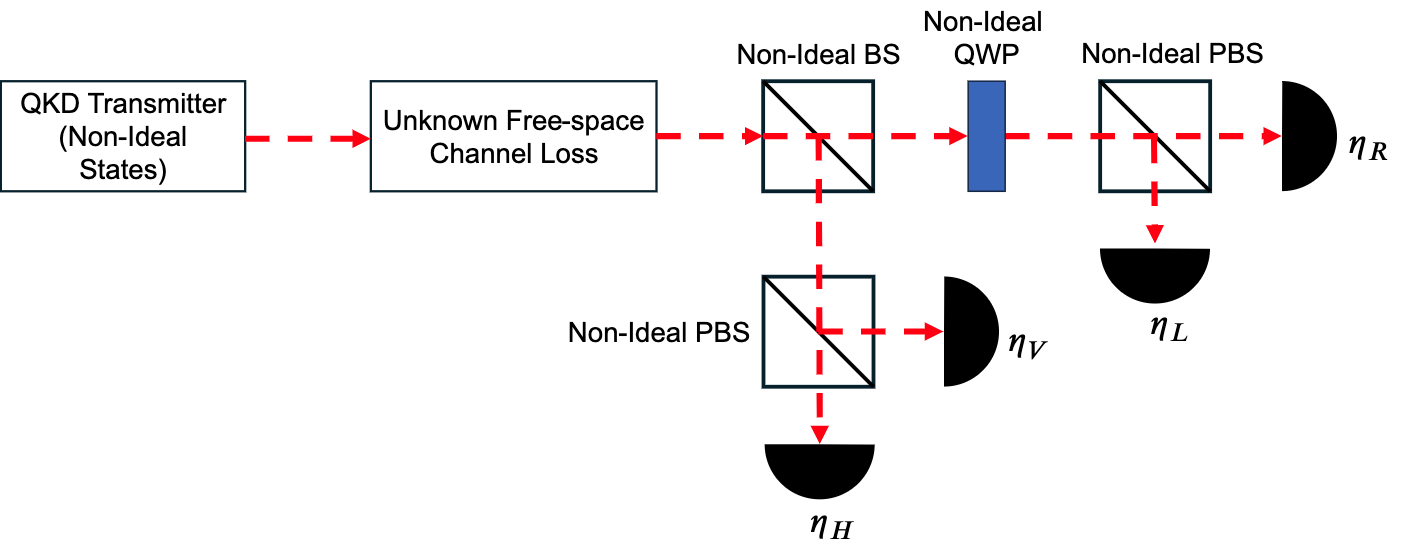} 
    \caption{\label{fig:Non-Ideal_Setup} The QKD receiver setup consists of non-ideal optical components.} 
\end{figure}

An overview of the non-ideal parameters for the QKD model is presented in Fig. \ref{fig:Non-Ideal_Setup} and we discuss their origin here. Ideally, the QKD transmitter outputs perfect states: $\ket{R}, \ket{L}$, and $\ket{H}$; however, the produced states are slightly different due to imperfections, which we characterize using quantum state tomography. The transmitted QKD states are subject to loss, possibly dephasing, and/or a random unitary transformation, as they propagate through the free-space channel. In addition, the receiver optics setup includes a 50:50 beam splitter (BS) to perform random passive basis selection and a quarter waveplate (QWP) to convert polarization.  The BS and QWP have a design wavelength of \unit[670]{nm}, whereas the peak wavelength of the source is at \unit[656.5]{nm} \cite{Daniel_QKD_Source}, which causes the splitting ratio and retardance to be slightly different.  Also, the polarizing beam splitter (PBS) has an imperfect extinction ratio; mostly transmitting $\ket{H}$ while some $\ket{H}$ is reflected, and mostly reflecting  $\ket{V}$, while some $\ket{V}$ is transmitted. Finally, all components have some absorption and reflection from imperfect anti-reflection coatings. 

The receiver couples light from each state into multimode fibers (MMFs) using a $1"$-diameter collimator mounted in a tip/tilt mount. During our alignment process, each collimator is adjusted to be co-linear with the optical line-of-sight to maximize the single-photon collection efficiency. However, the collection efficiency for each collimator is slightly different. Once light is coupled into the MMF, it is routed to a 4-channel single-photon detector, which has slightly different detector efficiencies $\eta_R \neq \eta_L \neq\eta_H \neq\eta_V$. The different efficiencies in our entire setup implies that a single photon in the $\ket{R}$ path, for example, will likely have a different probability of detection in comparison to the other paths. 

\begin{figure} [hbt!]
    \centering 
    \includegraphics[height=4.5cm]{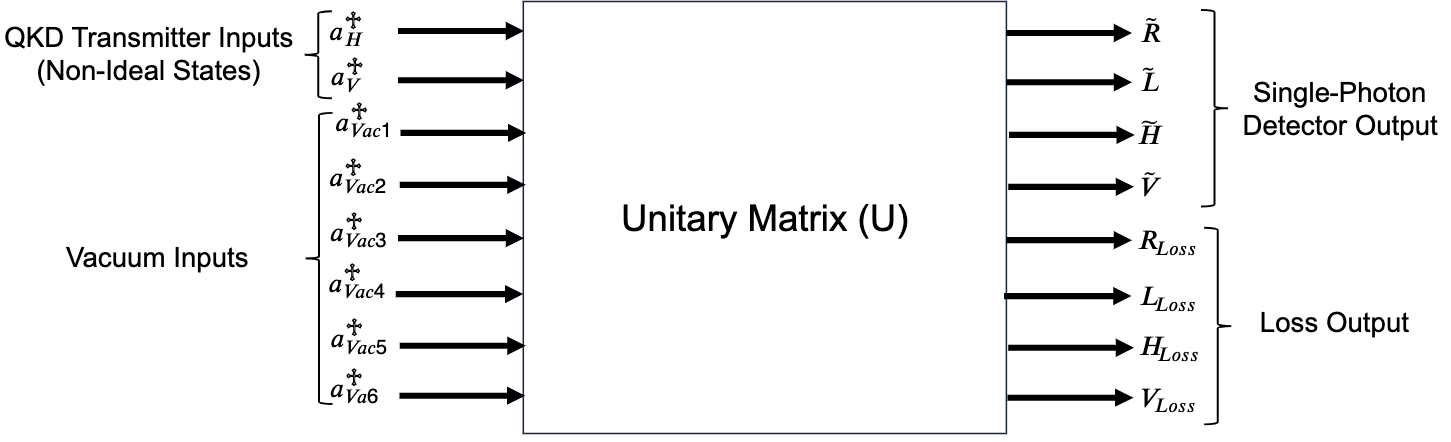} 
    \caption{\label{fig:QKD_Channel_Model} The unitary transformation modeling the entire QKD system.}
\end{figure}

There are many ways to build a model to account for these non-ideal properties. Initially, we attempted to characterize and model each component individually and aggregate the results into the system model. However, each optical component must be characterized separately, and we find that the results change slightly from test to test. Therefore, we opted for an adaptive approach that lumps all non-ideal behaviors into a single large unitary operator $U$, illustrated in Fig. \ref{fig:QKD_Channel_Model}, with the property
\begin{align}
    \label{eq:Unitary_Def}
    U^{\dagger}U = UU^{\dagger} &= \mathbb{I}. 
\end{align} 

We perform quantum state tomography on the transmitted states to begin building the device and channel models. Due to the nature of the finite-key model, we restrict the tomography to pure states, \textit{i.e.}, $\text{Tr}[\rho^2] = 1$, where $\rho$ is the density matrix of the state. We then propagate the non-ideal transmitted state via $U$, using creation operators $a^{\dagger}_H$ and $a^{\dagger}_V$ that operate on the vacuum state $\ket{0}$ to produce $a^{\dagger}_H\ket{0} = \ket{1}_H$ and $a^{\dagger}_V\ket{0} = \ket{1}_V$. The action of $U$ rotates the input single-photon states to one of the four output modes: $\tilde{R}, \tilde{L}, \tilde{H}, \tilde{V}$.

There are four additional output modes representing the effective loss in each channel: $R_{Loss}$, $L_{Loss}$, $H_{Loss}$, $V_{Loss}$. These account for loss anywhere in the system, including in the free-space channel, reflection or absorption in the optics, coupling loss and attenuation of the MMFs, and the non-unit efficiency of the single-photon detectors. Because there are 8 outputs from the transformation, we also need 8 inputs because the operator is unitary, therefore, we include vacuum inputs $a^{\dagger}_{vac}$. 

The device model including the channel is given by
\begin{align}
    \label{eq:Single_Unitary_Model}
    \begin{pmatrix}
            \tilde{R}\\
            \tilde{L}\\
            \tilde{H}\\
            \tilde{V}\\
            R_{\text{Loss}}\\
            L_{\text{Loss}}\\
            H_{\text{Loss}}\\
            V_{\text{Loss}}\\
        \end{pmatrix} &= U\cdot
        \begin{pmatrix}
            a^{\dagger}_H\\
            a^{\dagger}_V\\
            a^{\dagger}_{\text{vac1}}\\
            a^{\dagger}_{\text{vac2}}\\
            a^{\dagger}_{\text{vac3}}\\
            a^{\dagger}_{\text{vac4}}\\
            a^{\dagger}_{\text{vac5}}\\
            a^{\dagger}_{\text{vac6}}\\
        \end{pmatrix} \cdot \ket{0},
\end{align} 
where $U \in \mathbb{C}^{8x8}$, and $\ket{0}$ is the vacuum state. However, because there are six vacuum inputs to $U$, only a sub-matrix of $U$ is relevant to the model \cite{Kamin2024Phys.Rev.Res.}. Therefore, we decompose $U$ as
\begin{align}
    \label{eq:Single_Unitary_Model_with_G}
    U &= \begin{pmatrix}
            G_{8x2} | M_{8x6}
        \end{pmatrix},
\end{align} 
where one sub-matrix is given by
\begin{align}
    \label{eq:Sub-Matrix_G}
    G &= \begin{pmatrix}
            g_1 \; + \; i\cdot g_2, \; g_3 \; + \; i\cdot g_4\\
            g_5 \; + \; i\cdot g_6, \; g_7 \; + \; i\cdot g_8\\
            g_9 \; + \; i\cdot g_{10}, \; g_{11} + i\cdot g_{12}\\
            g_{13} + i\cdot g_{14}, \; g_{15} + i\cdot g_{16}\\
            g_{17} + i\cdot g_{18}, \; g_{19} + i\cdot g_{20}\\
            g_{21} + i\cdot g_{22}, \; g_{23} + i\cdot g_{24}\\
            g_{25} + i\cdot g_{26}, \; g_{27} + i\cdot g_{28}\\
            g_{29} + i\cdot g_{30}, \; g_{31} + i\cdot g_{32}
        \end{pmatrix},
\end{align} 
with
\begin{align}
    \label{eq:Sub-Matrix_G_ID}
    G^{\dagger}G &= \mathbb{I} ~~~~~~~ G \in \mathbb{C}^{8\times2},
\end{align}
and the sub-matrix $M \in \mathbb{C}^{8\times6}$ is unused. The sub-matrix $G$ consists of 16 complex numbers that are specified by 32 real parameters. 

Using this model, we predict the received MPNs given by 
\begin{align}
    \label{eq:gamma_hat_1}
    \hat{\gamma}_{RR} &=  \left|
    \begin{pmatrix}
            1, 0, 0, 0, 0, 0, 0, 0
    \end{pmatrix}
    \cdot G \cdot
    \begin{pmatrix}
            \ket{\psi_1}_H, \ket{\psi_1}_V, 0, 0, 0, 0, 0, 0
    \end{pmatrix}^T
    \right|^2 \cdot \mu_R\\
    \label{eq:gamma_hat_2}
    \hat{\gamma}_{RL} &=  \left|
    \begin{pmatrix}
            0, 1, 0, 0, 0, 0, 0, 0
    \end{pmatrix}
    \cdot G \cdot
    \begin{pmatrix}
            \ket{\psi_1}_H, \ket{\psi_1}_V, 0, 0, 0, 0, 0, 0
    \end{pmatrix}^T
    \right|^2 \cdot \mu_R\\
    \label{eq:gamma_hat_3}
    \hat{\gamma}_{RH} &=  \left|
    \begin{pmatrix}
            0, 0, 1, 0, 0, 0, 0, 0
    \end{pmatrix}
    \cdot G \cdot
    \begin{pmatrix}
            \ket{\psi_1}_H, \ket{\psi_1}_V, 0, 0, 0, 0, 0, 0
    \end{pmatrix}^T
    \right|^2 \cdot \mu_R\\
    \label{eq:gamma_hat_4}
    \hat{\gamma}_{RV} &=  \left|
    \begin{pmatrix}
            0, 0, 0, 1, 0, 0, 0, 0
    \end{pmatrix}
    \cdot G \cdot
    \begin{pmatrix}
            \ket{\psi_1}_H, \ket{\psi_1}_V, 0, 0, 0, 0, 0, 0
    \end{pmatrix}^T
    \right|^2 \cdot \mu_R\\
    \label{eq:gamma_hat_5}
    \hat{\gamma}_{LR} &=  \left|
    \begin{pmatrix}
            1, 0, 0, 0, 0, 0, 0, 0
    \end{pmatrix}
    \cdot G \cdot
    \begin{pmatrix}
            \ket{\psi_2}_H, \ket{\psi_2}_V, 0, 0, 0, 0, 0, 0
    \end{pmatrix}^T
    \right|^2 \cdot \mu_L\\
    \label{eq:gamma_hat_6}
    \hat{\gamma}_{LL} &=  \left|
    \begin{pmatrix}
            0, 1, 0, 0, 0, 0, 0, 0
    \end{pmatrix}
    \cdot G \cdot
    \begin{pmatrix}
            \ket{\psi_2}_H, \ket{\psi_2}_V, 0, 0, 0, 0, 0, 0
    \end{pmatrix}^T
    \right|^2 \cdot \mu_L\\
    \label{eq:gamma_hat_7}
    \hat{\gamma}_{LH} &=  \left|
    \begin{pmatrix}
            0, 0, 1, 0, 0, 0, 0, 0
    \end{pmatrix}
    \cdot G \cdot
    \begin{pmatrix}
            \ket{\psi_2}_H, \ket{\psi_2}_V, 0, 0, 0, 0, 0, 0
    \end{pmatrix}^T
    \right|^2 \cdot \mu_L\\
    \label{eq:gamma_hat_8}
    \hat{\gamma}_{LV} &=  \left|
    \begin{pmatrix}
            0, 0, 0, 1, 0, 0, 0, 0
    \end{pmatrix}
    \cdot G \cdot
    \begin{pmatrix}
            \ket{\psi_2}_H, \ket{\psi_2}_V, 0, 0, 0, 0, 0, 0
    \end{pmatrix}^T
    \right|^2 \cdot \mu_L\\
    \label{eq:gamma_hat_9}
    \hat{\gamma}_{HR} &=  \left|
    \begin{pmatrix}
            1, 0, 0, 0, 0, 0, 0, 0
    \end{pmatrix}
    \cdot G \cdot
    \begin{pmatrix}
            \ket{\psi_3}_H, \ket{\psi_3}_V, 0, 0, 0, 0, 0, 0
    \end{pmatrix}^T
    \right|^2 \cdot \mu_H\\
    \label{eq:gamma_hat_10}
    \hat{\gamma}_{HL} &=  \left|
    \begin{pmatrix}
            0, 1, 0, 0, 0, 0, 0, 0
    \end{pmatrix}
    \cdot G \cdot
    \begin{pmatrix}
            \ket{\psi_3}_H, \ket{\psi_3}_V, 0, 0, 0, 0, 0, 0
    \end{pmatrix}^T
    \right|^2 \cdot \mu_H\\
    \label{eq:gamma_hat_11}
    \hat{\gamma}_{HH} &=  \left|
    \begin{pmatrix}
            0, 0, 1, 0, 0, 0, 0, 0
    \end{pmatrix}
    \cdot G \cdot
    \begin{pmatrix}
            \ket{\psi_3}_H, \ket{\psi_3}_V, 0, 0, 0, 0, 0, 0
    \end{pmatrix}^T
    \right|^2 \cdot \mu_H\\
    \label{eq:gamma_hat_12}
    \hat{\gamma}_{HV} &=  \left|
    \begin{pmatrix}
            0, 0, 0, 1, 0, 0, 0, 0
    \end{pmatrix}
    \cdot G \cdot
    \begin{pmatrix}
            \ket{\psi_3}_H, \ket{\psi_3}_V, 0, 0, 0, 0, 0, 0
    \end{pmatrix}^T
    \right|^2 \cdot \mu_H,
\end{align} 
for all input/output combinations, where $X$ ($Y$) is the transmitted (received) states. Here, $\mu_R$, $\mu_L$, and $\mu_H$ are the transmitted MPN for states $\ket{R}$, $\ket{L}$, and $\ket{H}$, respectively, as measured from the QKD source collected once each day before the first QKD session.

We used the state tomography to find the single-photon component of the coherent states $\ket{\psi_1}$, $\ket{\psi_2}$, $\ket{\psi_3}$. 
Then, for the purpose of this unitary model, we set the complex amplitude of the coherent state equal to this single-photon component. With this, we assumed that the states are effectively attenuated well into the single-photon regime. We find the model parameters using  least-squares optimization using the cost function given by 
\begin{align}
    \label{eq:Objective_Function}
    F &= \text{min} [ 
    |\gamma_{RR} - \hat{\gamma}_{RR}|^2 + 
    |\gamma_{RL} - \hat{\gamma}_{RL}|^2 + 
    |\gamma_{RH} - \hat{\gamma}_{RH}|^2 + 
    |\gamma_{RV} - \hat{\gamma}_{RV}|^2 \\ \nonumber
    & \;\;\;\;\;\;\;\;\,+|\gamma_{LR} - \hat{\gamma}_{LR}|^2 +
    |\gamma_{LL} - \hat{\gamma}_{LL}|^2 + 
    |\gamma_{LH} - \hat{\gamma}_{LH}|^2 + 
    |\gamma_{LV} - \hat{\gamma}_{LV}|^2 \\ \nonumber
    & \;\;\;\;\;\;\;\;\,+|\gamma_{HR} - \hat{\gamma}_{HR}|^2 +
    |\gamma_{HL} - \hat{\gamma}_{HL}|^2 + 
    |\gamma_{HH} - \hat{\gamma}_{HH}|^2 + 
    |\gamma_{HV} - \hat{\gamma}_{HV}|^2 ], \\ \nonumber
\end{align} 
subject to \(G\) being an isometry, Eq.~\ref{eq:Sub-Matrix_G_ID}.

\begin{figure} [hbt!]
    \centering 
    \includegraphics[width=\textwidth]{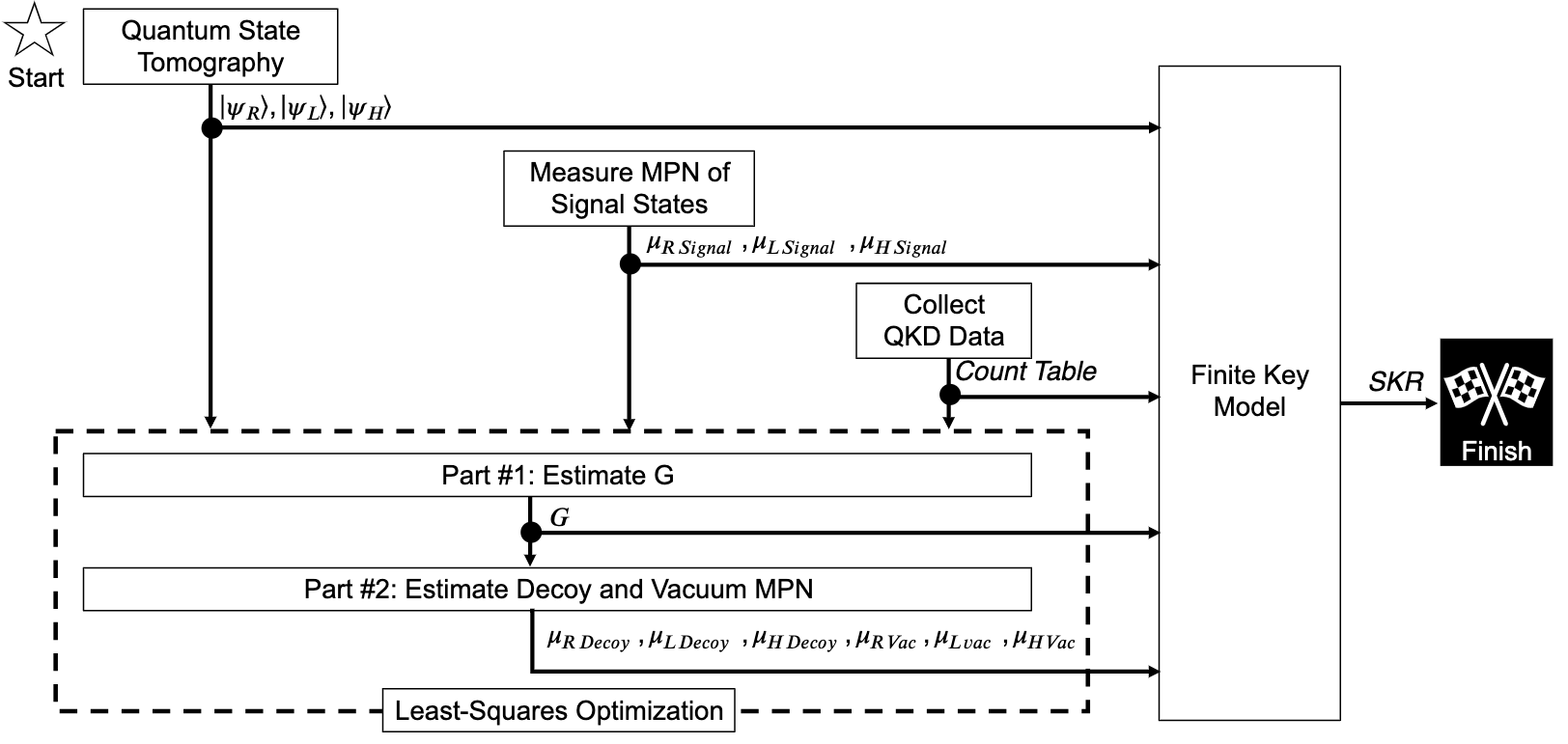} 
    \caption{\label{fig:Least-Squares_Modeling} Illustration of the QKD model and the procedure for estimating the model parameters.}
\end{figure}

An overview of the QKD modeling approach is outlined in Fig.~\ref{fig:Least-Squares_Modeling}. There are three inputs to the least-squares optimizer and the finite-key model: a) quantum state tomography; b) the MPNs of the signal states; and c) the QKD counts. We split the least-squares solver into two parts. Part 1 estimates $G$ using the transmitted states measured by quantum state tomography, the MPN of the signal states, and the QKD counts. The signal states provide higher counts, and hence reduced statistical error, than the decoy or vacuum states, while $G$ is independent of the signal strength. Therefore, we only use the signal counts to estimate $G$. Part 2 estimates the decoy and vacuum MPN using the estimated $G$ from Part 1. The ``vacuum intensity" level setting does not produce single photons from the source, \textit{i.e.}, we do not apply a voltage signal to the resonant-cavity light-emitting diodes. Thus, the received MPN for the vacuum counts is used to estimate the stray-light background and dark counts.  

\begin{figure} [hbt!]
    \centering 
    \includegraphics[height=12cm]{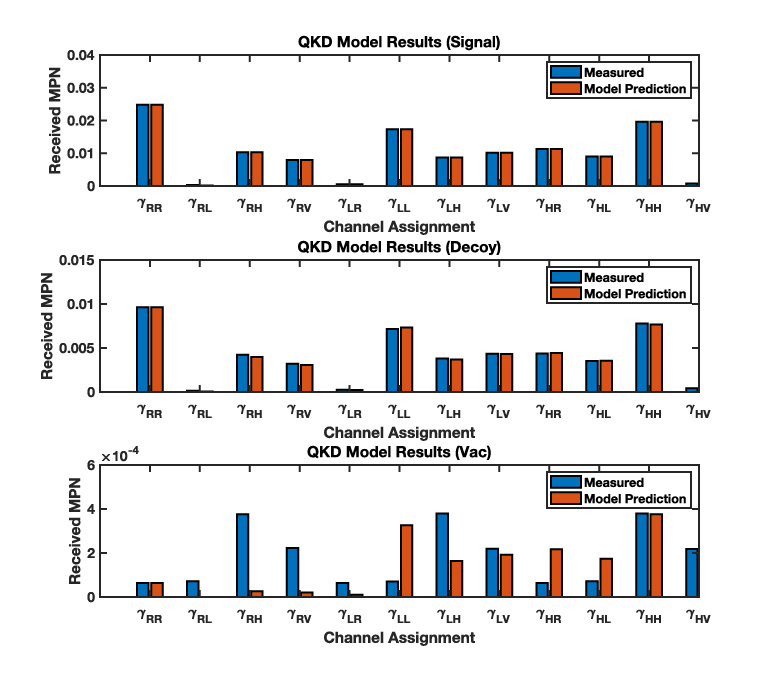} 
    \caption{\label{fig:Model_vs_Prediction} Least-squares solver model prediction compared to the experimental measurements for the Drone-to-Drone QKD data (Flight \#1, Nov 2, 2022).}
\end{figure}

\begin{figure} [hbt!]
    \centering 
    \includegraphics[height=12cm]{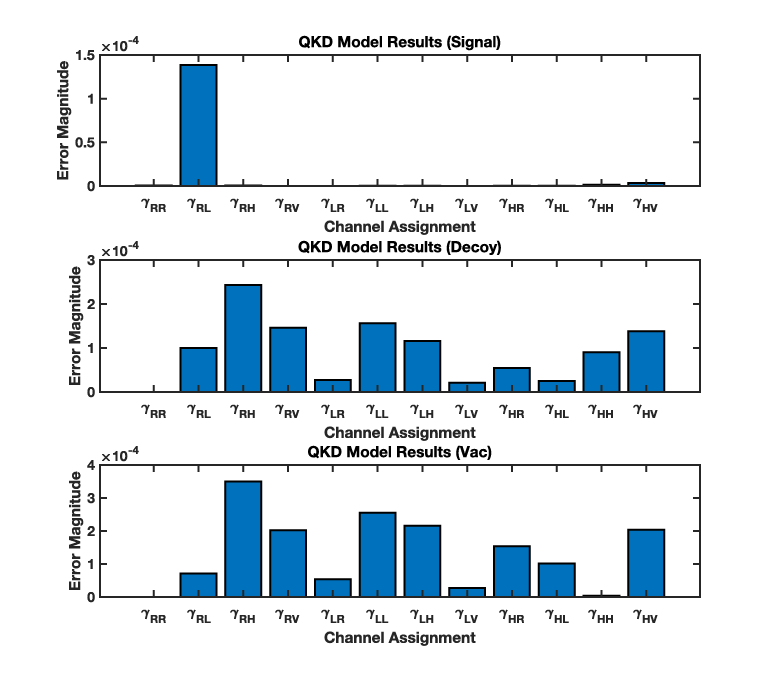} 
    \caption{\label{fig:Residuals} Least-squares solver model error for the Drone-to-Drone QKD data (Flight \#1, Nov 2, 2022).}
\end{figure}

To assess the performance of the least-squares optimization, we compare the estimated received MPN with the measured received MPN; see Fig. \ref{fig:Model_vs_Prediction} for our Drone-to-Drone QKD flight data (Flight \#1, Nov 2, 2022). The model errors for this data set are presented in Fig.~\ref{fig:Residuals}. The model performs well for the signal and decoy intensities.

\section*{Supplementary Note 12. Safety Considerations}
\label{chap:Note_12}

We conduct V2V-QKD highway tests using at least three personnel: one drives each vehicle, and the last manages QKD data collection. The beacon lasers for the PAT system are another safety consideration. First, we do not use visible-light sources, to prevent distracting highway drivers during the night-time experiments. Thus, we remove the 520-nm beacon laser from the TX and change the beacon laser in the RX to one with a wavelength of \unit[850]{nm}. 

We follow standard American National Standard Z136.1 \textit{Laser Safety Standard} \cite{american2022ansi} for the remaining beacon. From Table 5a. Maximum permissible exposure (MPE) for point source ocular exposure to a laser beam, the MPE for a continuous-wave laser in the range \unit[700 -- 1050]{nm} is given by 
\begin{equation}
\mathrm{MPE} = C_A \times 10^{-3}~~~~\mathrm{W/cm}^2
\end{equation}
for an exposure time between 10 and \unit[$3\times 10^4$]{s}, where the longest interval is longer than a key-exchange session.  The wavelength-dependent coefficient correction factor $C_A$ can be found in Figure 8a. Correction factor $C_A$ used to determine the MPE for wavelengths from \unit[0.400 to 1.400]{$\mu$m} and is equal to $\sim$2 at \unit[850]{nm}.  Thus, 
\begin{equation}
\mathrm{MPE} = 2\times 10^{-3}~\mathrm{W/cm}^2 = 2~\mathrm{mW/cm}^2.
\end{equation}

The continuous-wave beacon laser emits a diverging beam, set to reduce the PAT error, and has a power of \unit[1.9]{mW}.  We estimate this corresponds to an apparent beam size of approximately six inches to one foot at the opposite platform. We measure the laser beacon's power as \unit[0.21]{mW}, located 1-car width away from the laser beacon using a free-space power meter (Thorlabs, PN: S120C), which has a sensor diameter of \unit[9.5]{mm} (area = \unit[0.71]{$\text{cm}^2$}).  The corresponding power density is \unit[0.3]{mW/cm$^2$}, which is well below the MPE limit, and therefore our system using NIR beacons is eye-safe for nearby drivers. 

\section*{Supplementary Note 13. Daylight Operations}
\label{chap:Note_13}

\begin{figure} [hbt!]
    \centering 
    \includegraphics[width=\textwidth]{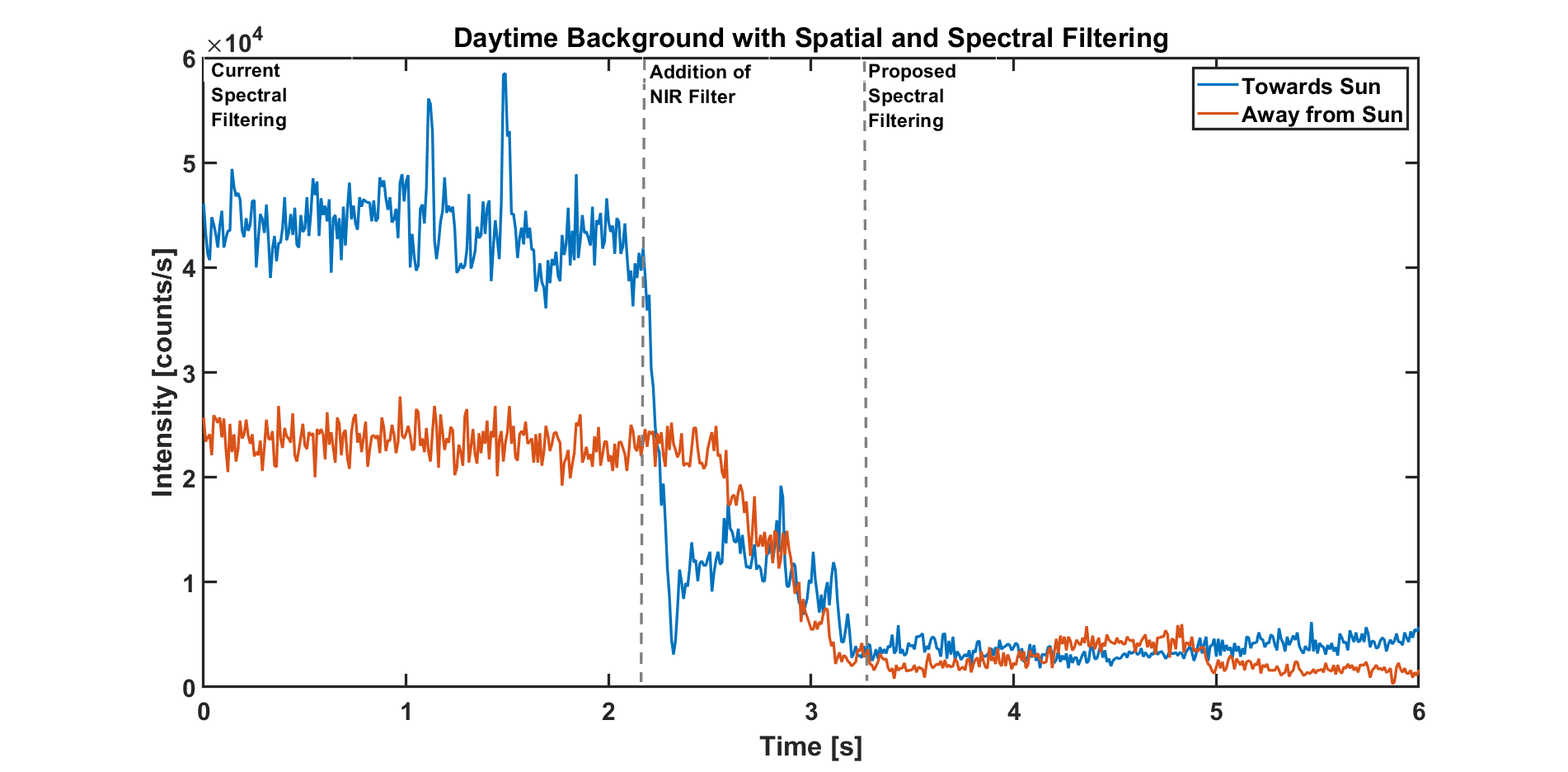} 
    \caption{\label{fig:Background_Counts} Background single-photon counts during daylight using improved spectral and spatial filtering; see text for details.}
\end{figure}

Our current system demonstrates QKD between mobile platforms (drones and vehicles) while operating at night. During daylight without any additional filtering, background counts saturate the single photon detectors, e.g., \unit[$>$2M]{counts/s}. To enable daylight QKD operations, additional spectral and spatial filtering is required to reduce the background counts from the sun. Towards this goal, we make several improvements to our system and measure the background counts of our QKD receiver during daylight in different configurations, as shown in Fig. \ref{fig:Background_Counts}. The first improvement consists of moving from MMF to few-mode fiber (Thorlabs, PN: 780HP), which should have approximately 4 modes at our QKD wavelength of \unit[656]{nm}. Next, we add an NIR spectral filter and measure the background light during daylight looking towards (but not directly at) the sun and away from the sun for the following configurations: a) existing spectral filtering used for night time QKD, and b) addition of a NIR spectral filter (Semrock, PN: NIR01-1550/3-25). With these two modifications, we achieve a maximum background of approximately \unit[6,000]{counts/s} and an average of approximately \unit[3,000]{counts/s}, (reducing the average daylight background counts by \unit[$>$28]{dB}) which is comparable to our background counts during night time QKD operations. Higher performance filters exist (Thorlabs PN: FESH1000), which have a similar attenuation (\unit[$>5$]{OD}) at NIR wavelengths with negligible loss (\unit[$<0.01$]{dB}) at the QKD wavelength. Coupling light into few-mode fibers is more difficult than into our current high Numerical Aperture (NA) MMF; however, it provides a nice middle ground than coupling into single-mode fiber (SMF), which is adversely affected by wavefront distortions. Our current system is designed for coupling light into MMF; however, with several upgrades including improved mode matching and updates to the PAT, we expect to achieve similar coupling performance into few-mode fiber, thus realizing a similar signal-to-noise (SNR) during daylight. Finally, using a 1-ns coincidence timing window \cite{zhang2025daylight} (at an operational repetition rate of 100 MHz) would further reduce the effective noise rate by the factor $10^{-9}\cdot10^8=0.1$ (an additional \unit[10]{dB} reduction to the daylight background counts), bringing the estimated improvement to SNR \unit[$>38$]{dB} during daylight.

\section*{Supplementary Note 14. Drone Power Tether}
\label{chap:Note_14}

LiPo batteries are commonly used to provide flight power in the drone industry. Their high peak discharge current is excellent to provide peak power when needed; however, a significant disadvantage is the limited total energy capacity, which restricts flight times for nearly all drones. To increase the effective flight time, we develop a simple custom drone power tether to provide indefinite flight time to our drone.

\begin{figure} [hbt!]
    \centering 
    \includegraphics[height=8cm]{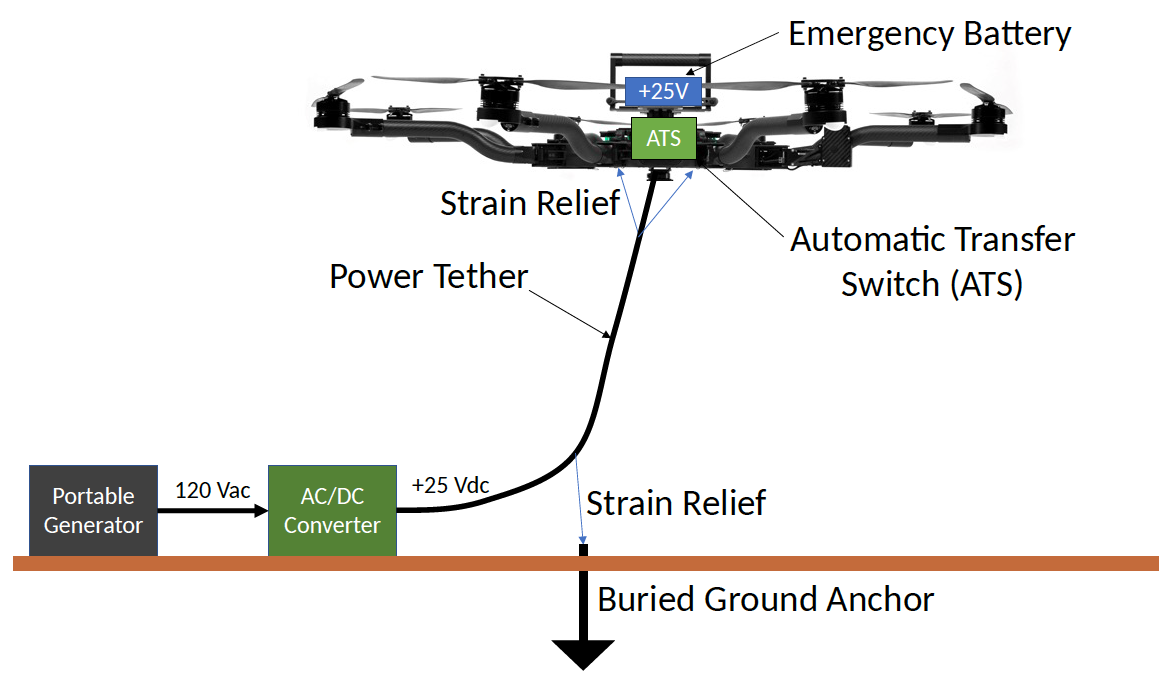}\caption{\label{fig:Drone_Tether} Drone power tether system provides indefinite flight time from a portable generator and has appropriate strain relief, and an emergency battery (and automatic transfer switch) in case the primary power source is disrupted.} 
\end{figure}

The design concept of the power tether is presented in Fig. \ref{fig:Drone_Tether}. A gasoline portable generator produces \unit[$120/240$]{VAC}, which is then converted to \unit[$+25$]{V} using an AC/DC converter. The \unit[$+25$]{V} travels to the drone through a power tether consisting of a 50-ft long 8 AWG copper-clad aluminum cable (GS Power, PN: ZA8AWG50), with a weight of \unit[0.12]{kg/m}, and is secured with 16-inch extension spring strain relief (McMaster-Carr, PN: 9630K75) to a 30-inch buried ground anchor (McMaster-Carr, PN: 6300A26). Additionally, the power tether connects to the drone with 7.5-inch extension springs (McMaster-Carr, PN: 9630K93) for strain relief, and finally a locking carabiner (McMaster-Carr, PN: 3370T51). The strain relief both at the ground and on the drone ensures that the power tether remains connected to the AC/DC converter and to the drone, respectively. The \unit[$+25$]{V} from the power tether enters the drone through a compact Automatic Transfer Switch (ATS), which also receives a power input from an emergency battery on board the drone. The output of the ATS provides uninterrupted \unit[$+25$]{V} to power the drone.

The onboard emergency battery is critical for safe operation of the drone when operating using the power tether. If there is any disruption in the function of the portable generator, AC/DC converter, or if the power tether breaks, then the drone would immediately lose power and fall out of the sky. The emergency battery should ideally be small, but capable of providing necessary power to safely land the drone in the event that the power from the tether is interrupted. 

Based on characterization testing, we select an appropriate AC/DC converter (Mean Well, PN: RST-5000-24) which can provide \unit[$+24$]{VDC} nominal voltage (which we tune to \unit[$+25$]{VDC}), $200$ Amps of current, and \unit[$4.8$]{kW} of power. The RST-5000-24 requires \unit[$240$]{VAC} input voltage. The \unit[$200$]{Amps} is greater than any of the continuous current estimates from the drone's datasheet, and corresponds to $100\%$ full load from the characterization testing. To provide sufficient power to the AC/DC converter, we select a \unit[$6$]{kW} portable generator (PowerStroke, PN: 1000026133), which has \unit[$240$]{VAC} output and has a greater power capacity than is required by the AC/DC converter. For the emergency battery we select a \unit[$1,800$]{mA-hr} LiPo battery (Tattu, PN: TAA18004S95X6) which has a \unit[$95$]{C} discharge rating, nominally \unit[$+25$]{V} at full charge, and capable of providing \unit[$171$]{Amps}. This is sufficient for keeping the drone in the air and immediately landing, if tether power is disrupted. 

An ATS switch is required to manage power from two independent input sources (the power tether and emergency battery) to the drone. Simply placing the emergency battery in parallel with the power tether at the drone is not appropriate, since the power tether could source an (unrestricted) large current to charge the emergency battery, likely damaging it. Additionally, if there is a shorting fault in the power tether, the emergency battery would send a large current backwards towards the power tether potentially causing damage to the AC/DC converter in addition to cutting power to the drone and causing it to crash. The ATS must prevent the power tether and emergency battery from interacting with each other. Moreover, the ATS must switch between the power tether and the emergency battery extremely fast \unit[$(< 1$]{$\mu\text{s})$}, such that the drone remains in the air and the flight controller does not reboot (which would also cause a crash). Finally, the ATS must operate within a minimal size and weight footprint. 

\begin{figure} [hbt!]
    \centering 
    \includegraphics[height=8cm]{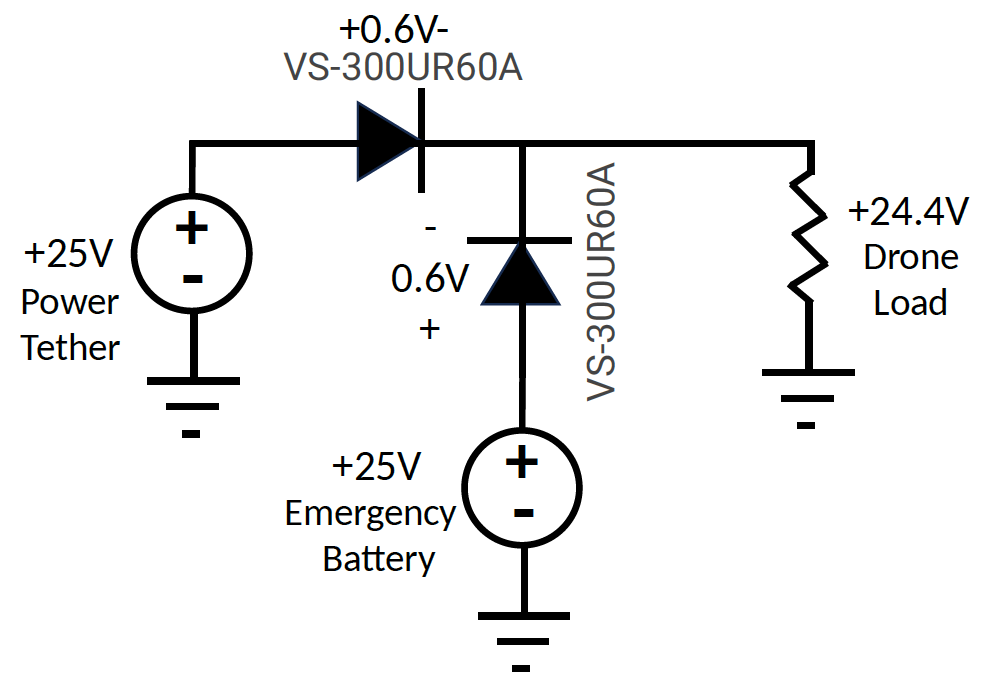}\caption{\label{fig:ATS} The Automatic Transfer Switch (ATS) provides uninterruptible power to the drone while in flight using power from either the power tether \hbox{\emph{or}} the on-board emergency battery.} 
\end{figure}

\begin{figure} [hbt!]
    \centering 
    \includegraphics[height=8cm]{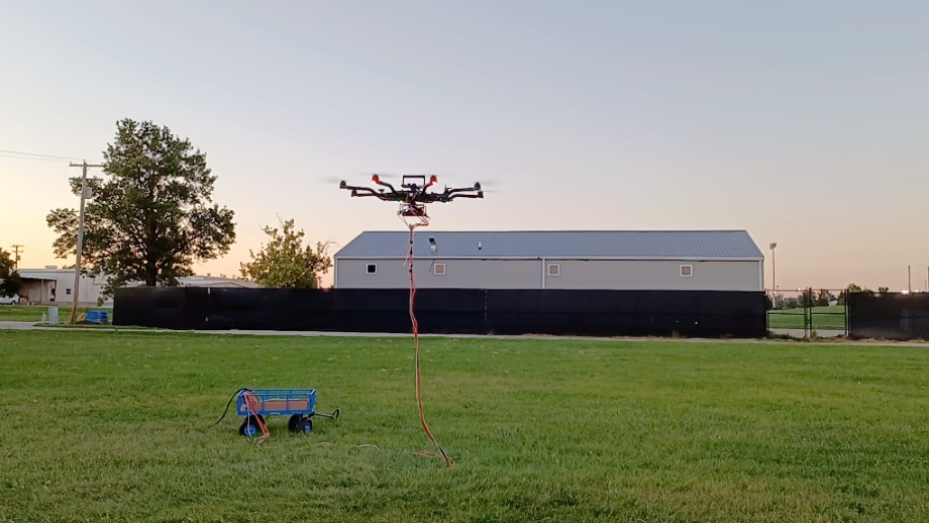}\caption{\label{fig:Full_Power_Tether_Operation} Drone power tether in action in the Drone Test Field, University of Illinois. We placed the AC/DC converter in a cart for easy transport. (Photo courtesy Andrew Conrad.)} 
\end{figure}

The design of our ATS is capable of isolating each input power supply, providing fast switching and handling large current; it consists of only two components, which reduces the size and weight. We design the ATS using two high-current power diodes (Vishay, PN: VS-300UR60A); see Fig \ref{fig:ATS}. These power diodes can safely handle \unit[$300$]{Amps} and have a breakdown voltage of \unit[$600$]{V}. Crucially, these power diodes have standard recovery times \unit[$>500$]{ns}, and are very fast with respect to providing power to the drone. The forward voltage of the power diode is \unit[$0.6$]{V}, providing \unit[$+24.4$]{V} to the drone under normal operation from the power tether. We successfully test our ATS on the ground, \emph{before} flight testing. Specifically, we provide power while the drone was secured on the ground from our drone tether through our ATS with the emergency battery also connected. We cut power from the power tether, and the drone's motors kept flying without interruption, seamlessly switching to the emergency battery. After performing this successful ground testing, we test the drone in flight using the power tether. An image of our full drone tether in action is presented in Fig. \ref{fig:Full_Power_Tether_Operation}. 

The maximum height that we fly our drone using the power tether is approximately 10 meters. As the drone flies higher, it must support a greater weight of the power tether. Our power tether architecture transmits \unit[$+25$]{VDC} (low voltage) to the drone. However, this approach requires using a large gauge wire $8$ AWG to support higher currents. Alternatively, if we transmitted \unit[$240$]{VAC} up to the drone and converted to \unit[$+25$]{VDC} on the drone, we could use a smaller gauge power tether, reducing the weight/length of the cable and increasing the maximum flight height using the tether. The disadvantage to sending \unit[$240$]{VAC} up the power tether, is that an AC/DC converter with high current capacity is required on the drone - increasing the system complexity. Drone power tethers using higher-voltage architectures are commercially available. The primary advantage of our low-voltage approach is simplicity, speed of implementation, and high power (\unit[4.8]{kW}) for our drone-QKD effort.

\section*{Supplementary Note 15. Operation During Relative Motion}
\label{chap:Note_15}

In our work, our system demonstrates QKD between mobile platforms in several configurations: drone-to-drone, drone-to-vehicle, and vehicle-to-vehicle at \unit[5]{mph} and \unit[70]{mph}. During these tests, our QKD systems operate in the presence of platform vibrations from the drones and vehicles, but experience very little relative motion between each other. While multiple quantum link applications can be achieved with lower relative motion (e.g., hovering drones, drones following vehicles, and nearby vehicles traveling similar speeds on the highway), additional use cases can be realized if the systems can sustain quantum links between platforms experiencing higher relative motion (e.g., cars approaching an intersection, intersatellite communication between platforms in different orbits, etc.). Towards addressing this limitation, we collect quantum link data in the vehicle-to-vehicle configuration while the vehicles are experiencing significant relative motion between each other, using our existing system as-is. This data investigates vehicles accelerating with respect to each other reaching relative speeds exceeding \unit[1]{m/s} while maintaining a quantum link at a simulated four-way traffic intersection. 

\begin{figure} [hbt!]

    \centering 
    \includegraphics[height=12cm]{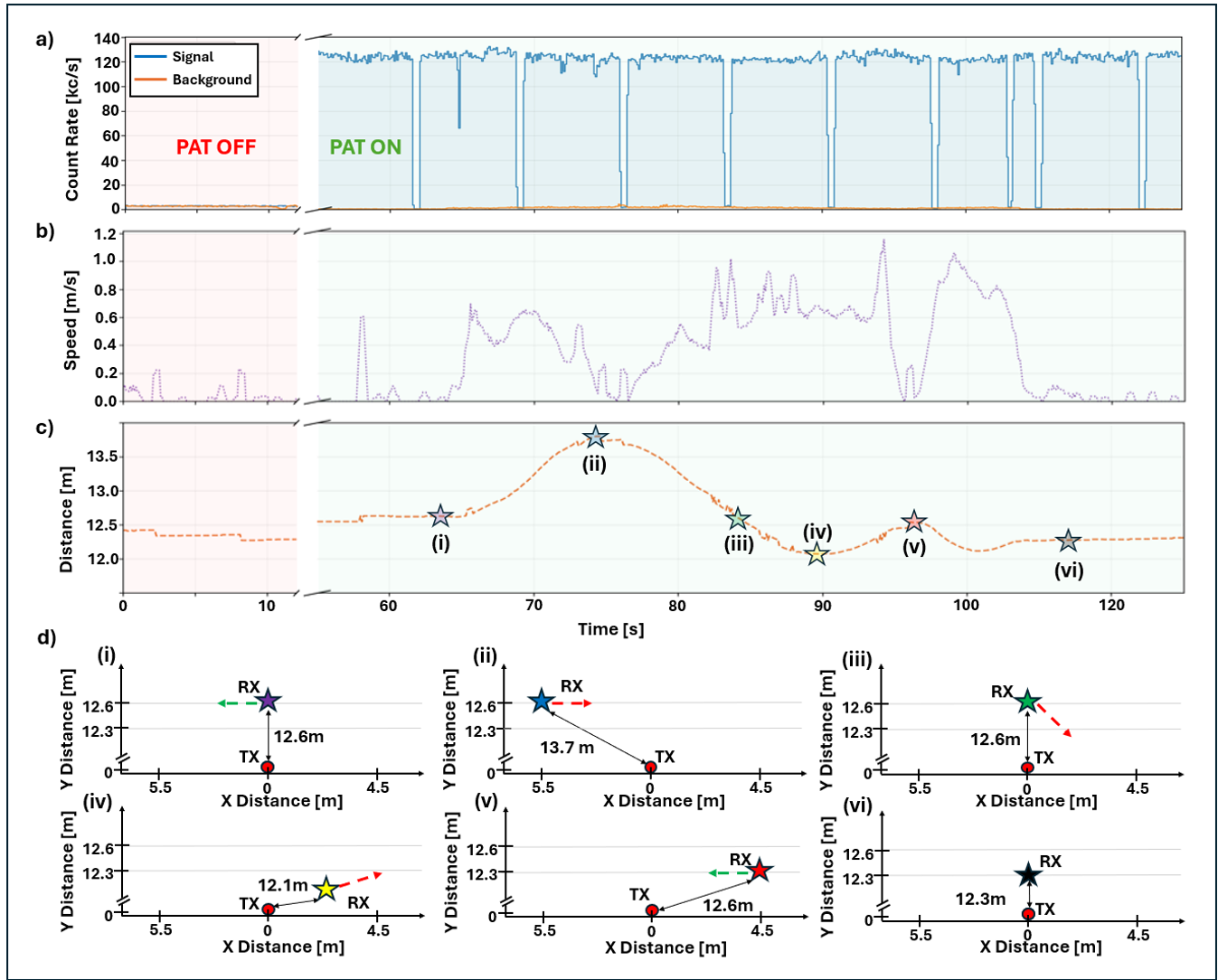}\caption{\label{fig:Rel_Motion} Quantum Link under relative vehicle motion at traffic intersection.
    (a) Single-photon count rates, (b) relative speed between vehicles, (c) relative distance between vehicles, (d)
    relative location of vehicles.} 
\end{figure}

The setup consists of the quantum transmitter located on the top of a vehicle, while the quantum receiver is located in the back seat of a vehicle traveling in an orthogonal direction (on the other street of the intersection). The transmitter vehicle is stationary while the receiver is moving in relative motion at different speeds and distances. We measure the PAT performance in different modes of operations (e.g., PAT controls off, Gimbal Control + Fast Steering Control on), see Fig. \ref{fig:Rel_Motion} (a). We record the relative speed and distances between platforms using GPS receivers in each vehicle, see Fig. \ref{fig:Rel_Motion} (b), and (c), respectively. The testing evolution includes the receiver vehicle moving forwards, backwards, closer and further away from the transmitter vehicle, see Fig. \ref{fig:Rel_Motion} (d).

\end{document}